\documentclass[twocolumn]{aastex701}

\newcommand{\fracb}[2]{\left(\frac{#1}{#2}\right)}
\newcommand{\fracbs}[2]{\left[\frac{#1}{#2}\right]}
\usepackage{amsmath}

\usepackage{multirow}
\usepackage{hyperref}
\usepackage[noabbrev]{cleveref}  
\usepackage{float}  

\usepackage{xcolor}
\definecolor{blazeorange}{rgb}{1.0, 0.4, 0.0}
\definecolor{seagreen}{rgb}{0.18, 0.55, 0.34}
\definecolor{rufous}{rgb}{0.66, 0.11, 0.03}
\definecolor{royalfuchsia}{rgb}{0.79, 0.17, 0.57}
\definecolor{scarlet}{rgb}{1.0, 0.13, 0.0}
\definecolor{royalpurple}{rgb}{0.47, 0.32, 0.66}
\definecolor{darkblue}{rgb}{0, 0, 0.66}

\begin{document}

\title{The Deep Newtonian Regime in Late-Time Blast Waves: Inevitable Transition and Distinct Flux Signatures}

\author[0000-0003-1214-0521]{Sk. Minhajur Rahaman}
\email{rahaman.minhajur93@gmail.com}
\affiliation{Astrophysics Research Center of the Open University (ARCO), The Open University of
Israel,  P.O Box 808,\\ Ra’anana 4353701, Israel}

\author[0000-0001-8530-8941]{Jonathan Granot}
\email{granot@openu.ac.il}
\affiliation{Astrophysics Research Center of the Open University (ARCO), The Open University of
Israel, P.O Box 808,\\ Ra’anana 4353701, Israel}
\affiliation{Astroparticule et Cosmologie (APC,) 
CNRS – Université Paris Cité, F-75013 Paris, France}
\affiliation{Department of Physics, The George Washington University, 725 21st Street NW, Washington, DC 20052, USA}

\author[0000-0001-7833-1043]{Paz Beniamini}
\email{pazb@openu.ac.il}
\affiliation{Astrophysics Research Center of the Open University (ARCO), The Open University of
Israel, P.O Box 808,\\ Ra’anana 4353701, Israel}

\affiliation{Department of Natural Sciences, The Open University of Israel, P.O Box 808, Ra’anana
4353701, Israel}

\affiliation{Department of Physics, The George Washington University, 725 21st Street NW, Washington, DC 20052, USA}


\begin{abstract}
In many astrophysical transients, outflows drive shocks into the ambient medium, accelerating electrons to non-thermal energy distributions that produce broadband synchrotron emission. At late times, even initially collimated relativistic jets evolve into quasi-spherical Newtonian blastwaves. As the shock decelerates, the post-shock internal energy per particle decreases; below a critical velocity $\beta_{\rm DN} \approx 0.2$, only a fraction $\xi_{\rm e} < 1$ of electrons are accelerated to relativistic energies, defining the deep Newtonian (DN) regime. We develop a unified analytic framework for synchrotron emission in this phase, applicable to both single-velocity and stratified ejecta.
For gamma-ray burst afterglows in a uniform medium, the DN transition occurs at $t_{\rm DN} \approx 3.7\,E_{51}^{1/3} n_0^{-1/3}$~yr, yielding a shallower decay by $\delta\alpha = 6(p-2)/5$ relative to standard Newtonian predictions. For kilonova remnants ($E_0 = 10^{50.5}$~erg, $M_{\rm ej} = 0.1\,M_\odot$), the DN phase begins prior to deceleration; neglecting it underestimates radio flux by factors of $\sim 3$--$5$ during coasting and even more thereafter. Magnetar-boosted remnants ($E \sim10^{52}$~erg) should reach $\sim$\,10\,--\,100\,$\mu$Jy at 3~GHz at $\sim$\,40\;Mpc, though limits on GW170817 already disfavor a long-lived millisecond magnetar. In core-collapse supernovae in a wind medium ($\rho\!\propto\!r^{-k}$), the peak luminosity remains constant during coasting, while $\nu_{\rm pk} \propto t^{-1}$; for SN~2023ixf, we find $k = 1.29 \pm 0.14$. The DN 
spectral energy distribution typically satisfies $\nu_m\!<\!\nu_{\rm sa}\!<\!\nu_c$, peaking at sub-GHz frequencies where LOFAR and SKA-low are most sensitive. Even non-detections place robust constraints on ambient density and outflow energetics.
\end{abstract}



\section{Introduction}
Outflows in astrophysical transients including gamma-ray bursts (GRBs), kilonova remnants (KNRs), magnetar giant flares (MGFs), core-collapse supernovae (CCSNe) and superluminous supernovae (SLSNe) drive forward shocks into the ambient medium, accelerating non-thermal electrons and producing broadband synchrotron (afterglow) emission. An ubiquitous late-time feature in these outflows is that the shock eventually decelerates,  becomes Newtonian and approaches spherical symmetry, even if it was initially launched as a highly collimated relativistic jet. In the Newtonian spherical phase, the post-shock internal energy per electron decreases, and when the outflow speed drops below a critical value ($\beta_{\rm DN}$; see Equation~(\ref{eq:beta_DN})) there is enough internal energy to accelerate only a small fraction ($\xi_{\rm e}<1$) of the electrons to relativistic energies. This defines the onset of the \emph{deep Newtonian} (DN) regime. A self-consistent physical treatment must take into account the modifications in the particle energy distribution and the subsequent emission properties in this phase. Although the DN phase has been noted in several works on GRB \citep{Huang03,Sironi13,BGG2022} , MGF \citep{Granot06} , KNR \citep{Kathirgamaraju19,Acharya25} , CCSNe \citep{Duran16}  - in a vast majority of the relevant scientific literature, the DN phase is overlooked, despite its potential to modify both the flux normalization and the late-time rise and decay (see \S \ref{sec:conc}).

Currently, a generalized treatment of the DN regime is lacking. Existing studies are object specific and for simplicity typically assume a single ejecta velocity and a uniform ambient density, whereas theoretical models and simulations indicate stratified ejecta and non-uniform ambient media. The broadband DN spectral evolution is poorly explored. The DN phase during the coasting stage also remains poorly explored, and the self-absorption frequency, $\nu_{\rm sa}$, is generally only approximately estimated. As we demonstrate here, the low-frequency spectral turnover and peak spectral luminosity occur at $\nu_{\rm sa}$, making its accurate determination essential for late-time observational strategies.

In this work, we present a unified framework for the DN regime, applicable to both single-velocity and stratified ejecta interacting with a variety of ambient density profiles. We compute the synchrotron spectrum across all relevant dynamical phases, providing a tool to assess detectability and diagnostic potential. The paper is organized as follows: In \S\ref{sec:DN_desc}, we describe DN shock dynamics and the resulting spectra and light curves. In \S\ref{sec:astro_appl}, we explore the parameter space for transients expected to enter the DN phase. Finally, in \S \ref{sec:conc} , we summarize our main results and discuss observational strategies.

\section{DN regime}\label{sec:DN_desc}
In this section we motivate the DN phase from shock hydrodynamics and explore the temporal effects on the light curves and the spectrum.

We first provide a brief summary of the DN phase and the motivation for the prescription adopted in this work. The DN phase was first introduced phenomenologically by \cite{Huang03} in studies of late-time GRB afterglows, where it was recognized that the commonly adopted electron distribution, $dN/d\gamma_{\rm e}\propto\gamma_{\rm e}^{-p}$, is appropriate for relativistic electrons with $\gamma_{\rm e}\gg1$, but becomes inadequate as $\gamma_{\rm e}$ approaches unity. They therefore introduced the phenomenological prescription $dN/d\gamma_{\rm e}\propto(\gamma_{\rm e}-1)^{-p}$ corresponding to a power-law electron kinetic energy distribution. In the non-relativistic limit, $\gamma_{\rm e}-1\simeq\beta_{\rm e}^{2}/2$, such that this becomes a power law in $\beta_{\rm e}^{2}$. 
Consequently, in this prescription, both the particle number and energy content are dominated by the minimum electron velocity, $\beta_{\rm e,min}\propto\beta_{\rm sh}$, such that the accelerated-particle number fraction $\xi_{\rm e}$ and energy fraction $\epsilon_{\rm e}$ both decrease as the shock velocity $\beta_{\rm sh}$ decreases. This behavior is illustrated in the right panels of Figures~\ref{Fig:DN_distri} and \ref{fig:Distri_eps}, and is further quantified in Figure~\ref{fig:Distri_eps} in Appendix~\ref{app:DN_momenta}. Thus, in the DN regime, both the number of relativistic particles,  and the energy carried by them, declines substantially as the shock decelerates. As discussed below, however, for the prescription adopted in this work, $\epsilon_{\rm e}$ can be treated as constant in the DN phase.

A physically motivated alternative was subsequently proposed by \cite{Granot06} in their study of the radio afterglow of the 27 December 2004 giant flare from the magnetar SGR\;1806$-$20 (see \S \ref{sec:GRB_MGF} for details). They showed that the observed radio emission could be reproduced by assuming that a constant fraction $\epsilon_{\rm e}$ of the injected energy is transferred to accelerated particles characterized by an effective energy $\gamma_m-1\sim1$, while the fraction of accelerated particles, $\xi_{\rm e}$, decreases as the shock becomes increasingly non-relativistic. This prescription is equivalent to a power-law distribution in the accelerated particles' dimensionless momentum, $dN/du_{\rm e}\propto u_{\rm e}^{-p}$ for $u_e\geq u_{\rm e,min}$ where $u_{\rm e}\equiv\gamma_{\rm e}\beta_{\rm e}$. In this case for $u_{\rm e,min}  \ll 1$ the particle number density of particles is dominated by $u_e\sim u_{\rm e,min}$ while the energy density is dominated by $u_e\sim1$ or $\gamma_{\rm dn} \sim 2-3$ (see left panel of Figure \ref{Fig:DN_distri} and Figure~\ref{fig:Distri_eps} in Appendix~\ref{app:DN_momenta}). Such a momentum-space power law is also naturally motivated by the theory of diffusive shock acceleration \citep[e.g.][]{Bell78,BO78,BE87}. In this formulation, the transition to the DN regime is not imposed through an ad hoc modification of the electron energy distribution at sub-relativistic
values of $\gamma_{\rm e}$; instead, the power-law distribution is maintained in momentum space as the characteristic particle energy becomes non-relativistic. \cite{Sironi13} subsequently adopted the same prescription and showed that late-time radio afterglows of GRBs, which are expected to enter the DN phase, are more naturally described by a power-law distribution in momentum rather than in Lorentz factor.

Particle-in-cell (PIC) and hybrid kinetic simulations have been instrumental in uncovering the microphysics of mildly relativistic to non-relativistic collisionless shocks \citep{Spitkovsky08,Caprioli14b,Bykov22}. These simulations have demonstrated that such shocks produce complex particle distributions consisting of a thermal Maxwellian core, a suprathermal population, and a developing non-thermal power-law tail. However, the primary strength of current kinetic simulations lies in describing the \textit{initial stages of particle injection and acceleration}, rather than obtaining fully converged asymptotic particle distributions relevant for late-time astrophysical emission.

The formation of a steady-state non-thermal population is intrinsically a long-timescale process. While kinetic simulations successfully capture shock formation, particle reflection, and suprathermal particle generation, the development of the high-energy tail requires repeated shock crossings and sustained non-linear interaction with self-generated turbulence over many acceleration cycles. Both the particle momentum distribution and the magnetic fields in the shock upstream and downstream have not yet reached a steady state in these simulations, and keep evolving together. Current kinetic simulations typically follow the evolution of the shock for a limited number of ion gyroperiods, $t\sim10^{2}-10^{4}\Omega_{\rm ci}^{-1}$, depending on the numerical method, the dimensionality and the shock parameters. Over such timescales, the maximum particle energy, normalization of the non-thermal tail, and fraction of particles participating in acceleration can continue to evolve beyond the accessible simulation duration.

Moreover, computational limitations often require reduced ion-to-electron mass ratios $m_{\rm p}/m_{\rm e}\ll1836$, which decrease the separation between electron and ion kinetic scales and can modify electron heating and injection physics \citep{Ligorini21a,Ligorini21b,Rassel23}. Furthermore, many kinetic investigations are performed in one or two spatial dimensions (see \citealt{Gupta24} for a detailed discussion), where the restricted magnetic-field geometry can suppress the full three-dimensional turbulent cascade and alter particle scattering. Consequently, the acceleration efficiency and spectral properties obtained in reduced-dimensionality simulations may not fully represent the asymptotic behavior of real astrophysical shocks. Given these limitations, in this work we adopt the prescription developed by \cite{Granot06} and \cite{Sironi13}. If future PIC or hybrid kinetic simulations demonstrate a different asymptotic behavior in the DN regime, the prescription adopted here can be revised accordingly.

For the remainder of this analysis, all primed quantities are defined in the comoving frame. The LF of non-thermal electrons, denoted by $\gamma_\mathrm{e}$, is left unprimed to align with standard notation in the literature, despite representing a comoving quantity. For the rest of the work, we adopt the usual convention $Q_\mathrm{x} = Q/10^{x}$ in cgs units.

\subsection{Microphysical parameters }

We assume the distribution of shock-accelerated non-thermal electrons in the comoving frame follows a power-law of the form 
\begin{equation}
    \frac{dn'}{d\gamma_\mathrm{e}} \propto \gamma_\mathrm{e}^{-p}\quad\ \ \textrm{for}\ \ \gamma_\mathrm{m}<\gamma_e<\gamma_\mathrm{M}\;,
\end{equation}
where $n'$ is the number density, $p$ is the \textit{power-law index}, bounded between a minimum LF $\gamma_\mathrm{m}$ and a maximum LF $\gamma_\mathrm{M}$. These non-thermal electrons constitute a fraction $\xi_\mathrm{e}$ of the total number of electrons and carry a fraction $\epsilon_\mathrm{e}$ of the comoving internal energy density $e'_{\rm int}$. 

The average LF of this distribution is given by
\begin{equation}
    \langle \gamma_\mathrm{e} \rangle \equiv \frac{ \int^{\gamma_\mathrm{M}}_{\gamma_\mathrm{m}} \gamma_\mathrm{e} \, \frac{dn'}{d\gamma_\mathrm{e}} \, d\gamma_\mathrm{e}}{ \int^{\gamma_\mathrm{M}}_{\gamma_\mathrm{m}} \frac{dn'}{d\gamma_\mathrm{e}} \, d\gamma_\mathrm{e}} = \frac{\gamma_\mathrm{m}}{G(p)}\;,
    \label{avg_LF}
\end{equation}
where the dimensionless function $G(p)$ captures the ratio $\gamma_\mathrm{m} / \langle \gamma_\mathrm{e} \rangle$ and is defined as
\begin{equation}
\begin{split}
    G(p) \equiv \frac{\gamma_\mathrm{m}}{\langle \gamma_\mathrm{e} \rangle} &= \frac{(p-2)}{(p-1)} \left[ \frac{1 - (\gamma_\mathrm{M}/\gamma_\mathrm{m})^{1-p}}{1 - (\gamma_\mathrm{M}/\gamma_\mathrm{m})^{2-p}} \right] \\
    &\approx \frac{p - 2}{p - 1}, \quad \text{for } p > 2 \text{ and } \gamma_\mathrm{M} \rightarrow \infty\;.
\end{split}
\end{equation}

The LF $\gamma_\mathrm{m}$ is determined by requiring that the kinetic energy of the non-thermal electrons equals $\epsilon_\mathrm{e} e'_{\rm int}$. This yields
\begin{equation}
    \frac{\gamma_\mathrm{m}}{G(p)} - 1 = \epsilon_\mathrm{e} \xi_\mathrm{e}^{-1} \left( \frac{m_\mathrm{p}}{m_\mathrm{e}} \right)(\Gamma_{\rm ud} - 1)\;,
    \label{min_LF}
\end{equation}
where $\Gamma_{\rm ud}$ is the relative LF between the upstream and downstream fluids. A straightforward way to see this is to note that a fraction $\epsilon_{\rm e}$ of the mean internal energy per proton, $(\Gamma_{\rm ud}-1)m_{\rm p}c^{2}$, is transferred to $\xi_{\rm e}$ power-law electrons with a mean internal energy $(\langle \gamma_{\rm e} \rangle - 1)m_{\rm e}c^{2}$. Assuming $\gamma_\mathrm{m} \gg 1$, the expression simplifies to
\begin{equation}
    \gamma_\mathrm{m} \approx 1836 \, G(p) \, \epsilon_\mathrm{e} \, \xi^{-1}_{\mathrm{e0}} (\Gamma_{\rm ud} - 1)\;. \label{eq:gamma_m_UR}
\end{equation}

As the shock propagates into the ambient medium, the available internal energy per baryon  $\Gamma_{\rm ud} - 1$ gradually decreases (in the Newtonian limit, $\Gamma_{\rm ud} - 1 \approx \frac{1}{2} \beta^2_{\rm ud} \ll 1$ where $\beta_{\rm ud}$ is the relative upstream to downstream normalized velocity). This gives
\begin{equation}
    \gamma_\mathrm{m} \approx 918 \, G(p) \, \epsilon_\mathrm{e} \, \xi^{-1}_{\mathrm{e0}} \beta^2_{\rm ud}\;. \label{eq:gamma_m_NR}
\end{equation}

For an initially constant $\xi_\mathrm{e}=\xi_\mathrm{e0}$, equation~(\ref{eq:gamma_m_NR}) shows that this would eventually lead to $\gamma_{\rm m}-1<1$, suggesting that an increasing fraction of particles becomes nonrelativistic. Indeed, the DN regime is established when $\gamma_\mathrm{m} = \gamma_\mathrm{dn} = \sqrt{2}$, a fixed fiducial value corresponding to mildly relativistic electrons (which corresponds to $u_e=\gamma_e\beta_e = 1$). For the rest of the analysis, we assume a fixed value of $\gamma_{\rm dn} = \sqrt{2}$. Beyond this point for $p>2$, it is the accelerated electron fraction, $\xi_\mathrm{e}$, rather than $\gamma_\mathrm{m}$ that must evolve. Subscripts in lowercase letters, dn, refer to the properties of the particle distribution, whereas uppercase letters, DN, denote the properties of the bulk outflow. The proper speed of the particle distribution $u_{\rm e} \equiv \gamma_{\rm e} \beta_{\rm e}$ has a subscript `e' while the proper speed of the outflow $u \equiv \Gamma \beta$ is without this subscript.

In the DN regime, we have
\begin{equation}
\begin{split}
    \frac{ \gamma_\mathrm{dn}}{G(p)} -1 &\approx \frac{1}{2} \beta^2_{\rm ud} \left( \frac{m_\mathrm{p}}{m_\mathrm{e}} \right) \epsilon_\mathrm{e} \xi^{-1}, \\
    \Rightarrow  \xi  &\ \approx 2.8 \times 10^{-3} \;\epsilon_\mathrm{e,-1}^{-1} \tilde{f}(p)\left( \frac{\beta_{\rm ud}}{10^{-2}} \right)^2\;, 
\end{split} 
\end{equation}
where $f(p)\equiv G(p)/[\gamma_\mathrm{dn}-G(p)]\approx (p-2)/[(\sqrt{2}-1)(p+\sqrt{2})]$ for $p>2$ and $\gamma_\mathrm{M}\rightarrow\infty$, and $\tilde{f}(p)\equiv f(p)/f(2.5)$.

For a fixed $\epsilon_\mathrm{e}$, the scaling $\xi \propto \beta^2_{\rm ud}$ arises because the shock strength $(\Gamma_{\rm ud} - 1) \approx \frac{1}{2} \beta^2_{\rm ud}$ sets the internal energy per baryon. As the shock weakens, the available energy per electron diminishes. To maintain a power-law distribution with an average LF $\langle \gamma_\mathrm{e} \rangle \sim 3\gamma_{\mathrm{dn}}$ (e.g., for $p = 2.5$), the energy must be concentrated in a decreasing subset of the electron population, implying a decreasing $\xi_{\mathrm{dn}}$. The critical bulk proper speed $u_{\rm DN}$ at the DN transition can be defined by equating $\xi_{\mathrm{dn}} = \xi_{\mathrm{e0}}$, leading to
\begin{equation}\label{eq:beta_DN}
\begin{split}
  u_{\mathrm{DN}} &\ \approx \beta_\mathrm{DN} \approx  0.19 \; \tilde{f}(p)^{-\frac{1}{2}} \; \xi_{\rm e0}^{\frac{1}{2}} \ \epsilon_{\rm e,-1}^{-\frac{1}{2}}\;.   
\end{split}
\end{equation}

For simplicity, throughout this work we assume $\xi_\mathrm{e0} = 1$. Thus, the scaling of $\xi_{\mathrm{e}}$ across the entire dynamical range is given by
\begin{equation}\label{eq:xi_e}
    \frac{\xi_{\mathrm{e}}}{\xi_{\mathrm{e0}}} = \min \left( 1, \frac{\beta_{\rm ud}^2}{\beta^2_{\rm DN}} \right) =
    \begin{cases}
     \ 1\;, & \text{for } \beta_{\rm ud} \geq \beta_{\rm DN}\;, \\
     \ \frac{\beta_{\rm ud}^2}{\beta^2_{\rm DN}}\;, & \text{for } \beta_{\rm ud} \leq \beta_{\rm DN}\;.
    \end{cases}
\end{equation}

In this work, we consider a forward shock propagating into a stationary upstream medium. From this point onward, we therefore drop the subscript $ud$ from the outflow velocity.

\subsection{Adiabatic blastwave evolution}
\label{sec:adiabaticevolution}
\begin{figure*}
     \includegraphics[scale=0.45]{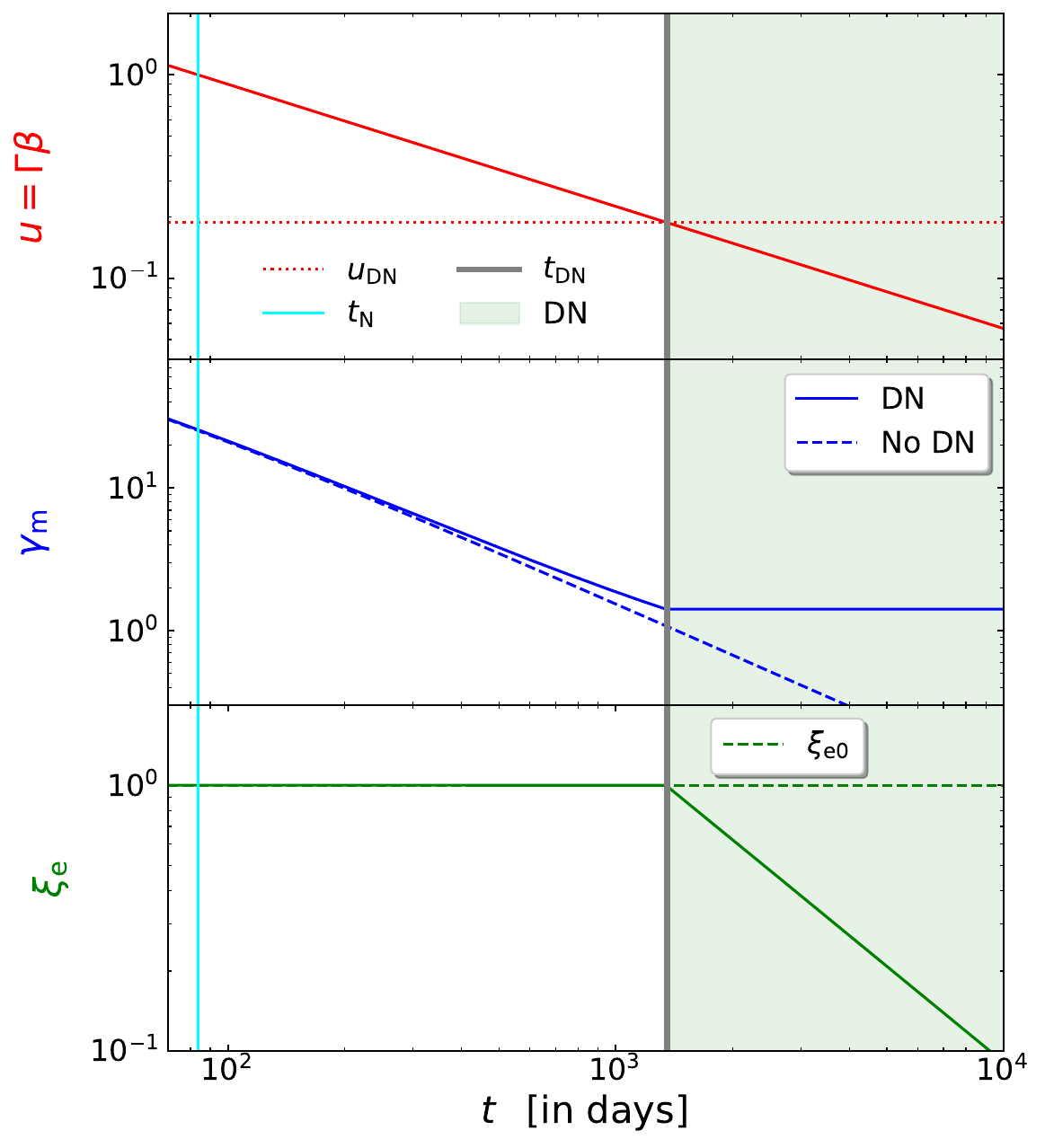}\hspace{0.3cm}    
\includegraphics[scale=0.45]{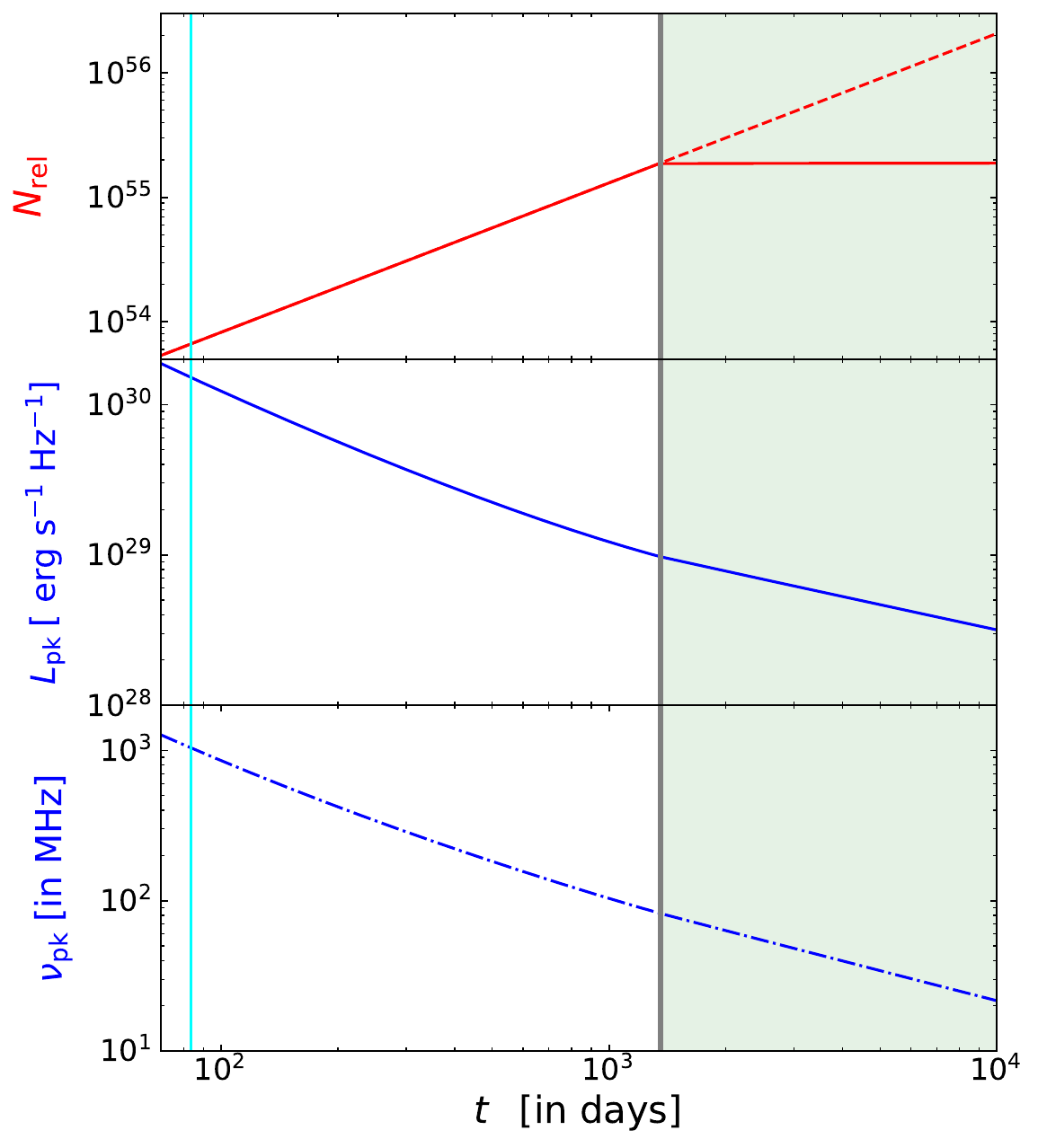}  
    \caption{
\textbf{Illustration of the temporal evolution of the hydrodynamics, particle microphysics, and synchrotron spectrum in the DN regime for a spherical blast wave expanding into a constant-density medium} for an initially relativistic outflow ($u_0\gg1\Leftrightarrow t_{\rm dec}\ll t_{\rm N}$). The assumed parameters are $E_{51}=1$, $n_{0}=1$ ($k=0$), $\epsilon_{\rm e}=0.1$, $p=2.5$, $\epsilon_{\rm B}=0.01$, and $\xi_{\rm e0}=1$. In both panels, the vertical cyan and gray lines denote the Newtonian transition time (Equation~(\ref{eq:t_n})) and the DN transition time (Equation~(\ref{eq:t_dn})), respectively, while the shaded green region marks the DN regime. 
\textit{Left panel:} The top subpanel shows the normalized outflow proper speed $u$ (solid red); the horizontal dashed red line indicates the critical proper speed below which the flow enters the DN regime. The middle subpanel shows the minimum electron Lorentz factor $\gamma_{\rm m}$  in solid blue with DN correction while the dashed blue shows the extrapolation of Equation~(\ref{eq:gamma_m_UR}) to Newtonian bulk speed. The bottom subpanel shows the particle acceleration efficiency $\xi_{\rm e}$ (solid green).
\textit{Right panel:} The top subpanel shows the number of relativistic electrons contributing to the emission (solid red; Equation~(\ref{eq:rel_elec})), while the total number of shocked electrons is shown by the dashed red curve (Equation~(\ref{eq:tot_elec})). The middle and bottom subpanels show the peak luminosity (solid blue; Equation~(\ref{eq:L_pk})) and corresponding peak frequency (dotted blue; Equation~(\ref{eq:nu_pk})), respectively.}\label{fig:DN_HydroSpec}    \end{figure*}

For illustration purposes, we consider the adiabatic evolution (Appendix \ref{app:rad_phase} gives the conditions under which adiabatic evolution holds) of a spherical forward blast wave due to a single velocity ejecta launched with $u_{0}$ propagating into a cold, stationary external medium characterized by a radial density profile $\rho(R) = A R^{-k}$, where $k = 0$ and $2$ corresponds to a uniform medium and a stellar wind profile respectively. We will consider synchrotron emission from the shocked fluid. 

The dynamics of the shocked fluid are described in terms of the (normalized) proper speed $u \equiv \Gamma \beta$, where $\Gamma$ is the bulk Lorentz factor and $\beta$ is the dimensionless velocity in units of $c$. We assume a spherically symmetric blast wave, with a true kinetic energy $E$ equal to the isotropic-equivalent kinetic energy $E_{\rm iso}$. In realistic systems such as GRBs the ultra-relativistic ejecta with $u_{0} \gg 1$ are initially \emph{collimated}, and during this phase the deceleration time is set by the isotropic-equivalent energy $E_{\rm iso}$ rather than the true energy $E$. For narrow jets $E_\mathrm{iso}$ can substantially exceed $E$. For 
such collimated outflows, typically within a few dynamical timescales after jet-break, the jet undergoes lateral expansion and gradually transitions toward a quasi-isotropic flow as it becomes sub-relativistic \citep[e.g.][]{Panaitescu&Meszaros1999,Panaitescu&Kumar00,Granot07,Granot_Ramirez-Ruiz_2012}. The detailed physics governing this transition are complex and model dependent; however, once the outflow becomes non-relativistic, it is already very close to spherical symmetry, and the subsequent hydrodynamics are determined by the true explosion energy $E$ \citep{GRL05,Granot12,DeColle12,Duffell18}. Moreover, beaming corrections to the radiation are at most mild at this stage, such that the we are sensitive to emission from the entire shocked region.

For adiabatic evolution (negligible energy gains or radiative losses), the explosion energy $E$ is approximately constant 
and both the Blandford-McKee and Sedov-Taylor phase can be summarized succinctly as: 
\begin{equation}\label{eq:E(u,R)}
    M(<\!R)\, c^2\, u^2(R) = C_{u}^2 E\;. 
\end{equation}
 Equation~(\ref{eq:E(u,R)}) captures the correct scaling of the blast-wave evolution, with the factor $C_{u}$ accounting for the normalization required to recover the exact Blandford--McKee and Sedov--Taylor self-similar solutions. Since these correction factors are of order unity and primarily affect the normalization, we neglect them throughout most of the text for simplicity. The corrected velocity and radius can be obtained by applying the corresponding factors $C_{\rm u}$ and $C_{\rm R}$
 (see Appendix~\ref{app:blastwave_correction} for details).

The swept-up mass depends on the stratification of the external medium and (for $k < 3$) can be expressed as
\begin{equation}
    M(<\!R) = \int_0^R 4\pi r^2 \rho(r)\,dr = \frac{4\pi A}{3 - k} R^{3 - k}\;,
\end{equation}
which implies
\begin{equation}
u(R) = \sqrt{\frac{(3-k)E}{4\pi Ac^2R^{3-k}}}=u_0\,\min\left[1,\fracb{R}{R_{\rm dec}}^{-\frac{3-k}{2}}\right]\;,
\end{equation}
where $R_{\rm dec}$ is the deceleration radius, given by $u(R_{\rm dec})=u_0$, i.e.
\begin{equation}
R_{\rm dec} = \fracbs{(3-k)E}{4\pi Ac^2 u_0^2}^{\frac{1}{3-k}}\;.
\end{equation}

For non-relativistic sources like a supernova (SN), the circumstellar medium is wind-like up to a wind termination radius $R_{\rm TS} \lesssim 0.3$ pc \citep{Moriya14,Matsuoka25} beyond which the density can be assumed to be constant. The proper speed at launch, the ratio of the launch speed to the critical DN proper speed, and the deceleration radius for non-relativistic sources ($u_{0} < u_{\rm DN} \ll 1$) can be expressed as
\begin{align}
 &\ u_{0} = \frac{1}{c} \sqrt{\frac{2 E}{M_{\rm ej}} }  \approx 1.93 \times 10^{-2} \; E_{51}^{\frac{1}{2}}  \; M_{3}^{-\frac{1}{2}}\;, \\
    &\ \frac{u_{0}}{u_{\rm DN}}   \approx 0.1 \; \tilde{f}(p)^{\frac{1}{2}} \; E_{51}^{\frac{1}{2}}  \; M_{3}^{-\frac{1}{2}}\;    \xi_{\rm e0}^{-\frac{1}{2}} \ \epsilon_{\rm e,-1}^{\frac{1}{2}}\;,\label{eq:uratio_SNR} \\ 
&\ R_\mathrm{dec}  = \left[ \frac{3 M_{\rm ej}}{8 \pi n m_{\rm p}} \right]^{\frac{1}{3}}  \approx  2.4  \; M_{3}^{\frac{1}{3}} \, n_{0}^{-\frac{1}{3}} \, \text{pc}\;,\label{eq:coast} 
\end{align} 
where $M_{3} = M_{\rm ej}/ 3 M_{\odot}$.

For $u \leq 1$, the observed time of emission is given as (see Appendix~\ref{app:t-R} for physical motivation)
\begin{equation}
\begin{split}
    t \approx \frac{2}{5-k} \frac{R}{u c}\;.
\end{split}
\end{equation}

At the Newtonian transition time $t_\mathrm{N}$ we have $u=1$
and, as stated before, the outflow 
approaches spherical symmetry at this stage. The Newtonian transition occurs at a radius $R_N = R_{\rm dec}(u_0\!=\!1)$ corresponding to an observed time
\begin{equation}\label{eq:t_n}
\begin{split}
   t_\mathrm{N} &\ \equiv \frac{2}{5-k}\frac{R_{\rm N}}{ c}   \approx \begin{cases}
      84 \; E_{51}^{\frac{1}{3}}\; n_{0}^{-\frac{1}{3}}  \; \text{days} \;   \hspace{0.5cm}        (k=0)\;, \vspace{0.2cm}\\          
      46\; E_{51} \; A_{\star}^{-1} \; \text{days} \; \hspace{0.7cm}          (k=2)\;.  
    \end{cases} \\ 
\end{split}
\end{equation}

Thus, $u$ evolves in the adiabatic phase as 
\begin{equation}
    u(t) = \fracbs{R(t)}{R_{\rm N}}^{-\frac{3-k}{2}} =
    \begin{cases}
        \fracbs{4(4-k)t}{(5-k)^2 t_\mathrm{N}}^{-\frac{(3-k)}{2(4-k)}} 
        & t \ll t_\mathrm{N},  \vspace{0.2cm}
        \\
        \left( \frac{t}{t_\mathrm{N}} \right)^{-\frac{(3-k)}{(5-k)}} & t \geq t_\mathrm{N}.
    \end{cases}
\end{equation}

The DN transition time is (for $u_{\rm 0} > u_{\rm DN}$)
\begin{equation}\label{eq:t_dn}
\begin{split}
    &\ t_{\rm DN}  \equiv  \frac{2}{5-k} \frac{R_{\rm DN}}{u_{\rm DN} c}
    \\ 
    & \quad\quad\approx   \begin{cases}
        3.7 \; E_{51}^{\frac{1}{3}}\; n_{0}^{-\frac{1}{3}} \;  \tilde{f}(p)^{\frac{5}{6}} \; \xi_{\rm e0}^{- \frac{5}{6}} \; \epsilon_{\rm e,-1}^{\frac{5}{6}}   \;\;  \text{yr} \hspace{0.3cm}  (k=0),  \\ 
        18.7 \; E_{51}\; A_{\star}^{-1} \; \tilde{f}(p)^{\frac{3}{2}} \; \xi_{\rm e0}^{- \frac{3}{2}} \; \epsilon_{\rm e,-1}^{\frac{3}{2}}  \; \; \text{yr} \;  \hspace{0.2cm}(k=2). \\  
     \end{cases}
\end{split}
\end{equation}

For the ordering  $u_\mathrm{0} > 1 > u_\mathrm{DN}$ we can define the ratio of the Newtonian to the  DN transition time as 
\begin{equation} \label{eq:tdn_tn}
\begin{split}
     \frac{t_{\rm DN}}{t_{\rm N}} &\ = u_{\rm DN}^{-\frac{5-k}{3-k}} = \left[ \frac{f(p)}{2 }  \frac{m_{\rm p}}{m_{\rm e}}   \frac{\epsilon_{\rm e}}{\xi_{e0}} \right]^{\frac{5 - k}{2(3 - k)}} \\ &\  \approx \begin{cases}
      15.7 \;   \tilde{f}(p)^{\frac{5}{6}} \; \xi_{\rm e0}^{-\frac{5}{6}} \; \epsilon_{\rm e,-1}^{\frac{5}{6}}  \hspace{0.5cm}   (k=0)\;,   \vspace{0.2cm}\\
      143  \;  \tilde{f}(p)^{\frac{3}{2}} \; \xi_{\rm e0}^{-\frac{3}{2}} \; \epsilon_{\rm e,-1}^{\frac{3}{2}}  \hspace{0.5cm}     (k=2)\;. \\ 
     \end{cases}
\end{split}
\end{equation}

Equation~(\ref{eq:tdn_tn}) shows that measuring the ratio of $t_\mathrm{DN}$ to $t_\mathrm{N}$  directly constrains $\beta_{\rm DN}$ and the microphysical parameters. This can in principle be measured in GRB afterglows (see \S \ref{sec:GRB_MGF}).

The swept-up mass of the ambient medium required to transition to the DN regime is
\begin{equation}
    \begin{split}
     M_\mathrm{DN} &\ \equiv M(<\! R_\mathrm{DN}) = \frac{E}{u_{\rm DN}^2 c^2} = \frac{E}{2 c^2}  \frac{1}{f(p)} \frac{\xi_{\rm e0}}{\epsilon_{\rm e}} \frac{m_{\rm e}}{m_{\rm p}} \\
     &\ \approx 1.5 \times  10^{-2}\;  E_{51} \;  \tilde{f}(p)^{-1}  \; \xi_{\rm e0}\; \epsilon_{\rm e,-1}^{-1} \; M_{\odot}\;,
    \end{split}
\end{equation}
which is independent of the stratification index of the ambient medium $k$. Physically, this means that the DN regime is realized once the swept-up mass of the ambient medium exceeds a critical mass $M_{\rm DN}$. An outflow decelerates when the swept-up ambient mass becomes comparable to the mass of the ejecta. Therefore, if the ejecta mass satisfies $M_{\rm ej} > M_{\rm DN}$, the outflow enters the DN regime before the deceleration time, i.e., at times $t < t_{\rm dec}$ which is the coasting phase if the outflow is launched at a single velocity. This is typically the case for supernova remnants, whose ejecta masses are on the order of a few $M_{\odot}$. Kilonova remnants, with characteristic ejecta masses $M_{\rm KNR} \sim 0.1\,M_{\odot}$, present a particularly interesting scenario. For outflows with $E_{51} \lesssim 1$, the KNR is already in the DN phase before $t_{\rm dec}$. However, for a magnetar-boosted kilonova remnant with $E_{51} \sim 10$, one obtains $M_{\rm DN} \sim 0.1\,M_{\odot}$, implying that the DN regime is reached just after deceleration, i.e., at $t_{\rm DN} \gtrsim t_{\rm dec}$ (see \S \ref{sec:KNR}). In contrast, 
GRB jets and magnetar giant-flare outflows have ejecta masses far below $M_{\rm DN}$, so their transition to the DN regime occurs only after deceleration, at times $t > t_{\rm dec}$ (see \S \ref{sec:GRB_MGF}).

The number of electrons in the shocked fluid increases as (for $t \geq t_{\rm DN}$) 
\begin{equation}\label{eq:tot_elec}
\begin{split}
        &\ N_\mathrm{e} \equiv \frac{M(<\!R)}{m_{\rm p}} = \frac{E}{m_{\rm p} u^2 c^2} = \frac{E}{m_{\rm p} u_{\rm DN}^2 c^2}  \left( \frac{t}{t_{\rm DN}}\right)^{\frac{2(3-k)}{(5-k)}}   \\
        &\ \approx 1.8 \!\times\! 10^{55} \, \tilde{f}(p) \,\epsilon_{\rm  e,-1} \, \xi_{\rm e0}^{-1}   \times  
        \begin{cases}
              \left( \frac{t}{t_{\rm DN}} \right)^{\frac{6}{5}} \; \hspace{0.4cm}  (k=0),                   \\
             \left( \frac{t}{t_{\rm DN}} \right)^{\frac{2}{3}} \; \hspace{0.4cm}   (k=2).   
        \end{cases}
\end{split}
\end{equation}

Using equations~(\ref{eq:xi_e}) and (\ref{eq:tot_elec}) the number of relativistic electrons in the DN phase can be estimated as (for $t \geq t_{\rm DN}$) 
\begin{equation}\label{eq:rel_elec}
\begin{split}
     N_{\rm rel} &\ = \xi_\mathrm{e} N_{\rm e} =  \frac{\xi_{\rm e0} E}{m_{\rm p} u_{\rm DN}^2 c^2} = \frac{f(p) \epsilon_{\rm e} E}{2 m_{\rm e} c^2}  \\
     &\ \approx  1.8 \times 10^{55}   \;   E_{51} \; \tilde{f}(p) \; \epsilon_{\rm  e,-1}\;.         
\end{split}
\end{equation}

Equations (\ref{eq:tot_elec})–(\ref{eq:rel_elec}) show that during the DN phase, although the total number of electrons swept into the shocked fluid increases with time, the number of relativistic electrons stays constant and equals the total number of electrons at the DN transition, so that $N_{\rm rel}$ in Eq.~(\ref{eq:rel_elec}) can be thought of as the maximum number of electrons accelerated to relativistic velocities in the blast wave. This number is independent of the ambient medium stratification index $k$.  The reason is as follows: in the adiabatic DN regime, relativistic electrons carry a fixed fraction, $\epsilon_{e}\epsilon_{\rm int} E$ of the outflow energy (where a fraction $\epsilon_{\rm int}\approx\frac{1}{2}$ of the total energy is in internal energy), while maintaining a constant  average Lorentz factor to $\langle \gamma \rangle \approx  \gamma_{\rm dn}/G(p)$. Thus, since the number of relativistic electrons is the ratio of the two it does not change with time.

Appendix \ref{app:expr} shows that the slow cooling synchrotron regime ($\nu_{\rm m} \ll \nu_{\rm c}$) is naturally realized in the DN phase for a broad parameter space. The DN cooling break lies in the optical / UV band (see equation~(\ref{eq:DN_cool})).  Appendix \ref{app:Spec1} shows that in the DN regime the self-absorption frequency $\nu_{\rm sa}$ cannot fall below the minimal synchrotron frequency $\nu_{\rm m}$. This is because $\nu_{\rm m}$ lies very close to the cyclotron frequency (within a factor of 2 when $\gamma_{\rm dn} = \sqrt{2}$), making it difficult for $\nu_{\rm sa}$ to reach lower values. Consequently, in the DN phase, the slow-cooling synchrotron spectrum naturally satisfies the ordering $\nu_{\rm m} < \nu_{\rm sa} < \nu_{\rm c}$. In this “spectrum~2” regime (following the terminology of \citealt{GranotSari02}), the peak luminosity and corresponding peak frequency are at the self-absorption frequency. The peak/self-absorption frequency (see Appendix \ref{app:adia} ) can be written as  (for $t>t_{\rm DN}$ )
\begin{equation}\label{eq:nu_pk}
\begin{split}
      &\ \nu_{\rm pk} = \nu_{\rm sa} =  \left[ \frac{1}{m_{\rm e}} \frac{L_{\rm \nu,max}}{8 \pi^2 R^2} \;  \gamma_{\rm dn}^{p-1} \nu_{\rm B}^{\frac{p}{2}}\right]^{\frac{2}{p+4}} \hspace{0.5cm} \text{(for $\nu_{\rm sa} > \nu_{\rm m}$) } \\
      &\ \approx 
      \begin{cases}
           115  \; E_{51}^{\frac{2}{3(p+4)}}\; n_{0}^{\frac{14+3p}{6(p+4)}} \; \tilde{f}(p)^{-\frac{(2+3p)}{6(p+4)}} \;\xi_{\rm e0}^{\frac{14+3p}{6(p+4)}} \\
           \ \ \times\;   \epsilon_{\rm e,-1}^{-\frac{(2+3p)}{6(p+4)}}  \; \epsilon_{\rm B,-2}^{\frac{p+2}{2(p+4)}} \; t_{0.6}^{-\frac{3p+14}{5(p+4)}} \text{MHz} \hspace{0.2cm} (k=0)\;,              \vspace{0.30cm}\\ 
           12 \; E_{51}^{-1} \; A_{\star}^{\frac{14+3p}{2(p+4)}} \; \tilde{f}(p)^{-\frac{(10+3p)}{2(p+4)}} \xi_{\rm e0}^{\frac{14+3p}{2(p+4)}} \;  \\
           \ \ \times\;  \epsilon_{\rm e,-1}^{-\frac{(10+3p)}{2(p+4)}} \; \epsilon_{\rm B,-2}^{\frac{p+2}{2(p+4)}} \;t_{1.3}^{-\frac{3p+14}{3(p+4)}}\; \text{MHz} \hspace{0.3cm} (k=2)\;, 
      \end{cases}
\end{split}
\end{equation}
where $t_{0.6} = t/3.7$ yr and $t_{1.3} = t/18.7$ yr. The peak spectral luminosity (see Appendix \ref{app:adia} ) is given as (for $t>t_{\rm DN}$) 
\begin{equation}\label{eq:L_pk}
\begin{split}
        &\ L_{\rm pk} = L_{\nu_{\rm sa}} = L_{\rm \nu,max} \left( \frac{\nu_{\rm sa}}{\nu_{\rm m}} \right)^{\frac{1-p}{2}} \hspace{0.5cm} \text{(for $\nu_{\rm sa} > \nu_{\rm m}$)} \\ 
       &\ \approx \begin{cases}
             7.3\!\times\! 10^{28} \; E_{51}^{\frac{2p+13}{3(p+4)}}\, n_{0}^{\frac{2p+13}{6(p+4)}}\,  \tilde{f}(p)^{\frac{17-2p}{6(p+4)}}\,  \xi_{\rm e0}^{\frac{2p+13}{6(p+4)}} \,\epsilon_{\rm e,-1}^{\frac{17-2p}{6(p+4)}} \\ 
             \times  \epsilon_{\rm B,-2}^{\frac{2p+3}{2(p+4)}} \; t_{0.6}^{-\frac{(2p+13)}{5(p+4)}} \text{erg}\;\text{s}^{-1}\;\text{Hz}^{-1}  \hspace{0.2cm} (k=0)\,,             \\ \\ 
             8 \times 10^{27} \;  A_{\star}^{\frac{2p+13}{2(p+4)}}  \; \tilde{f}(p)^{-\frac{(3+2p)}{2(p+4)}}  \xi_{\rm e0}^{\frac{2p+13}{2(p+4)}} \; \epsilon_{\rm e,-1}^{-\frac{(3+2p)}{2(p+4)}}\; \epsilon_{\rm B,-2}^{\frac{2p+3}{2(p+4)}} \; \\
            \times t_{1.3}^{-\frac{(2p+13)}{3(p+4)}} \text{erg}\;\text{s}^{-1}\;\text{Hz}^{-1}   \hspace{0.2cm} (k=2)\;.               \\ 
        \end{cases}
\end{split}
\end{equation}

The spectral luminosity at frequency $\nu=\nu_{\rm GHz}\;$GHz in PLS G ($\nu_{\rm sa} < \nu < \nu_{\rm c}$) of spectrum 2 following notation by \cite{GranotSari02}  is given as (for $t>t_{\rm DN}$)
\begin{equation}
\begin{split}
   &\  L_{\rm \nu,G} \equiv  L_{\nu_{\rm sa}} \left( \frac{\nu}{\nu_{\rm sa}} \right)^{\frac{(1-p)}{2}} =  L_{\rm \nu,max} \left( \frac{\nu}{\nu_{\rm m}} \right)^{\frac{(1-p)}{2}} 
   \\
   &\ \approx  \begin{cases}
        1.5 \times 10^{28} \,  \nu_{\rm GHz}^{\frac{(1-p)}{2}}  \,  E_{51} \, n_{0}^{\frac{p+1}{4}} \, \tilde{f}(p)^{\frac{3-p}{4}}\,  \xi_{\rm e0}^{\frac{p+1}{4}}\,\epsilon_{\rm e,-1}^{\frac{3-p}{4}} \\
        \times \epsilon_{\rm B,-2}^{\frac{p+1}{4}} \,t_{0.6}^{-\frac{3(p+1)}{10}}  \;  \text{erg}\;\text{s}^{-1}\;\text{Hz}^{-1} \;\     (k=0)\, \vspace{0.30cm}\\  
         3 \times 10^{26}  \, \nu_{\rm GHz}^{\frac{(1-p)}{2}}  \,\,E_{51}^{\frac{1-p}{2}} \, A_{\star}^{\frac{3(p+1)}{4}} \, \tilde{f}(p)^{-\frac{(3p-1)}{4}} \;  \xi_{\rm e0}^{\frac{3(p+1)}{4}} \,\epsilon_{\rm e,-1}^{-\frac{(3p-1)}{4}} \\  \times \epsilon_{\rm B,-2}^{\frac{p+1}{4}} \, t_{1.3}^{-\frac{(p+1)}{2}} \;  \text{erg}\;\text{s}^{-1}\;\text{Hz}^{-1}   \;\ (k=2)\,. 
   \end{cases}
\end{split}
\end{equation}

\begin{table*}
\centering
    \caption{Temporal slopes $\alpha_{\rm G} = \frac{d \ln F_{\rm \nu,G}}{d \ln t}$  in PLS G  of the flux density $F_{\rm \nu,G} \propto t^{\alpha_{\rm G}}$ for different relative ordering of $(u_{0},u_{\rm DN})$} \label{tab:temp_scalings}
    \begin{tabular}{ccccc} \hline 
      Scenario  & $t<t_{\rm dec}$  & $t_{\rm dec} < t < t_{\rm N}$ & $t_{\rm N}< t < t_{\rm DN}$ &  $t>t_{\rm DN}$ \\ \hline 
       $u_{0} \gg 1> u_{\rm DN}$              &   $3 - \frac{k(p+5)}{4}$ & $-\frac{(3-k)(p-1)}{4}$ 
       &  $-\frac{15p-21 -4k(p-2)}{2(5-k)}$
       &  $-\frac{3(p+1)}{2(5-k)}$ \\ \\ 
     $k=0$    &   $3$  & $-\frac{3(p-1)}{4}$ &  $-\frac{3( 5p-7)}{10}$  &     $-\frac{3(p+1)}{10}$    \vspace{0.1cm}\\ 
     $k=0,p=2.5$ &  3 & -$\frac{9}{8}$   & $-\frac{33}{20}$ & $-\frac{21}{20}$     \\                           \\
     $k =2$    &   $-\frac{p-1}{2}$  & $-\frac{p-1}{4}$ &  $- \frac{7p-5}{6}$   & $-\frac{p+1}{2}$   \vspace{0.1cm}\\ 
     $k=2,p=2.5$ & $-\frac{3}{4}$ & $-\frac{3}{8}$ &  $-\frac{25}{12}$ & $-\frac{7}{4}$ \vspace{0.1cm}\\ \hline
    \end{tabular}  \\ 
    \vspace{0.3cm}
\centering     
    \begin{tabular}{cccc} \hline 
      Scenario & $t<t_{\rm dec}$  & $t_{\rm dec}< t < t_{\rm DN}$ &  $t>t_{\rm DN}$ \\ \hline 
       $1>u_{0} > u_{\rm DN}$               & $3-\frac{k(p+5)}{4}$ & $-\frac{15p-21 -4k(p-2)}{2(5-k)}$
       &  $-\frac{3(p+1)}{2(5-k)}$ \\ \\ 
     $k=0$    & 3 &$-\frac{3( 5p-7)}{10}$  &     $-\frac{3(p+1)}{10}$    \vspace{0.1cm}\\ 
      $k=0,p=2.5$  & 3 & $-\frac{33}{20}$ & $-\frac{21}{20}$                    \\   \\ 
     $k =2$    & $-\frac{p-1}{2}$  & $- \frac{7p-5}{6}$   & $-\frac{p+1}{2}$  \vspace{0.1cm}\\
     $k=2,p=2.5$ & $-\frac{3}{4}$ & $-\frac{25}{12}$ & $-\frac{7}{4}$ \vspace{0.1cm}\\ \hline
    \end{tabular} \\ 
    \vspace{0.3cm}
\centering
    \begin{tabular}{cccc} \hline 
     Scenario    & $t< t_{\rm dec}$       & $t>t_{\rm dec}$ \\ \hline 
      $u_{0}< u_{\rm DN}$           &  $3-\frac{k(p+5)}{4}$ &  $-\frac{3(p+1)}{2(5-k)}$ \vspace{0.1cm}\\ 
      $k = 0$  &  $3$    & $- \frac{3(p+1)}{10}$ \vspace{0.1cm}\\
      $k=0,p=2.5$ &  3 &  $-\frac{21}{20}$ \\ \\
      $k = 2 $ &   $-\frac{p-1}{2}$ & $ -\frac{p+1}{2}$ \vspace{0.1cm}\\ 
       $k=2,p=2.5$  & $-\frac{3}{4}$    & $-\frac{7}{4}$    \vspace{0.1cm}\\ \hline 
    \end{tabular}
\end{table*}

Figure~\ref{fig:DN_HydroSpec} summarizes the temporal evolution of the hydrodynamics, particle microphysics, and synchrotron spectral properties of a spherical blast wave propagating into a constant-density medium. Since the peak of the spectral luminosity $L_{\nu}$ occurs at the synchrotron self-absorption frequency, the most observationally relevant power-law segment (PLS) of the slow-cooling synchrotron spectrum is PLS~G ($\nu_{\rm sa} < \nu < \nu_{\rm c}$). This segment extends with a spectral slope $(1-p)/2$ from radio to optical frequencies and, therefore, dominates the observable emission over a broad frequency range.

Table~\ref{tab:temp_scalings} summarizes the temporal decay indices $\alpha_{\rm G}$ (defined as $F_{\nu_{\rm G}} \propto t^{\alpha_{\rm G}}$) in PLS~G within the Newtonian and DN regimes for different orderings of $u_{\rm 0}$ and $u_{\rm DN}$, and for an ambient medium characterized by a density stratification index $k$. An additional quantity of interest is the deviation of the DN temporal slope from the value obtained by extrapolating the Newtonian solution. As an illustration, for $k=0$ and $t>t_{\rm dec}$ this difference for PLS~G is given by
\begin{equation}
\delta \alpha_{\rm G} \equiv \alpha_{\rm G,DN} - \alpha_{\rm G,N}
= \frac{6}{5}(p-2)\;,
\label{eq:slope_DN}
\end{equation}
which implies that for $p>2$ the DN temporal decay is shallower than the extrapolated Newtonian prediction.

In the limit $p \rightarrow 2$, Equation~(\ref{eq:beta_DN}) indicates that the onset of the DN phase occurs at earlier times as compared with larger $p$ values, while Equation~(\ref{eq:slope_DN}) shows that the difference between the DN and Newtonian temporal slopes decreases. However, this limit must be treated with caution, since $\beta_{\rm DN} \rightarrow \infty$ as $p \rightarrow 2$, and the underlying approximation ceases to be valid. This highlights a trade-off: larger values of $p$ delay the onset of the DN phase through a smaller $\beta_{\rm DN}$, but enhance the contrast between the DN and Newtonian temporal slopes, whereas smaller values of $p$ produce the opposite effect.

In this section, we consider an outflow with $u_{0} > u_{\rm DN}$.
In many astrophysical scenarios like kilonovae and supernovae, the outflow is in the DN phase in the coasting phase itself. The corresponding expressions for these scenarios are summarized in Appendix \ref{app:coast}.

\section{Astrophysical transients}\label{sec:astro_appl}

We turn next to systematically exploring the synchrotron parameter space of astrophysical transients in the DN regime and examining the temporal evolution of the light curves and the spectrum. Our primary objective is to assess the observability of this late-time dynamical phase and identify the physical conditions under which it becomes accessible to current and future radio facilities. Table \ref{tab:tran_param} summarizes the fiducial parameter values adopted for different astrophysical transients.

\begin{table*}
\centering
\caption{Fiducial parameters for spherical outflows from different astrophysical transients: $M_{\rm ej}$ and $E$ refer to the true (not isotropic equivalent) mass and kinetic energy associated with the outflow, respectively, $u_\mathrm{ej}$ is the outflow proper speed, 
$\epsilon_\mathrm{e}$ and $\epsilon_\mathrm{B}$ represent the fractions of internal energy channeled into shock accelerated electrons and magnetic field, respectively in the shocked external medium,
and $\xi_\mathrm{e0}$ represents the fraction of number of non-thermal to total electrons in the pre-DN phase.  }
\begin{tabular}{lcccccc}
\hline\hline
Object & $u_\mathrm{ej}$ &$M_{\rm ej}$ [$M_\odot$] & $E$ [erg] & $\epsilon_\mathrm{e}$ & $\epsilon_\mathrm{B}$ & $\xi_{e0}$ \\
\hline
 Gamma-ray burst (GRB) & $10^2\!-\!10^3$ & $10^{-6}\!-\!10^{-5}$ & $10^{50}\!-\!10^{52}$ & $10^{-2}\!-\!10^{-0.5}$ & $10^{-5}\!-\!10^{-2}$ & 1 \\
 Kilonovae remnant (KNR)  & $10^{-1}$ & $10^{-1.5}$ & $10^{50.5}$ & $10^{-1}$ & $10^{-2}$ & 1 \\ 
 Magnetar Giant Flare (MGF)
 & $1\!-\!10^{2}$ & $10^{-11}\!-\!10^{-7}$ & $10^{45}$ & $10^{-1}$ & $10^{-2}$ & 1\\
 Supernovae remnant (SNR)  & $10^{-1.7}$ & $10^{0.5}$ & $10^{51}$ & $10^{-4}$$-$$10^{-2}$ & $10^{-3}$$-$$10^{-1}$ & 1    \\ 
 Superluminous Supernovae (SLSNe) & $10^{-1.2}$    & $10^{0.5}$   &  $10^{52}$ &  $10^{-2}\!-\!10^{-1}$ & $10^{-3}\!-\!10^{-2}$ & 1 \\ 
\hline
\end{tabular}
\label{tab:tran_param}
\end{table*}

\begin{table*}\label{tab:astro_appl}
    \centering
    \caption{Summary of the temporal slopes $\alpha_{\rm G} = \frac{d \ln F_{\nu,{\rm G}}}{d \ln t}$ in PLS G for specific astrophysical scenarios in the Newtonian and DN regimes. Here $s$ denotes the power-law index of the cumulative energy profile of the ejecta (see Equation~(\ref{eq:cum_kin})), $k$ is the stratification index of the ambient medium, and $p$ is the power-law index of the non-thermal electron energy distribution. The expressions for the single-velocity ejecta case can be obtained in the limit $s \rightarrow \infty$.   }    
    \begin{tabular}{ccccc}\hline 
      Scenario   & Object       & $k$   & $\alpha_{\rm G}$ & Relevant Section \\ \hline 
      $u_{\rm 0} \gg u_{\rm DN}$ &  Gamma-ray burst (GRB)  &  0    &       \hspace{0.2cm}$\frac{21-15p}{10}$  \hspace{0.7cm} for $t_{\rm N} < t <t_{\rm DN}$  & \S \ref{sec:GRB_MGF} \vspace{0.1cm}\\ 
       &     &      &  \hspace{-0.35cm}$-\frac{(3p+1)}{10}$ \hspace{0.65cm} for $t>t_{\rm DN}$\hspace{0.35cm}   \\  \\ 
       
      $ u_{\rm 0}< u_{\rm DN}$ &      Kilonovae remnant (KNR) &  0  & \vspace{0.2cm}$\frac{21+6s-15p}{2(5+s)}$ \hspace{0.55cm} for $t_{\rm N}<t<t_{\rm DN}$ \vspace{0.0cm}\\  
       &  &   & \hspace{0.10cm}$\frac{6s-3(p+1)}{2(5+s)}$ \hspace{0.57cm} for $t_{\rm DN}<t<t_{\rm dec}$  & \S \ref{sec:KNR}   \vspace{0.1cm}\\
       &   &    & $-\frac{(3p+1)}{10}$ \hspace{0.55cm} for $t>t_{\rm DN}$ \hspace{0.50cm} \\  \\ 
       $ u_{\rm 0} > u_{\rm DN}$ &  Extreme kilonovae remnant (EKNR) & 0 & \hspace{-0.8cm}$\frac{21+6s-15p}{2(5+s)}$ \hspace{0.6cm} for $t<t_{\rm dec}$ \hspace{0.9cm} & \S \ref{sec:KNR} \vspace{0.1cm}\\ 
       &    &   &   \hspace{0.5cm}$\frac{21-15p}{10}$  \hspace{0.7cm} for $t_{\rm dec}< t <t_{\rm DN}$  \vspace{0.1cm}\\ 
       &     &      &  $-\frac{(3p+1)}{10}$ \hspace{0.62cm} for $t>t_{\rm DN}$ \hspace{0.55cm}  \\  \\ 
       $ u_{\rm 0}< u_{\rm DN}$          & Radio Supernovae (Pre-deceleration)& 2  & \hspace{-1.1cm}$\frac{s(1-p)-3(1+p)}{2(s+3)}$ \hspace{0.3cm} for $t<t_{\rm dec}$  & \S \ref{sec:Early_SN} \\ \\ 
       $\; \; u_{0} < u_{\rm DN}$ &  Superluminous Supernovae (SLSNe)   & 0 & \hspace{-0.7cm}$\frac{21+6s-15p}{2(5+s)}$ \hspace{0.40cm} for $t<t_{\rm DN}$ & \S \ref{sec:Early_SN}  \vspace{0.1cm}\\ 
       &  &   &  \hspace{0.4cm}$\frac{6s-3(p+1)}{2(5+s)}$ \hspace{0.4cm} for $t_{\rm DN}<t<t_{\rm dec}$     \vspace{0.1cm}\\
       &   &    & \hspace{-0.4cm}$-\frac{(3p+1)}{10}$ \hspace{0.47cm} for $t>t_{\rm DN}$  \vspace{0.1cm}\\ \hline 
    \end{tabular}
    \label{tab:flux_index}
\end{table*}

\subsection{Gamma-ray burst and magnetar giant flare afterglow}\label{sec:GRB_MGF}

\begin{figure}
    \centering
    \includegraphics[scale=0.4]{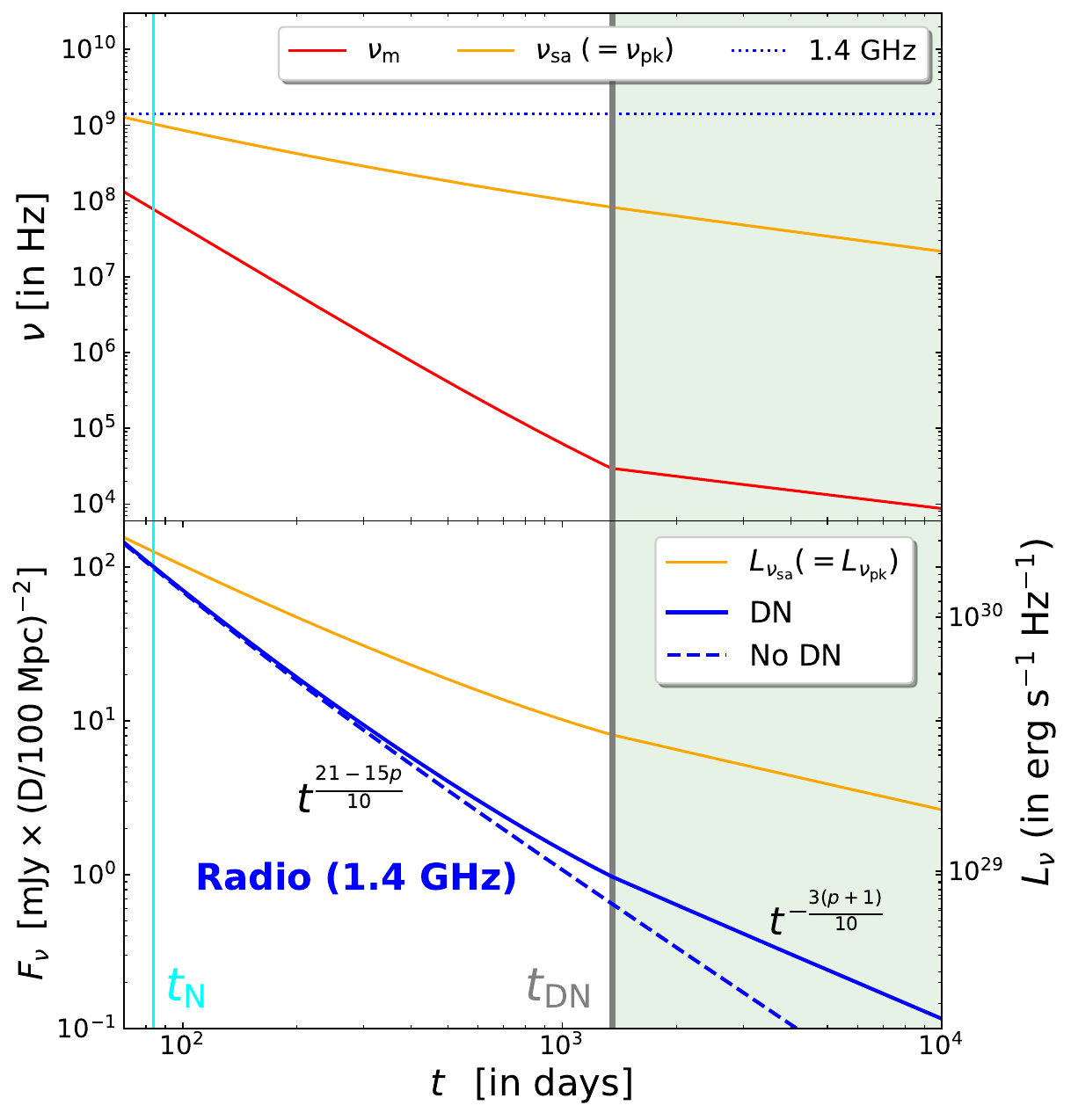}
    \caption{\textbf{Illustration of the DN phase in a gamma-ray burst radio afterglow} for an initially relativistic outflow ($u_0\!\gg\!1\Leftrightarrow t_{\rm dec}\!\ll\!t_{\rm N}$). The assumed parameters are $E_{51}=1$, $n_{0}=1$, $\epsilon_{\rm e}=10^{-1}$, $p=2.5$, $\epsilon_{\rm B}=10^{-2}$, and $\xi_{\rm e0}=1$. In all panels, the vertical cyan and gray lines indicate the onset of the Newtonian phase (Equation~(\ref{eq:t_n})) and the transition to the DN phase (Equation~(\ref{eq:t_dn})), respectively. The green shaded region to the right of the gray line denotes the DN regime. In the middle and bottom panels, the flux density (for a source distance of 100~Mpc) is shown on the left $y$-axis, while the corresponding spectral luminosity is shown on the right (twin) $y$-axis. \textit{Top panel:} The temporal evolution of the characteristic synchrotron frequencies is shown: the minimum frequency $\nu_{\rm m}$ (solid red), the self-absorption frequency $\nu_{\rm sa}$---which also corresponds to the peak frequency in this regime---(solid orange), and the observing frequency $\nu_{\rm obs}=1.4$~GHz (solid blue). \textit{Bottom panel:} The peak flux density (left $y$-axis) and the peak spectral luminosity (right $y$-axis) are shown. For the spectral ordering $\nu_{\rm sa}>\nu_{\rm m}$, the peak occurs at $\nu_{\rm sa}$. The flux density and spectral luminosity at $\nu_{\rm obs}=1.4$~GHz are shown as functions of time (solid blue). The dashed blue line in the DN regime indicates the extrapolation of the Newtonian-phase temporal slope into the DN phase (see text for details).}
    \label{fig:GRB_plot}
\end{figure}

As an illustrative example of an outflow that is initially highly collimated, later becomes Newtonian, and eventually enters the DN regime, we consider a GRB afterglow in a constant-density medium. By the time the outflow reaches the DN phase, its propagation effectively occurs in a constant density interstellar medium. Figure\;\ref{fig:GRB_plot} shows that for typical ISM parameters, the DN regime is realized on a timescale of $\sim $ a few years (see Equation~(\ref{eq:t_dn})), at which point the flux has decayed by roughly two orders of magnitude compared to the Newtonian transition, which typically occurs on timescales of $\sim$ a few months (see equation~(\ref{eq:t_n})). As discussed in equation~(\ref{eq:slope_DN})  the DN phase is delayed for large $p$ values, and in that case only nearby sources are expected to be detectable in this phase.
The detection prospects  improve in a higher-density ambient medium, which both shortens the transition timescale and yields brighter emission. In the DN regime, the flux decay is shallower than what would be inferred by extrapolating from the Newtonian phase. In case such transitions are observed, the ratio of the Newtonian to the DN transition time can be used to constrain shock microphysics (see Equation~(\ref{eq:tdn_tn})). This late-time flattening of the decay slope is a universal late-time feature for any outflow evolving in the adiabatic phase (see Table \ref{tab:flux_index}).

As another illustrative example of an outflow transitioning to the DN phase, we consider the radio nebula produced from interaction of the 27 December 2004 giant flare from the Galactic magnetar SGR 1806$-$20 \citep{Borkowski04,Gaenslar05,Cameron05} with a bow-shock shell formed by the magnetar wind interacting with the interstellar medium. This event is of particular interest, as it constitutes the first observational example where the DN regime was inferred directly from the data \citep{Granot06}. The highest velocity portion (leading edge) of the outflow had an apparent expansion speed of $\beta_{\rm app} \approx 1$ \citep{Gelfand05,Taylor05,TG06}. Attributing this to its proper speed $u$ implies a true expansion velocity of $\beta\approx0.7$. The mean apparent speed measured for this source from radio observations, has a median value of $\langle\beta_{\rm app}\rangle \sim 0.4$ \citep{Taylor05} which is representative speed for the bulk material. The radio nebula powered by this outflow implied a one-sided outflow that had an axis ratio of $\sim$\,2:1. The outflow's expansion speed was measured to decelerate at
$t_{\rm dec} \approx 33$ days, its isotropic-equivalent kinetic energy was inferred to be
$E_{\rm k,iso} \sim 10^{46}\;$erg and a relatively low 
ISM density of $n_0 \lesssim 0.3\;$cm$^{-3}$ is preferred on energetic grounds \citep{Granot06}. 
The true outflow energy was inferred to be $E\gtrsim 10^{44.5}\;$erg. The giant flare outflow collides inelastically with the bow-shock nebula shell, producing initial radio emission from both reverse and forward shocks. After the shock crossing, the outflow enters the adiabatic phase. At the deceleration time,  $t_{\rm dec} \approx 33\;$days, the radio emission is dominated by a different component attributed to the forward shock going into the external ISM, whose emission peaks at this time and then decays slowly as $F_\nu\propto t^{-1.05\pm0.05}$.
The emission rapidly transitions to the DN phase, as the deceleration time and DN transition time are nearly coincident, and the flow quickly approaches spherical symmetry upon deceleration.  Thus, post-deceleration the evolution is determined by the true energy $E$. For reasonable variations in the true energy of magnetar giant flares, the flux density at the DN transition remains detectable only for Galactic events.

\subsection{Kilonova remnant  afterglow}\label{sec:KNR}

\begin{figure*}
\begin{tabular}{c|c}
\includegraphics[scale=0.44]{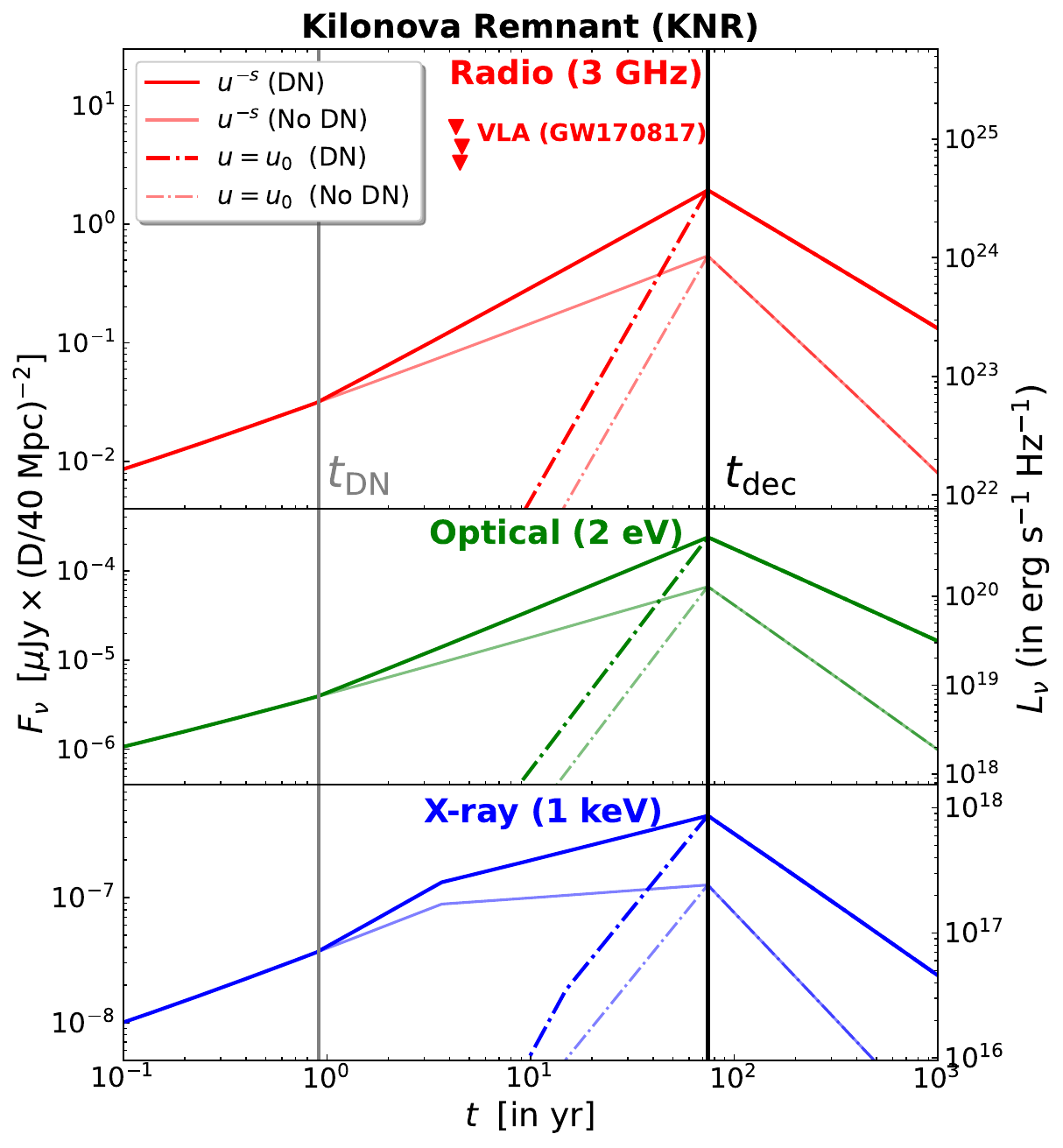}
   & \includegraphics[scale=0.44]{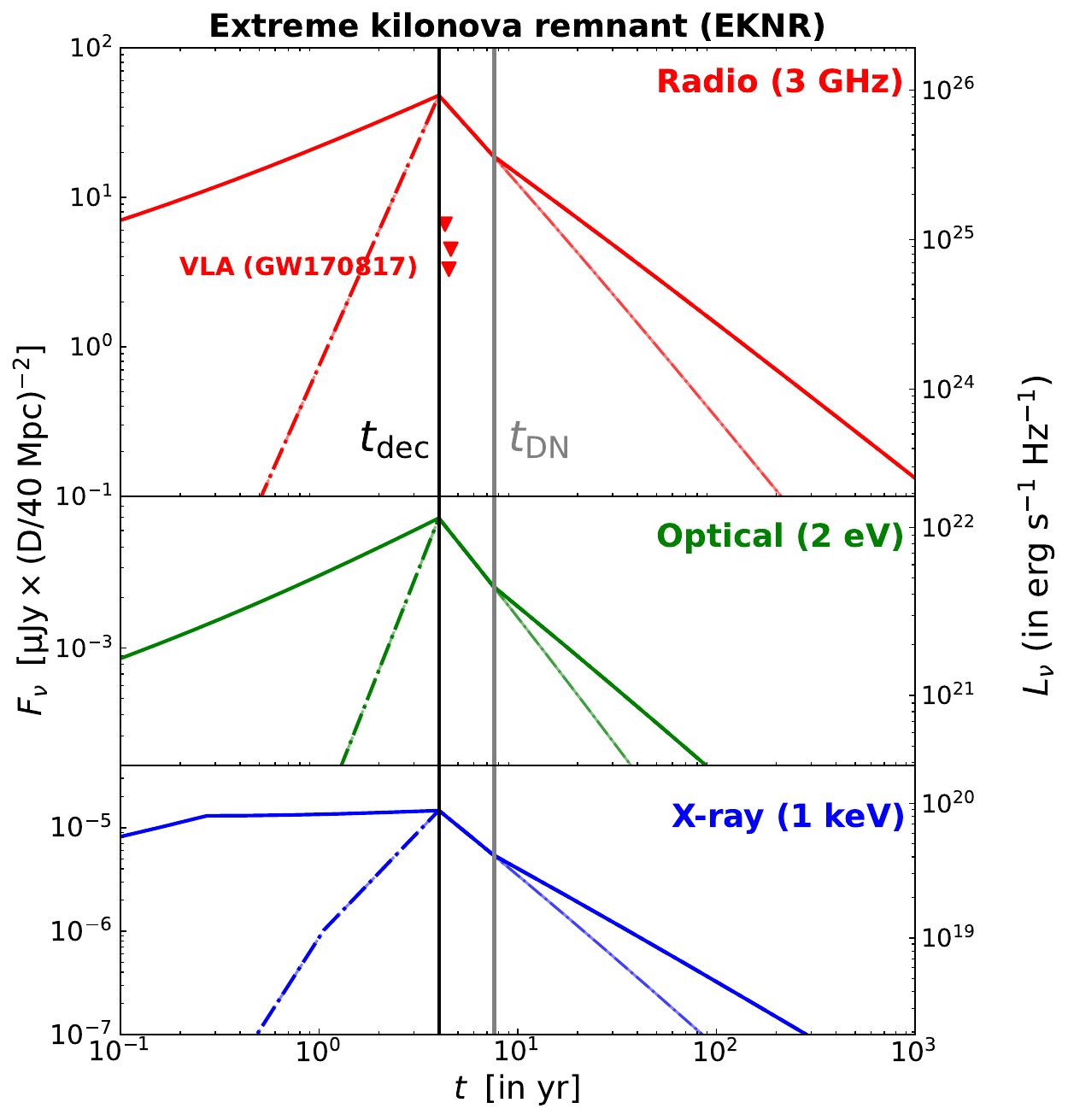} \\
\end{tabular}
    \centering
\caption{\textbf{Impact of DN phase on multi-wavelength afterglow from a kilonova remnant (KNR; left) and a magnetar-boosted extreme kilonova remnant (EKNR; right) at $D = 40$ Mpc}.  
Both panels assume $\epsilon_{\rm e}=0.1$, $\epsilon_{\rm B}=0.01$, $p=2.5$, and $\xi_{\rm e0}=1$, giving a critical proper speed $u_{\rm DN} \approx 0.2$. An ambient number density $n_\mathrm{0} = 0.01$ cm$^{-3}$ is assumed. Stratified ejecta use $s=5$, with $u_{\rm 0,max}=1.2$ and $u_{\rm 0,min}$ equal to the launch velocity of the single-velocity ejecta $u_{0}= \sqrt{\frac{2 E_{0}}{M_\mathrm{ej} c^2}} \approx$ 0.06 (KNR) and 0.3 (EKNR). The ejecta mass is $M_{\rm ej}=0.1\,M_\odot$, with total kinetic energy $E_0=10^{50.5}$ erg (KNR) and $10^{52}$ erg (EKNR).  The radio upper limits for the field associated with GW170817 in the upper left and right subpanels are taken from \citealt{Bala22}. Solid lines show stratified ejecta; dot--dashed lines show single-velocity ejecta. Thick lines include the DN phase; thin lines use the Newtonian approximation. Left y-axis shows flux density $F_{\nu}$ ($\mu$Jy ); right y-axis shows spectral luminosity $L_\nu$ (erg s$^{-1}$ Hz$^{-1}$). Grey and black vertical lines mark the DN transition and deceleration time, respectively; the cyan line (right panel) marks the Newtonian transition for EKNR. Top, middle, and bottom panels show radio (3 GHz), optical (2 eV; $r$-band), and X-ray (1 keV) light curves in red, green, and blue. For identical microphysics and distance, EKNR is significantly brighter across all bands (see text for details).}
\label{fig:KN_plots}
\end{figure*}

Simulations of binary neutron star (BNS) mergers indicate a stratified ejecta structure for the kilonova remnant, in which a small fraction of the mass carries most of the kinetic energy \citep{Kyutoku14,Radice16,Sekiguchi16,Shibata19,Kawaguchi22}. We model the outflow using a radially stratified cumulative kinetic energy distribution,
\begin{equation}\label{eq:cum_kin}
E(>\!u) = E_{\rm tot}\left(\frac{u}{u_{\rm min}}\right)^{-s}\quad\text{for}\ u_{\rm min} \leq u \leq u_{\rm max}\;,
\end{equation}
where $E_{\rm tot}$ is the total kinetic energy of the ejecta. Numerical simulations \citep{Kyutoku14} further suggest that the stratification index $s$ depends on the crustal equation of state and that the high-velocity component ($u \gg 1$) exhibits a flatter slope than the low-velocity component ($u \ll 1$). For concreteness, we focus on the low-velocity regime and adopt $s \simeq 5$.

For simplicity, several previous studies \citep{Nakar&Piran11,Acharya25} have adopted a single-velocity ejecta model. As we show below, ejecta stratification is important primarily during the pre-deceleration phase; after deceleration, in the adiabatic regime, the full ejecta energy has been transferred to the ambient medium. Consequently, the light curves obtained for a single-velocity outflow asymptotically approach those of a stratified ejecta at late times.

The fate of the merger remnant remains uncertain. One possible outcome is the formation of a long-lived millisecond magnetar \citep{Metzger14}; although observational evidence suggests that, if realized, this should be a rare outcome \citep{BL2021,Ricci2021,WBG2024}. If such a long-lived magnetar survives, energy injection during magnetar spin-down can significantly enhance the expansion velocity of the remnant. Following \cite{Acharya25}, we refer to such a system as an \emph{extreme kilonova remnant} (EKNR), and we explore the resulting multiwavelength light curves in this scenario.

Figure\;\ref{fig:KN_plots} shows the multi-wavelength kilonova afterglow light curves for both radially stratified ejecta and single-velocity ejecta. The light curve peaks at the deceleration time, which is the same for both types of ejecta. Differences between the two profiles arise only during the pre-deceleration phase: by the deceleration time, all material in the stratified ejecta has been decelerated, so the full available energy is in the shocked external medium during the adiabatic phase, and the light curves converge.  The radially stratified ejecta acts as a gradual energy injection process: faster material decelerates first, followed by slower material, until the deceleration time when even the slowest material has contributed. As each part of the ejecta decelerates, it transfers its energy to the shocked external medium through the $pdV$ work done across the contact discontinuity. Before this time, the single-velocity ejecta, in the coasting phase, show a flux density that grows steeply as $t^3$, since the number of relativistic electrons scales with the volume ($\propto R^3$) for a uniform external density. This is also the scaling of the energy of the shocked external medium, since the mean energy per particle is constant ($\approx u_0^2m_pc^2$). In contrast, the stratified outflow exhibits a shallower flux rise and higher flux density due to the continuous energy injection. For KNR, the DN phase begins before deceleration, whereas for EKNR, boosted ejecta speeds delay the DN phase until after deceleration. In both cases, the DN phase typically occurs within a decade; neglecting it and using the standard Newtonian approximation can underestimate the flux density by factors of $\sim3-5$ in the coasting phase and even larger factors in the post-deceleration phase. Moreover, accounting for the DN phase self-consistently produces a steeper flux rise and a shallower decay. Both of these effects increase the prospects for early detection. All else being equal, the magnetar boosted EKNR is substantially brighter at all wavelengths, particularly in the radio, improving prospects for detecting such afterglows within a decade following short GRBs. For EKNR, the predicted flux is expected to exhibit a shallow late-time decay once DN effects are included. The existing late-time radio upper limits for the field associated with GW170817 strongly disfavor an EKNR scenario and, by extension, the presence of a long-lived magnetar remnant (see the upper left and right subpanels of Fig.~\ref{fig:KN_plots}). More generally, the predicted shallow late-time decay in such systems motivates continued radio monitoring of future nearby short GRB sources down to sensitivities of order $\mu$Jy, both to detect EKNRs if they exist and to constrain the BNS merger driven magnetar formation rate.

\subsection{Early radio light curves of supernovae}\label{sec:Early_SN}

\begin{figure*}
    \centering
 \begin{tabular}{c|c}
\includegraphics[scale=0.45]{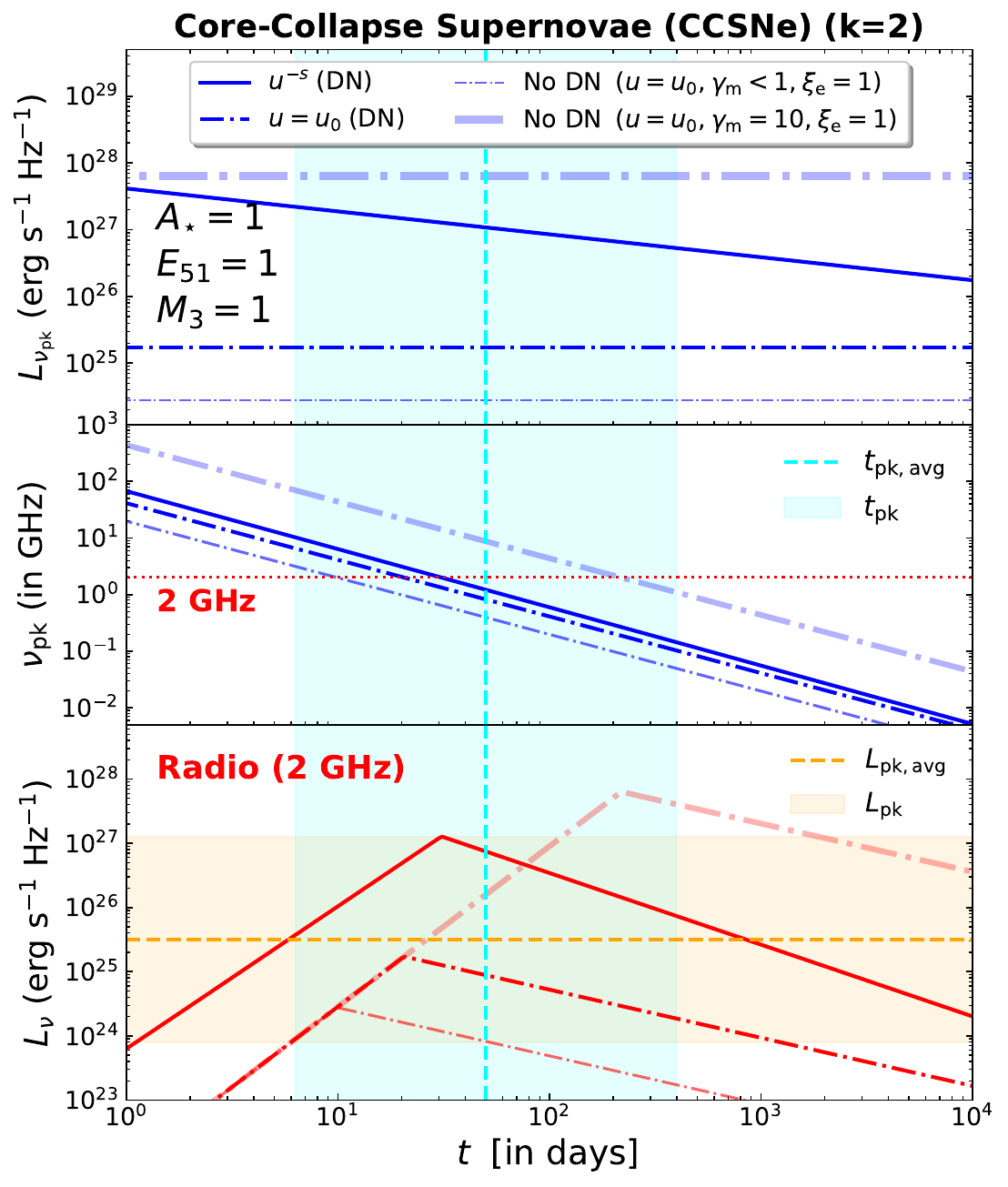}
   & \includegraphics[scale=0.45]{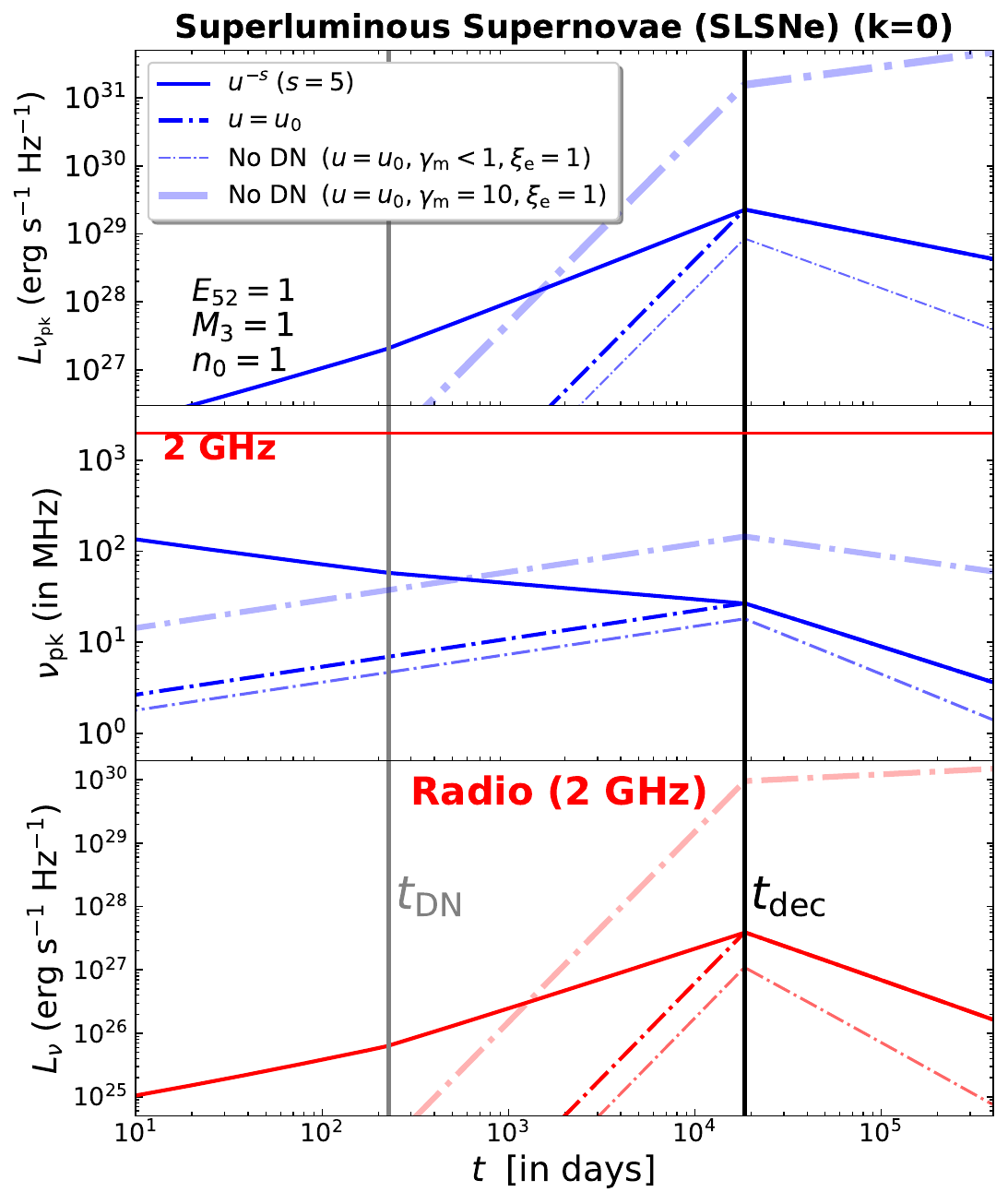} \\
    \includegraphics[scale=0.45]{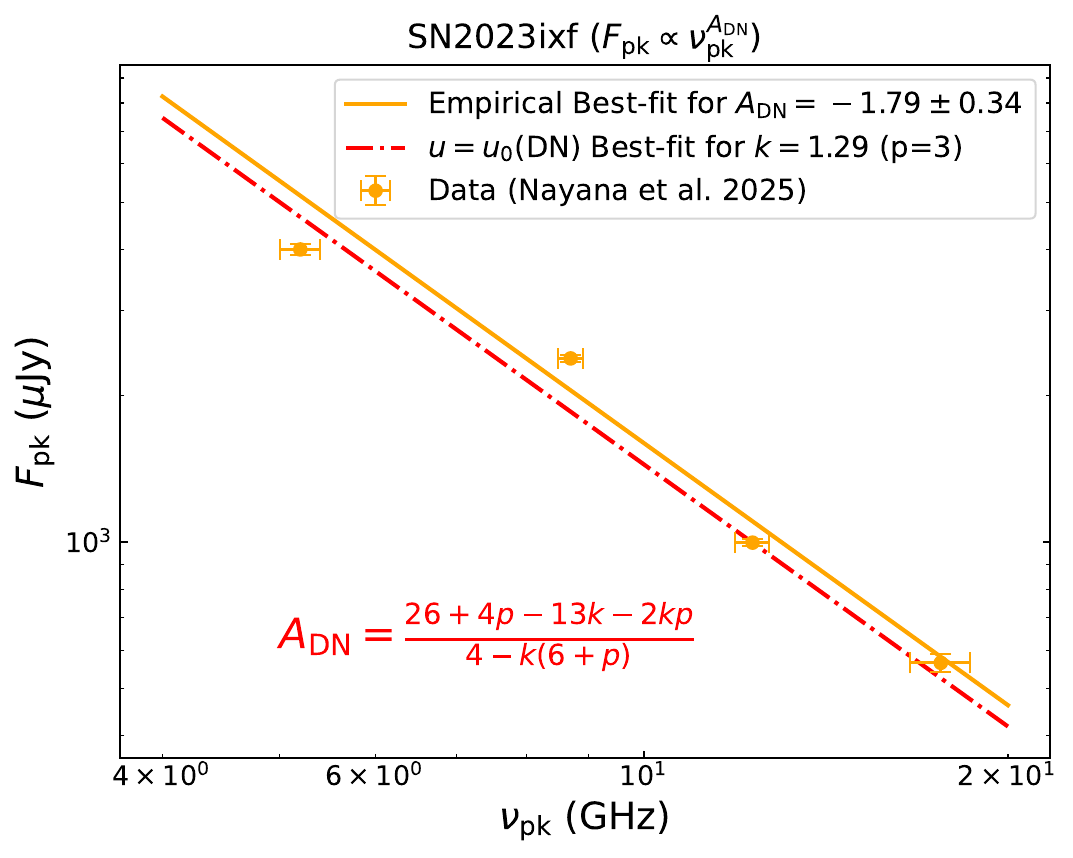}  & \includegraphics[scale=0.45]{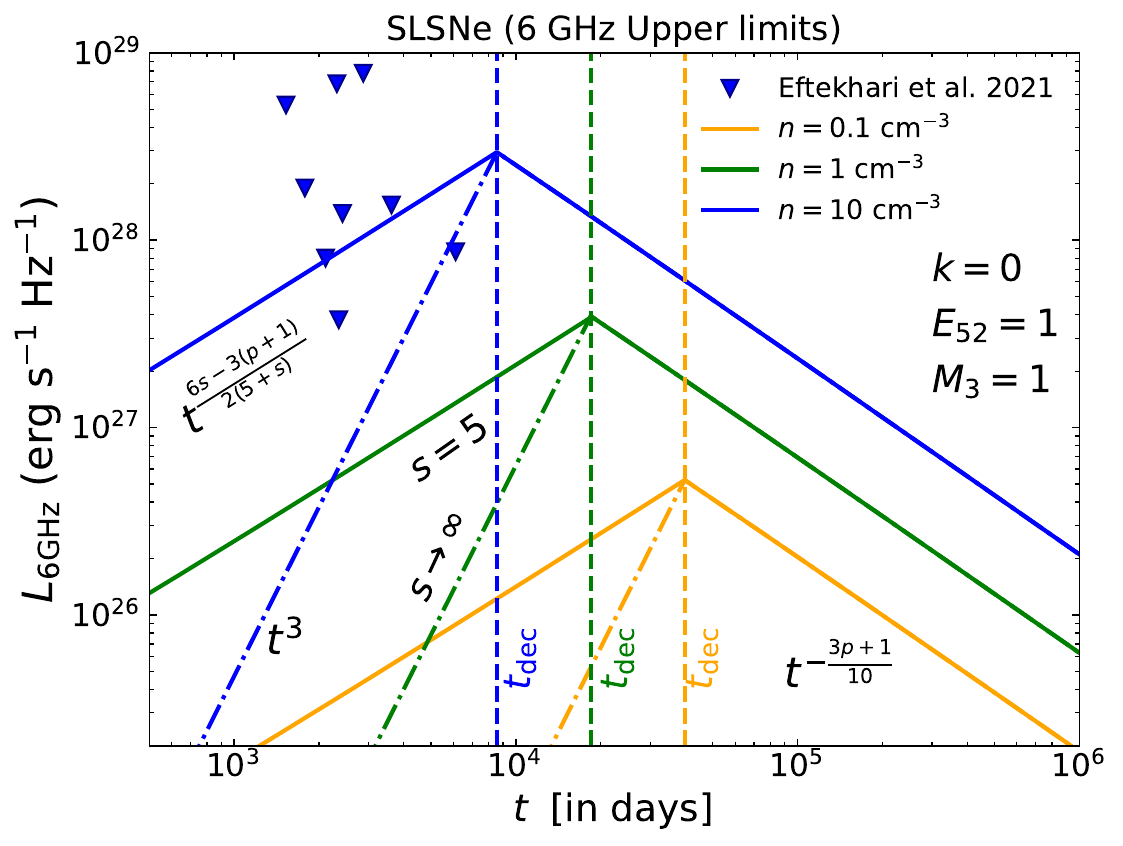} 
 \end{tabular}   
\caption{\textbf{Illustration of the DN phase on early radio light curves for core-collapse supernovae (CCSNe; left) and superluminous supernovae (SLSNe; right)}. Both upper panels show the peak spectral luminosity (top subpanel, blue), the corresponding peak frequency (middle subpanel, blue), and the spectral luminosity at 2 GHz (bottom subpanel, red). As in Figure~\ref{fig:KN_plots}, solid lines correspond to a radially stratified ejecta at launch with stratification index $s=5$, while dot--dashed lines indicate single-velocity ejecta. In both panels the thicker dot-dashed lines in lighter shade correspond to fixing $\gamma_{\rm m} = 10$ with $\xi_{\rm e} = 1$ and the narrower dot-dashed line in lighter shade corresponds to Newtonian extrapolation (Equation~(\ref{eq:gamma_m_NR})) with $\xi_{\rm e} = 1$. Microphysical parameters are $\epsilon_{\rm e}=0.1$, $\epsilon_{\rm B}=0.01$, $p=2.5$, and $\xi_{\rm e0}=1$, giving a critical proper speed $u_{\rm DN} \approx 0.2$.  The total energies are $(E_\mathrm{SN}, M_{\rm SNR}) = (10^{51}\;\mathrm{erg}, 3 M_\odot)$ for CCSNe and $(10^{52}\;\mathrm{erg}, 3 M_\odot)$ for SLSNe. The CCSNe expansion assumes a wind-like medium ($k=2$) with $A_\star=1$, whereas the SLSNe assume a uniform density medium with $n_0=1$ cm$^{-3}$. \textbf{Left panels: } \textit{left Upper panel:} The vertical cyan dotted line marks the average observed radio peak time of CCSNe, $t_{\rm pk,avg}=10^{1.7}$ days, while the cyan band indicates the observed range, $t_{\rm pk}=10^{0.8}$--$10^{2.6}$ days, in the 2--10 GHz radio band. Horizontal dashed orange lines indicate the average observed peak luminosity $L_{\rm pk} = 10^{25.5}$ erg s$^{-1}$ Hz$^{-1}$ and the orange band shows its spread $L_{\rm pk} = 10^{23.9} - 10^{27.1}$ erg s$^{-1}$ Hz$^{-1}$ (\citealt{Bietenholz21}). The middle subpanel the temporal evolution of the peak evolution, while  the bottom subpanel shows the 2 GHz light curves. \textit{Lower left panel:} Best-fit value for ambient medium stratification $k=1.29$ of the nearby CCSN SN2023xif based on $F_{\rm pk}$--$\nu_{\rm pk}$ data (\citealt{Nayana25}) and the corresponding DN fit using $p=3$ as inferred from optically thin spectra. \textbf{Right panels :} \textit{Upper right panel:} Grey and black vertical lines mark the DN transition time and deceleration time for the radial outflow. The temporal slopes in the bottom subpanel are summarized in Table \ref{tab:astro_appl}.  \textit{Lower right panel:} Constraints on the constant ISM density $n_0$ for SLSNe from 6 GHz upper limits (\citealt{Eftekhari21}), assuming $(E_\mathrm{SN}, M_{\rm SNR}) = (10^{52}\;\mathrm{erg}, 3 M_\odot)$.The dot-dashed lines correspond to single-velocity ejecta, while the solid lines correspond to radially stratified ejecta with stratification index $s=5$. Both curves merge at $t = t_{\rm dec}$ (shown as vertical dashed lines). The orange, green, and blue curves correspond to radio light curves at $6\,\mathrm{GHz}$ for ambient particle densities $n = (0.1, 1, 10)\,\mathrm{cm^{-3}}$. The time-axis is not extended beyond the regime where the adiabatic approximation breaks down as detailed in Appendix \ref{app:rad_phase} (see text for more details).}
\label{fig:SN_LC}
\end{figure*}

In this section, we examine the influence of the DN phase on the pre-deceleration light curves of supernova remnants. For illustrative purposes, we consider two representative cases: (i) superluminous supernovae (SLSN) remnant expanding into a constant-density medium ($k=0$), and (ii)  typical core-collapse supernovae (CCSNe) evolving in a wind-stratified environment ($k=2$).

The assumption of a constant-density ambient medium for SLSNe is motivated as follows. Only a small number of SLSNe have been detected in the radio band at early times. Recent observational studies aimed at detecting radio emission from SLSNe (e.g., \citealt{Eftekhari21}) adopt this simplifying assumption in order to place constraints on the circumburst density in the largely non-detection-dominated sample.

Equation~(\ref{eq:uratio_SNR}) shows that supernova remnants remain in the DN phase from the outset.  The wind-like environment typically continues up to a termination shock at $R \lesssim 0.3 $ pc beyond which the ambient medium can be assumed to have a constant density. The deceleration essentially happens in a constant density medium. For SNe with $u_{0} \ll 1$, the deceleration time is given as
\begin{equation}
    t_{\rm dec} = \frac{2}{5-k} \frac{R_{\rm dec}}{u_{0} c} \approx 165 \;  E_{51}^{-\frac{1}{2}} \; M_{3}^{-\frac{1}{6}} \; n_{0}^{-\frac{1}{3}} \; \text{yr}.
\end{equation}

The peak frequency is given as  (see Appendix \ref{app:coast})
\begin{align}
        \begin{split}
          & \nu_{\rm pk} = \nu_{\rm sa} \hspace{3cm} \text{for $t<t_{\rm dec}$} \\
           & \approx
            \begin{cases}
                  6.4 \times 10^{5} \;  \tilde{f}(p)^{\frac{2}{p+4}} \, n_{0}^{\frac{6+p}{2(p+4)}}  \, \\
                \times \epsilon_{\rm e,-1}^{\frac{2}{p+4}} \,
\epsilon_{\rm B,-2}^{\frac{2+p}{2(p+4)}} \,
\left( \frac{E_{\rm 51}}{M_{\rm 3}} \right)^{\frac{8 + p}{2(p+4)}} \,
t^{\frac{2}{p+4}}_{100} \; \; \text{Hz} \hspace{0.3cm} (k=0),    \vspace{0.3cm}\\ 
             4 \times 10^{8} \; \tilde{f}(p)^{\frac{2}{p+4}} \,
A_{\star}^{\frac{6+p}{2(p+4)}} \\
 \times \, \epsilon_{\rm e,-1}^{\frac{2}{p+4}} \,
\epsilon_{\rm B,-2}^{\frac{2+p}{2(p+4)}} \,
\left( \frac{E_{\rm 51}}{M_{\rm 3}} \right)^{\frac{1}{p+4}} \,
t_{100}^{-1} \; \;  \text{Hz} \hspace{0.3cm} (k=2),       \\
            \end{cases}
    \end{split}    \label{eq:coast_nupk}
\end{align}
while the peak spectral luminosity(see Appendix \ref{app:coast})
\begin{align}
\begin{split}
        &\ L_{\rm pk} = L_{\nu_{\rm sa}} \hspace{3cm} \text{for $t<t_{\rm dec}$}\\
        &\ \approx
        \begin{cases}
            1.5\!\times\!10^{19}\ \tilde{f}(p)^{\frac{5}{p+4}}\,
n_{0}^{\frac{2p + 13}{2(p+4)}}\,
\epsilon_{\rm e,-1}^{\frac{5}{p+4}} \vspace{0.15cm}\\
\times\,\epsilon_{\rm B,-2}^{\frac{2p + 3}{2(p+4)}}\!\fracb{E_{\rm 51}}{M_{\rm 3}}^{\frac{2p+13}{p+4}}
t_{100}^{\frac{2p+13}{p+4}} \; \frac{\text{erg}}{\text{s}\;\text{Hz}} \hspace{0.2cm} (k=0)\;,   \\ \\
            1.6 \times 10^{25} \; \tilde{f}(p)^{\frac{5}{p+4}} \, A_{\star}^{\frac{2p + 13}{2(p+4)}} \,
\epsilon_{\rm e,-1}^{\frac{5}{p+4}}  \vspace{0.15cm}\\
\times \, 
\epsilon_{\rm B,-2}^{\frac{2p + 3}{2(p+4)}} \,
\left( \frac{E_{\rm 51}}{M_{\rm 3}} \right)^{\frac{2p+13}{2(p+4)}} \; \frac{\text{erg}}{\text{s}\;\text{Hz}}\hspace{0.2cm} (k=2)\;, \\ 
        \end{cases}\label{eq:coast_Lpk}
\end{split}
\end{align}
where $t_{100} = t/100 \; \text{days}$.

Similar to KNR, the SNR can also have a radially stratified velocity distribution. We derive an expression for the cumulative energy stratification index $s$ for supernova ejecta in Appendix \ref{app:strat_SNR}.

Figure \ref{fig:SN_LC} shows the temporal evolution of the peak luminosity and corresponding peak frequency. For CCSNe in a wind-like medium, the decay of the self-absorption frequency (equivalent to the peak frequency for Spectrum~2) is very steep, shifting from tens of GHz to hundreds of MHz within days. The observed luminosity in a given radio band peaks when the self-absorption frequency passes through it. 
Equation~(\ref{eq:coast_Lpk}) shows that during the coasting phase in a wind-like external medium the peak luminosity remains constant, while Equation~(\ref{eq:coast_nupk}) indicates that the peak frequency scales as $t^{-1}$ (see Appendix\;\ref{app:coast} for derivation). This behavior can be understood physically as follows. In a wind-like medium, the number of radiating relativistic electrons scales as $N_{\rm rel} \propto R$, whereas the post-shock magnetic field scales as $B \propto R^{-1}$. Consequently, the maximum spectral luminosity is constant, $L_{\nu,\max} \propto N_{\rm rel} B \propto R^{0}$. In addition, both the characteristic synchrotron frequency $\nu_{\rm m}$ and the peak (self-absorption) frequency $\nu_{\rm sa}$ scale as $\nu_{\rm m}, \nu_{\rm sa} \propto R^{-1}$. Therefore, the peak luminosity,
$L_{\rm pk} = L_{\nu,\max}\left(\frac{\nu_{\rm pk}}{\nu_{\rm m}}\right)^{(1-p)/2} \propto R^{0}$ is also a constant, while the peak frequency decreases as $\nu_{\rm pk} \propto R^{-1} \propto t^{-1}$. The rise and decay of the flux are determined solely by the relative location of the observing band with respect to the self-absorption frequency. For the first $\sim$100 days, light curves are nearly identical whether the DN phase is accounted for or the Newtonian approximation is used; differences emerge at later times. Both stratified and single-velocity ejecta show similar behavior, though the stratified profile yields brighter emission. A combination of single-velocity and stratified ejecta can reproduce the observed spread in peak times and peak luminosities in the 2--10~GHz band (\citealt{Bietenholz21}).  

The lower left panel illustrates an application to the nearby CCSN SN2023ixf (\citealt{Nayana25}). Assuming single-velocity ejecta, the time dependence of the peak frequency can be removed, and the peak flux is expressed as a function of peak frequency. Using $p = 2\alpha + 1$ (where $\alpha$ is the spectral slope in $F_{\nu} \propto \nu^{-\alpha}$) from the PLS~G, the stratification index $k$ can then be fitted assuming single velocity ejecta. The best-fit value, $k=1.29 \pm 0.14  \;(2\sigma)$, is in excellent agreement with the constraints of free-free absorption \citep{Nayana25}.  
The upper right panel shows that for SLSNe in a constant-density medium, the peak frequency remains near $\sim 100$~MHz, making the low-frequency radio bands at LOFAR the most promising for late-time radio detections from SLSN sites. By comparison, the peak luminosity at GHz frequencies is nearly an order of magnitude lower. The lower right panel shows that deeper radio non-detections at GHz frequencies (\citealt{Eftekhari21}) can place upper limits on the ambient number density, assuming fixed supernova energy and ejecta mass. The lower right panel shows that a higher number density corresponds to a shorter deceleration time, predicting a higher flux at the same observing band at later times. Continued observations are recommended to test this prediction.

In the supernova literature, several  prescriptions have been adopted for the minimum electron Lorentz factor, $\gamma_{\rm m}$. One common approach is to use the Newtonian limit of the standard relativistic expression while assuming that all electrons are accelerated ($\xi_{\rm e}=1$) \citep{Soderberg05,Chevalier06,Kamble14}. However, this prescription yields $\gamma_{\rm m}<1$, which is unphysical. Moreover, in the Newtonian limit, this prescription implies $\gamma_{\rm m}-1 \propto \beta^2$, leading to a minimal synchrotron frequency $\nu_{\rm m} \propto \beta^5$. This steep velocity dependence propagates into the determination of the synchrotron self-absorption frequency and other inferred source properties. The use of this prescription is not unique to supernova studies and has also been adopted in several kilonova radio-emission models \citep{Piran13,Sadeh24}.

Alternatively, some studies \citep{Reynolds98,Reynolds21} adopt a fixed value, typically $\gamma_{\rm m}=10$ , while assuming different prescriptions for the accelerated-electron fraction, such as $\xi_{\rm e}=1$, $\xi_{\rm e} \propto \beta$, or $\xi_{\rm e} \propto \beta^2$. While fixing $\gamma_{\rm m}$ avoids the problem of $\gamma_{\rm m}<1$, not all choices of $\xi_{\rm e}$ are physically consistent. In Newtonian shocks, maintaining a fixed relativistic $\gamma_{\rm m}$ requires  $\xi_{\rm e} \propto \beta^2$. In particular, adopting $\xi_{\rm e}=1$ and $\gamma_{\rm m}=10$ leads to a significant overprediction of the radio flux. Consequently, many commonly adopted prescriptions either underpredict or overpredict the radio flux (see the upper panels of Figure \ref{fig:SN_LC}).

Other treatments \citep{Chevalier06} parameterize the relativistic electron population by assuming that its energy density scales with the post-shock internal energy density while adopting $\gamma_{\rm m}=1$. Although this approach is more closely related to the DN framework, it leaves  $\xi_{\rm e}$ unspecified.

We emphasize that $\gamma_{\rm m}$ should not be regarded as an independent free parameter. For a given shock velocity and set of microphysical parameters, $\gamma_{\rm m}$ is a derived quantity. Only when the formal Newtonian expression predicts $\gamma_{\rm m}<1$ does this extrapolation cease to be physically meaningful. At this point, one naturally transitions to the DN regime, in which $\gamma_{\rm m}$ is fixed at a mildly relativistic value and $\xi_{\rm e}$ is determined self-consistently from energy conservation.

\subsection{$\Sigma-D$ relation in evolved supernova remnants}\label{sec:Sigma_D}

\begin{figure*}
    \centering
\begin{tabular}{c}
\includegraphics[scale=0.45]{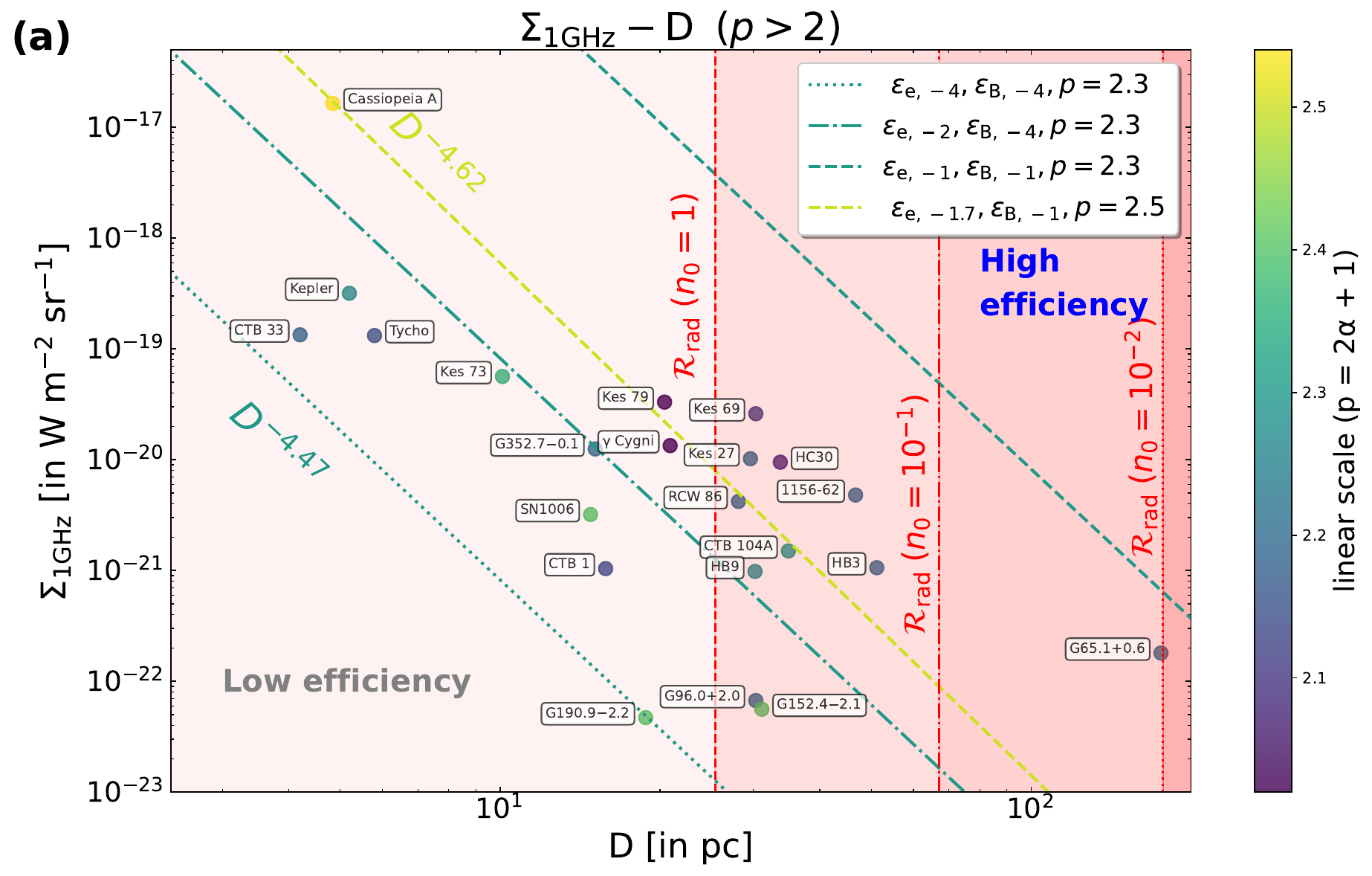} 
\end{tabular}
\includegraphics[scale=0.38]{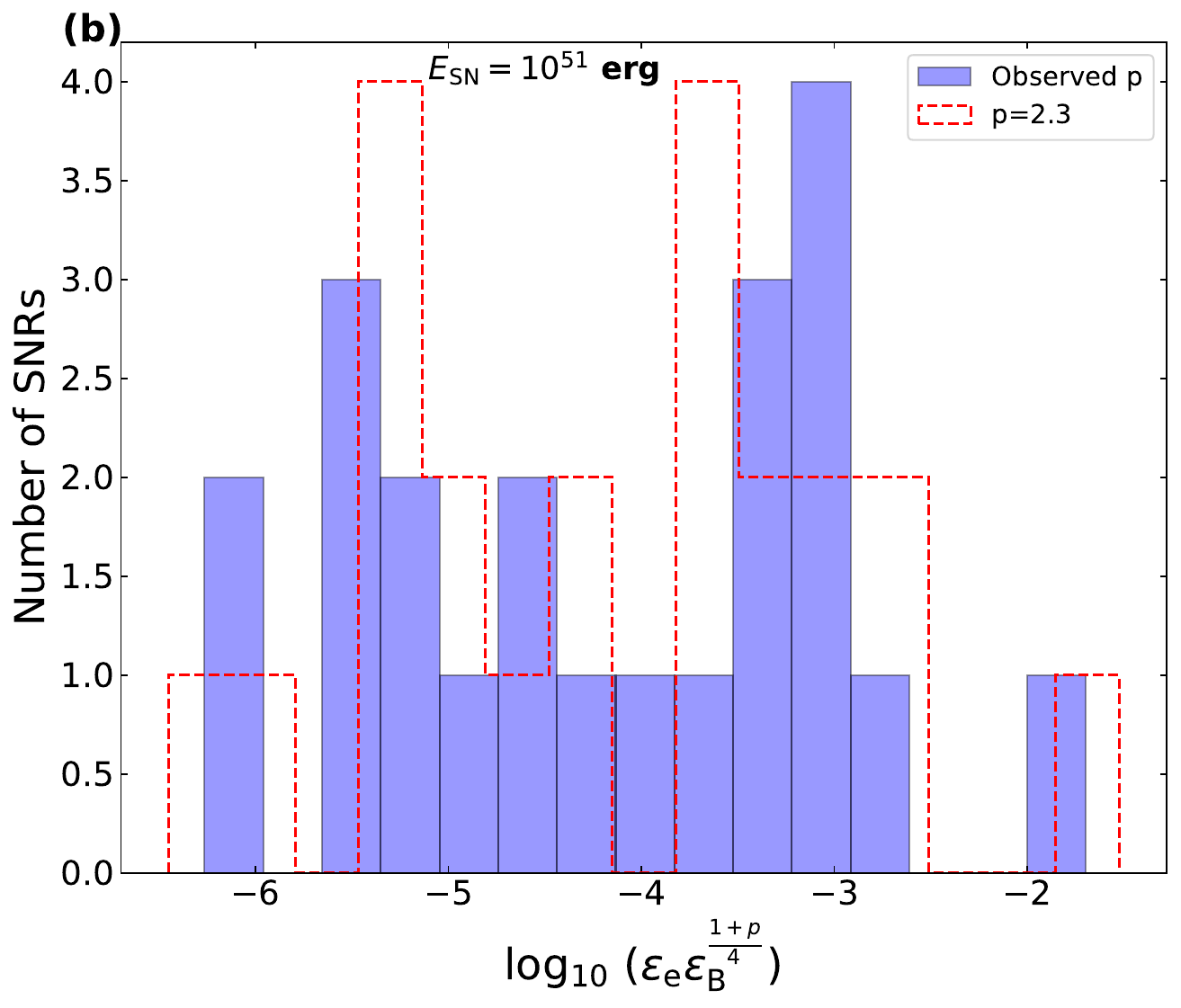}    
\hspace{0.5cm}
\includegraphics[scale=0.39]{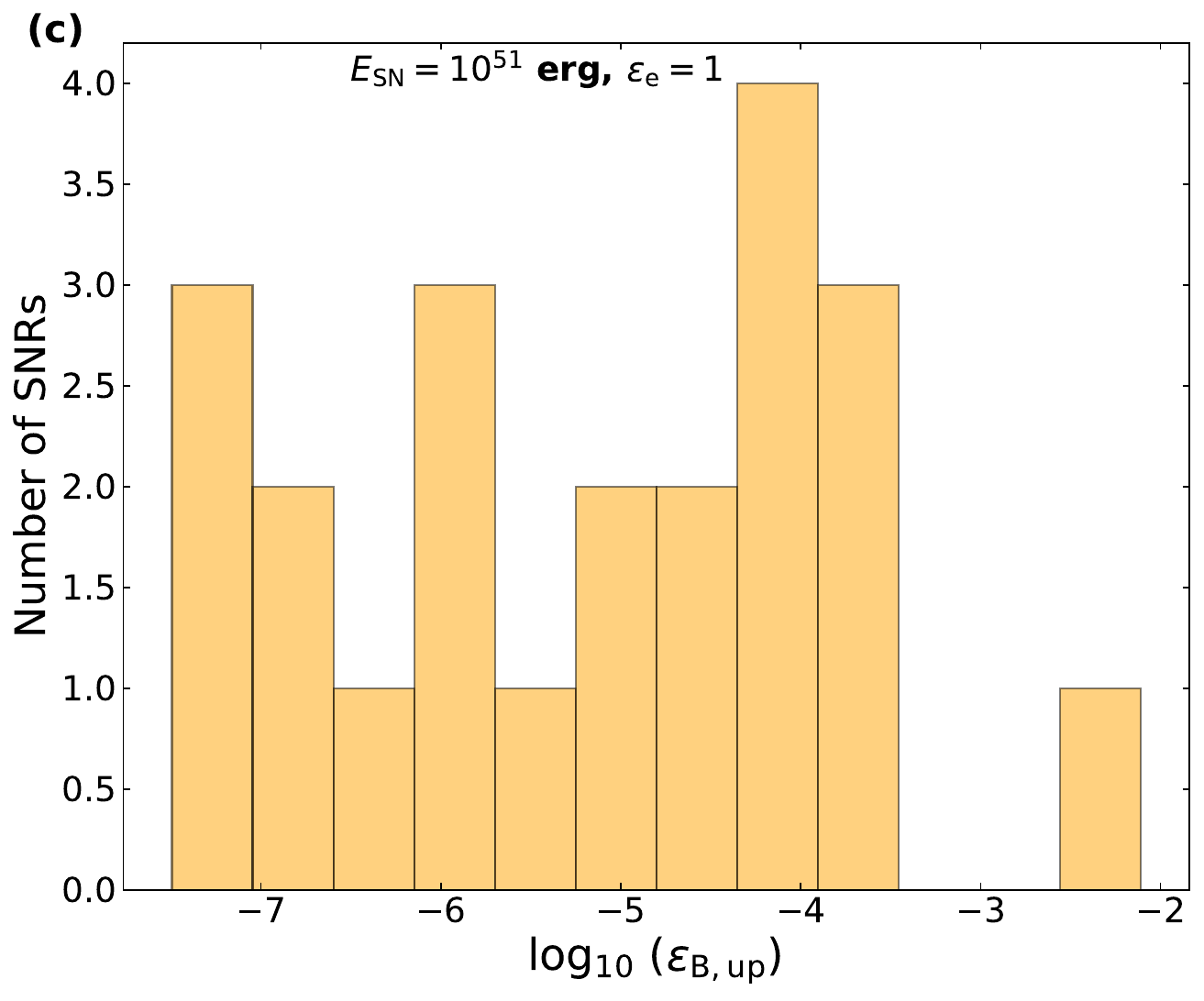} 
\caption{ \textbf{Illustration of the DN phase on the radio surface brightness–diameter ($\Sigma$–$D$) relation of evolved galactic SNRs}. For all panels, a uniform ambient medium with stratification index $k=0$ and an explosion energy of $E_{\rm SN} = 10^{51}$~erg is assumed. \textit{Upper panel (a):}  The $\Sigma$–$D$ relation at 1~GHz is shown for 22 Galactic SNRs with power-law index $p>2$ (see Table~\ref{tab:SigmaD}). Data points are color-coded by their inferred $p$ value, with the linear color scale shown on the right. The common names of individual remnants are labeled. The vertical dashed, dot-dashed, and dotted red lines mark the transition radii $\mathcal{R}_{\rm rad} = (25.4,\,67.1,\,177.0)\,\mathrm{pc}$, beyond which adiabatic evolution transitions to the radiative phase, for ambient densities $n_{0} = (1,\,10^{-1},\,10^{-2})$, respectively (see Appendix \ref{app:rad_phase}). The red shaded regions (lighter on the left and darker on the right) denote the radiative phase for intermediate ambient densities. Model DN curves are shown for different combinations of $(\epsilon_{\rm e}, \epsilon_{\rm B}, p)$, as indicated in the legend. For a given value of $p$, very low values of $(\epsilon_{\rm e}, \epsilon_{\rm B})$ are required to span the lower-left corner of the plot, whereas higher values populate the upper-right region. The deep green and greenish-yellow  lines correspond to $p = ( 2.3, 2.5)$, respectively, consistent with the color-coded colormap. \textit{Bottom panels:} Panel (b) shows the histogram of the parameter combination $\epsilon_{\rm e}\,\epsilon_{\rm B}^{\frac{1+p}{4}}$ assuming a fixed explosion energy of $E_{\rm SN}=10^{51}$~erg. Panel (c) shows the histogram of the upper limit on $\epsilon_{\rm B}$   which shows several order-of-magnitude–wide distribution of $\epsilon_{\rm B}$ (see text for discussion).
 }\label{fig:SN_SigmaD}
\end{figure*}

In this subsection, we explore the radio surface brightness - diameter relationship ($\Sigma-D$) for evolved and extended Galactic SNR (see Appendix \ref{app:SigmaD} for a derivation). In core-collapse scenarios,  the SNR reaches the deceleration phase in an effectively constant-density medium (since the total mass of the stellar wind, let alone its unshocked part, is not expected to exceed that of the SN ejecta). Moreover, the typical deceleration radius is $R_{\rm dec} \approx  2.4$ pc (see Equation~(\ref{eq:coast})),
supporting the use of a uniform-density profile ($k=0$) when examining the relation $\Sigma$ -- $D$ for typical diameters $D \gtrsim 5 \, \text{pc}$. Since in this work we focus on the adiabatic evolution, for the extended SNR sample considered in this section we explicitly indicate the transition radius beyond which this approximation breaks down and the SNR enters the radiative phase (see Appendix~\ref{app:rad_phase}).

For the typical range of $\epsilon_{\rm e} \approx 10^{-4} - 10^{-2}$ and $\epsilon_{\mathrm B} \approx 10^{-3} - 10^{-1}$  assumed in extended SNRs \citep{Reynolds21},  the $\Sigma-D$ relation in PLS G at $\nu = 1 $ GHz can be written as (see Appendix \ref{app:SigmaD}) (for $p = 2.3$ and $k =0$)  
\begin{equation}
\begin{split}
    \Sigma_\mathrm{1GHz}  \approx\!7.4\!\times10^{-20}
    \epsilon_{\rm e,-2}\epsilon_{\rm B,-3}^{\frac{1+p}{4}}E_{51}^{\frac{p+5}{4}} D_{10}^{-\frac{(3p+11)}{4}} \; \frac{\text{W}}{\text{m}^{2}\,\text{Hz}\,\text{Sr}}\,, 
\end{split}
\end{equation}
where $D_{10} = D/(10\,\text{pc})$.  

Figure \ref{fig:SN_SigmaD} shows the radio surface brightness (at 1\,GHz) – diameter relation for 22 Galactic supernova remnants with $p>2$ (see Table \ref{tab:SigmaD}). Using $p = 2\alpha + 1$ derived from the optically thin spectral index, we remove $p$ as a free parameter. For a fixed value of $p = 2.3$ and $E_{\rm SN} = 10^{51}\,\mathrm{erg}$, it can be seen that a very large range of $(\epsilon_{\rm e}, \epsilon_{\rm B})$ must be spanned in order to explain the large scatter in the data. Furthermore, for fixed $(\epsilon_{\rm e}, \epsilon_{\rm B})$, different values of $p$ correspond to different scalings of the surface brightness with diameter. For $p = (2.1, 2.3, 2.5)$, the surface brightness scales as $D^{-4.32}$, $D^{-4.47}$, and $D^{-4.62}$, respectively. This illustrates why the exponent obtained from the empirical fits may be less representative of the general population, as a wide variation in $(\epsilon_{\rm e}, \epsilon_{\rm B}, p)$ is required to account for the full spread of the observed data. Due to the steep dependence of the surface brightness on diameter, higher values of $(\epsilon_{\rm e}, \epsilon_{\rm B})$ are required to explain the surface brightness of SNRs with larger diameters (indicating very efficient particle acceleration and post-shock magnetic field amplification), while lower values are required for remnants with smaller diameters (indicating less efficient particle acceleration and post-shock magnetic field amplification). The lower panels demonstrate that, for a fixed explosion energy 
$E_{\rm SN} = 10^{51}\,\mathrm{erg}$, the combination 
$\epsilon_{\rm e}\,\epsilon_{\rm B}^{\frac{1+p}{4}}$ can be directly constrained and exhibits a broad scatter spanning several orders of magnitude. 
By taking $\epsilon_{\rm e}=1$ (the maximum physical value), one obtains an upper bound on 
$\epsilon_{\rm B}$, which likewise extends over several orders of magnitude. 
Such a wide range naturally translates into significant dispersion in the predicted synchrotron surface brightness, 
thereby contributing to the large scatter observed in the $\Sigma\text{--}D$ relation.

While this analysis assumes a fixed supernova energy, realistic explosions are expected to exhibit a spread in 
$E_{\rm SN}$ of up to an order of magnitude. However, this variation is subdominant compared to the much broader 
distribution inferred for the microphysical parameters. We therefore conclude that the dominant source of the 
observed scatter in the surface brightness--diameter relation is the intrinsic diversity in the microphysical 
parameters, rather than variations in the explosion energy.

\section{Conclusion}\label{sec:conc}

The deep-Newtonian (DN) phase represents a generic stage in the late-time evolution of astrophysical outflows, corresponding to the onset of a quasi-spherical Newtonian blast-wave. In this regime, the power-law relativistic electron population responsible for the synchrotron emission starts at mildly relativistic energies
and the emission typically occurs in the slow-cooling, self-absorbed synchrotron regime ($\nu_m<\nu_{\rm sa}<\nu_c$), with the flux density peaking at the self-absorption frequency, $\nu_{\rm sa}$, usually in the sub-GHz band. Low-frequency radio observations therefore provide the most direct probe of this phase.

For outflows launched with ultra-relativistic or mildly relativistic velocities, such as GRB jets and magnetar giant flares, the transition to the DN regime occurs after the deceleration phase. By contrast, for supernova and kilonova ejecta, the DN regime is already established during the coasting phase.  The essential difference between the standard Newtonian approximation and the DN regime lies in the scaling of the characteristic synchrotron frequency: $\nu_{\rm m} \propto \beta^5 \,\propto t^{-3}$ in the former, whereas $\nu_{\rm m} \propto \beta \,\propto t^{-3/5}$ in the latter. Neglecting the DN phase can lead to a significant underestimate of the radio flux, by factors of $\sim5$--$6$ during the coasting phase and by even larger factors after deceleration. 
Moreover, the temporal 
slopes of the light curve are modified, with a steeper flux rise and a shallower decay in the DN regime compared to the standard Newtonian case (see Equation~(\ref{eq:slope_DN})). 

Late-time, low-frequency radio observations of a broad class of astrophysical transients are therefore strongly motivated. Even non-detections can place meaningful constraints on the outflow energetics and the density of the surrounding medium. 

In the DN phase, the synchrotron self-absorption frequency, which traces the spectral peak, is typically found at low radio frequencies ($\sim 100~\mathrm{MHz}$--$1~\mathrm{GHz}$). Instruments such as LOFAR \citep{Haarlem13} and SKA \citep{Weltman20} are therefore well suited to probe this regime. For characteristic outflow energies of $E \sim 10^{51}~\mathrm{erg}$ and flux sensitivities of order $\sim 10~\mu\mathrm{Jy}$, this transition may be detectable in GRB afterglows within a decade for nearby events within a few hundred Mpc. 

For magnetar giant flares, this phase is likely observable only for nearby Galactic events. In the case of supernovae, there is no sharp transition, as the DN phase is effectively reached during the coasting phase; for nearby events, tracking the spectral peak evolution can provide a useful probe of the ambient density stratification. For superluminous supernovae, even radio non-detections can place stringent limits on the ambient density. Finally, for nearby short GRBs, the absence of a radio kilonova signal can constrain the presence and properties of a long-lived central magnetar engine.

\begin{acknowledgments}
SMR and JG thank the Yukawa Institute for Theoretical Physics at Kyoto University, where this work was nearly completed during the YITP long-term workshop YITP-T-26-02 on ``Multi-Messenger Astrophysics in the Dynamic Universe". SMR acknowledges useful discussions with A. J. Nayana.
PB is supported by a grant (no. 2024788) from the United States-Israel Binational Science Foundation (BSF), Jerusalem, Israel, by a grant (no. 1649/23) from the Israel Science Foundation, and by a grant (no. 80NSSC 24K0770) from the NASA Astrophysics Theory Program. 
\end{acknowledgments}

\bibliography{main}{}
\bibliographystyle{aasjournalv7}

\appendix

\section{Illustration of the DN phase from power-law distribution in particle momenta}\label{app:DN_momenta}

\begin{figure*}
\begin{tabular}{cc}
   \includegraphics[scale=0.42]{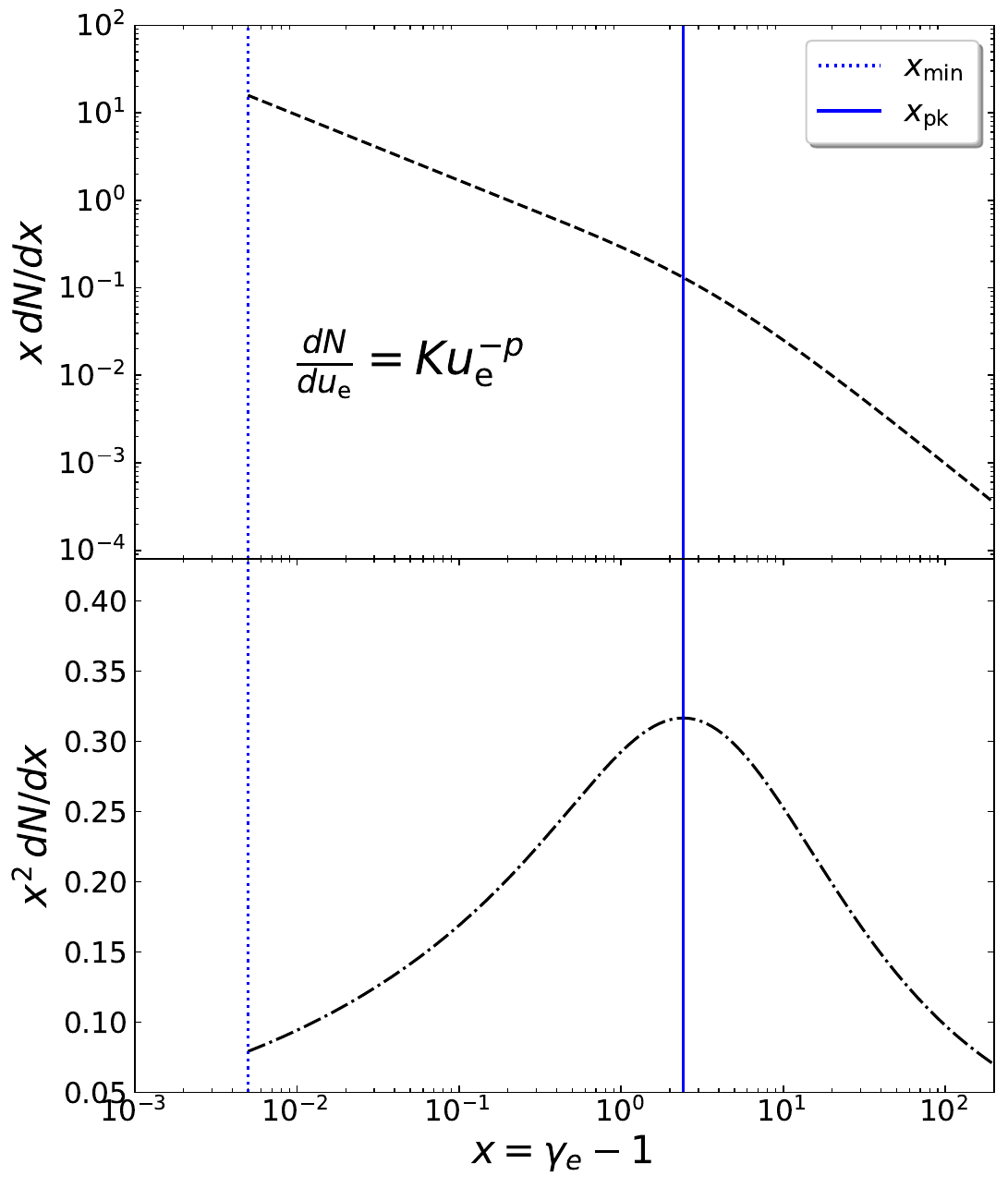}  & \includegraphics[scale=0.37]{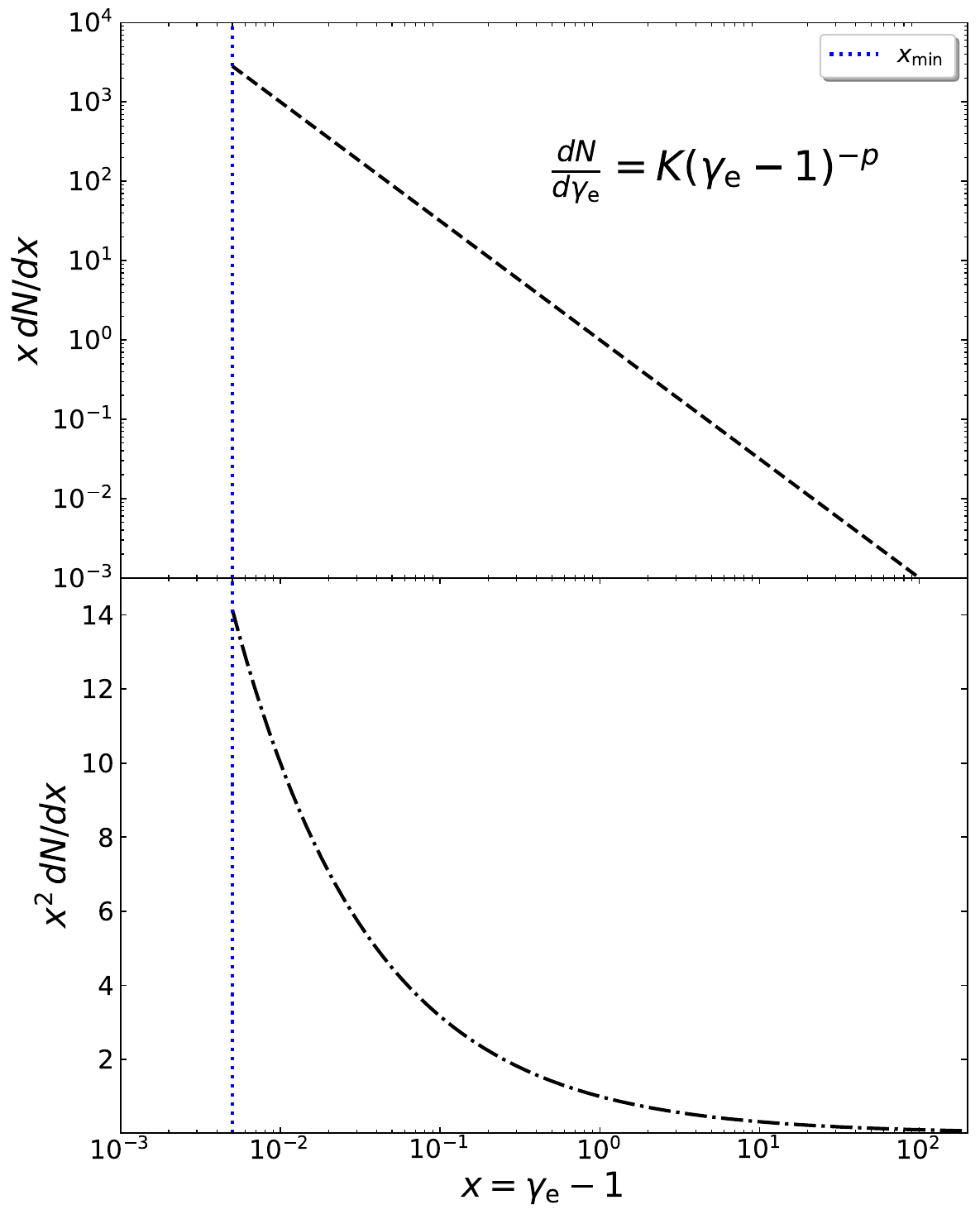}  \\ 
\end{tabular}
\caption{\textbf{Illustration of the DN regime from shock microphysics.} The lower and upper bounds on both panels  are$(u_{\mathrm{e,min}}, u_{\mathrm{e,max}}) = (10^{-1}, 10^{7})$ 
with power-law index $p = 2.5$. Both right and left plots show particle number (upper panel; dashed black curve) and energy 
(lower panel; dot-dashed curve) distributions as functions of the normalized 
kinetic energy $x \equiv \gamma_{\mathrm{e}} - 1$. Vertical dotted lines in both panels indicate $x_{\mathrm{min}} = \frac{1}{2} u_{\rm e,min}^2$. 
\textit{Left:} shows the number density and kinetic energy distribution for a power-law distribution of non-thermal particles, 
$dN/du_{\mathrm{e}}$, as a function of the normalized proper speed 
$u_{\mathrm{e}}$. The solid vertical lines indicate $x_{\mathrm{pk}}$ (Equation~\ref{eq:x_pk}), the location of the most energetic particles in the distribution. \textit{Right:} shows the number density and kinetic energy distribution for a power-law distribution of non-thermal particles, 
$dN/d\gamma_{\mathrm{e}}$, as a function of the normalized proper speed 
$\gamma_{\mathrm{e}} -1$.}\label{Fig:DN_distri}
\end{figure*}

In this appendix we consider a power-law distribution in the proper speed $u_{\rm e}$ of the non-thermal particles:
\begin{equation}
\frac{dN}{du_{\rm e}} = K \, u_{\rm e}^{-p}, \quad u_{\rm e} = \sqrt{\gamma_{\rm e}^2 - 1}\;.
\end{equation}
where $K$ is a normalization constant.

Following \cite{Sironi13} we define, 
\begin{equation}
x = \gamma_{\rm e} - 1 \quad \Rightarrow \quad u_{\rm e} = \sqrt{(x+1)^2 - 1} = \sqrt{2x+x^2}\;.
\end{equation}
The distribution in $x$ is obtained via
\begin{equation}
\frac{dN}{dx} = \frac{dN}{du_{\rm e}} \frac{du_{\rm e}}{dx}\;.
\end{equation}
Next we compute the derivative:
\begin{equation}
\frac{du_{\rm e}}{dx} = \frac{d}{dx} \sqrt{2x+x^2} = \frac{1+x}{\sqrt{2x+x^2}}\;.
\end{equation}
Thus, the distribution in $x$ becomes
\begin{equation}\label{eq:xdNdx}
\begin{split}
\frac{dN}{dx} = K \, (2x+x^2)^{-\frac{p}{2}} \cdot \frac{1+x}{\sqrt{2x+x^2}} 
= K (1+x)(2+x)^{-\frac{p+1}{2}}x^{-\frac{p+1}{2}}
\approx
\begin{cases}
            Kx^{-p} \hspace{1.0cm}  \text{for}\ x\gg1\;,                 \vspace{0.2cm}\\
            K(2x)^{-\frac{p+1}{2}}\hspace{0.45cm}  \text{for}\ x\ll1\;.   
\end{cases}
\end{split}
\end{equation}
The number of non-thermal electrons scales as
\begin{equation}
\begin{split}
N(x) \sim  x\frac{dN}{dx} = K(1+x)(2+x)^{-\frac{p+1}{2}}x^{\frac{1-p}{2}} 
\propto
\begin{cases}
            x^{1-p} \hspace{0.6cm}  \text{for}\ x\gg1\;,                 \vspace{0.2cm}\\
            x^{\frac{1-p}{2}}\hspace{0.6cm}  \text{for}\ x\ll1\;.   
\end{cases}
\end{split}
\end{equation}
\noindent
while the energy of the non-thermal electrons scales as 
\begin{equation}
\begin{split}
E(x) \sim  x^2\frac{dN}{dx} = K(1+x)(2+x)^{-\frac{p+1}{2}}x^{\frac{3-p}{2}} 
\propto
\begin{cases}
            x^{2-p} \hspace{0.6cm}  \text{for}\ x\gg1\;,                 \vspace{0.2cm}\\
            x^{\frac{3-p}{2}}\hspace{0.6cm}  \text{for}\ x\ll1\;.   
\end{cases}
\end{split}
\end{equation}
and peaks at the value
\begin{equation}\label{eq:x_pk}
    x_{\rm pk}=\frac{3-p+\sqrt{3-p}}{\,p-2\,} \; \hspace{1cm} \; \text{for $2 < p< 3$}\;. 
\end{equation}

Figure~\ref{Fig:DN_distri} and Eq.~(\ref{eq:xdNdx}) show that for $u_{\mathrm{e,min}} < 1$, 
the total number of particles is dominated by those near $u_{\mathrm{e,min}}$. However, as expressed in Eq.~(\ref{eq:x_pk}), the energy distribution 
peaks at $x_{\mathrm{pk}} > x_{\mathrm{min}}$, implying that while most 
particles reside at low energies, the bulk of the energy is carried by 
higher-energy mildly relativistic particles. As shown on the right panels no such feature exists if a power-law distribution is assumed in $(\gamma_{\rm e} -1)$.

It is instructive to consider the number and kinetic-energy fractions above
$\gamma_{\rm dn}=\sqrt{2}$ ($u_{\rm dn}=1$), defined as
\begin{equation}
\xi_{\rm e} \equiv \xi_{\rm e,pl} (>\gamma_{\rm dn})=\frac{N(>\gamma_{\rm dn})}{N_{\rm total}},\qquad
\epsilon_{\rm e} \equiv \epsilon_{\rm e,pl} (>\gamma_{\rm dn}) =\frac{E(>\gamma_{\rm dn})}{E_{\rm total}}.
\end{equation}
where $\xi_{\rm e,pl}$ and $\epsilon_{\rm e,pl}$ are the number and energy fraction for the full distribution functions for both prescriptions.

We compare power laws defined in kinetic energy,
$x=\gamma-1$, and in momentum,
$u=\sqrt{\gamma^2-1}$.

For $dn/dx=Kx^{-p}$, integration gives for $x_{\rm min} < x_{\rm dn}$
\begin{equation}
\xi_{\rm e}^{(x)}=
\left(\frac{x_{\min}}{x_{\rm dn}}\right)^{p-1},
\qquad
\epsilon_{\rm e}^{(x)}=
\left(\frac{x_{\min}}{x_{\rm dn}}\right)^{p-2},
\end{equation}
where
$x_{\min}=\sqrt{1+u_{\min}^2}-1$. For Newtonian shocks with shock velocity $\beta \ll 1$, for power-law in $x$ we have $x_{\min} \approx \frac{1}{2}u_{\min}^2 = \frac{1}{2} \beta_{\min}^2 =\frac{p-2}{p-1} \frac{m_p}{m_e} \frac{\epsilon_{\rm e,pl} }{\xi_{\rm e,pl} } \frac{1}{2}\beta^2$. Thus, for $u_{\min}\ll1$,
\begin{equation}
\xi_{\rm e}^{(x)}\propto u_{\min}^{2(p-1)},
\qquad
\epsilon_{\rm e}^{(x)}\propto u_{\min}^{2(p-2)}.
\end{equation}

For the momentum-space distribution $dn/du=Ku^{-p}$,
the number fraction is
\begin{equation}
\xi_{\rm e}^{(u)}=u_{\min}^{p-1} \propto \beta^2,  \hspace{1cm} \text{for $2<p<3$}
\end{equation}
while
\begin{equation}
\epsilon_{\rm e}^{(u)}=
\frac{\displaystyle\int_1^\infty
u^{-p}(\sqrt{1+u^2}-1)\,du}
{\displaystyle\int_{u_{\min}}^\infty
u^{-p}(\sqrt{1+u^2}-1)\,du}.
\end{equation}
Since the low-$u$ integrand scales as $u^{2-p}$,
\begin{equation}
\epsilon_{\rm e}^{(u)} = C_\mathrm{p}, \hspace{1cm} \text{for $2<p<3$},
\end{equation}
where $C_\mathrm{p}$ is a $p-$dependent constant. 

\begin{figure*}
    \centering
\includegraphics[scale = 0.5]{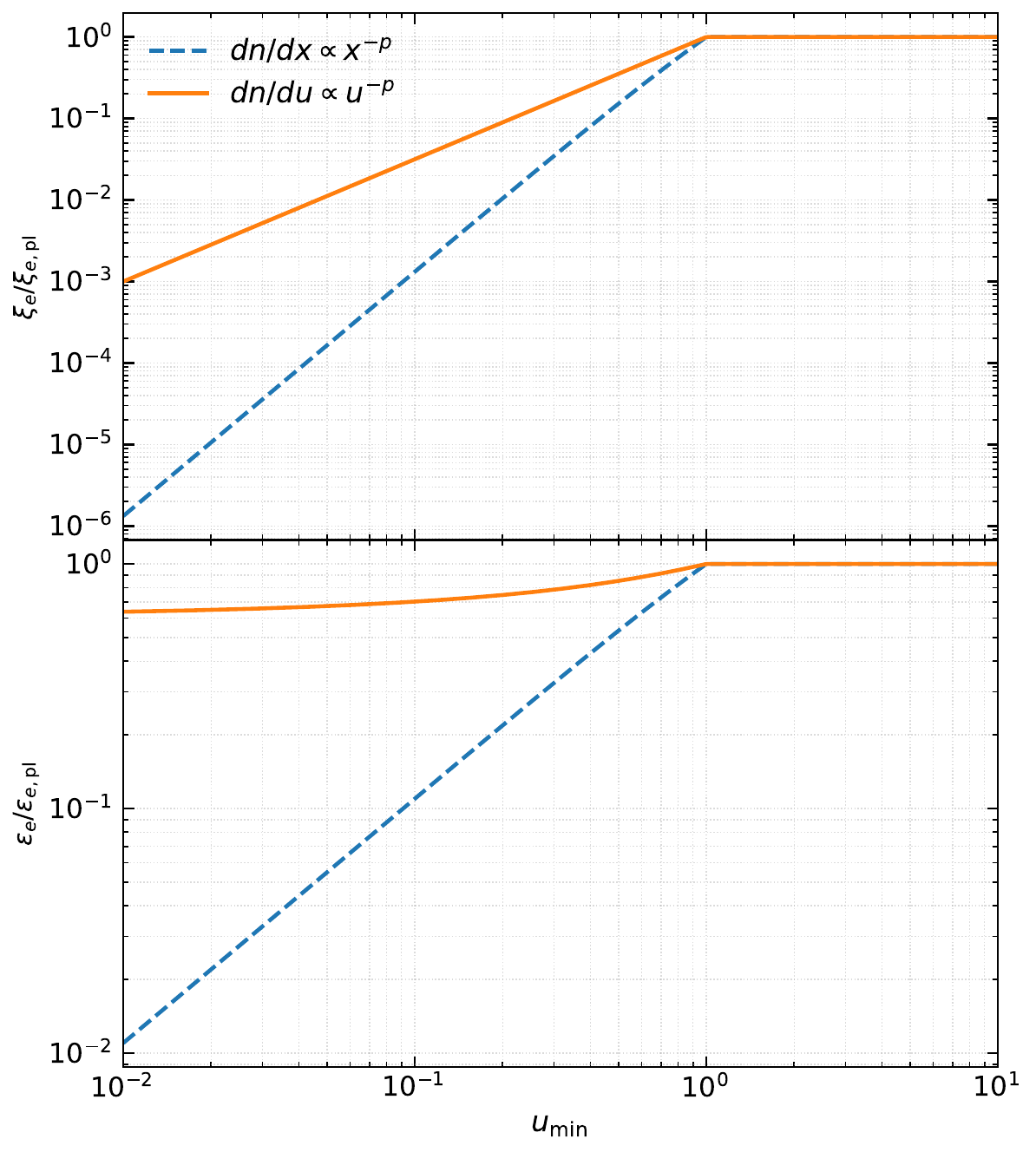}
\caption{Illustration of $\xi_e$ and $\epsilon_{\rm e}$ as a function of $u_{\min}$ for $p=2.5$. Here $\xi_{\rm e,pl}$ and $\epsilon_{\rm e,pl}$ correspond to number and energy fraction in the full-power law distribution. The top and middle panels show the accelerated-particle number fraction $\xi_{\rm e}$ and kinetic-energy fraction $\epsilon_{\rm e}$ respectively. Dashed blue and solid orange curves correspond to power-law distributions in kinetic energy, $dn/dx\propto x^{-p}$, and momentum, $dn/du\propto u^{-p}$, respectively.} \label{fig:Distri_eps}
\end{figure*}

Thus,  $\xi_{\rm e}$ remains dependent on $u_{\rm min}$ in both
parameterizations, $\epsilon_{\rm e}$ becomes insensitive to $x_{\rm min}$ for $2<p<3$ in the momentum-space formulation. In this work, we adopt $dn/du\propto u^{-p}$, for which $\epsilon_{\rm e}$
provides a robust measure of the energy carried by the
non-thermal particles. Figure~\ref{fig:Distri_eps} illustrates
these different cutoff dependences and highlights the robustness of
$\epsilon_{\rm e}$ in the momentum-space parameterization.

\section{Onset of the Radiative Phase for a Spherical Blastwave}\label{app:rad_phase}
\setcounter{equation}{0}

We consider a spherical blastwave of energy $E$ propagating into a uniform ambient medium of number density $n$. Post-deceleration the evolution is well described by the adiabatic Sedov--Taylor solution \citep{Sedov1959,Taylor1950}, until radiative losses become important. The cooling function is given by $\Lambda_{\rm net} = \Lambda_{\rm N} n_{\rm i} n_{\rm e} $ (where $n_{\rm i}$ and $n_{\rm e}$ are the number densities of ions and electrons, respectively). For a gas with internal energy $U$, the cooling timescale is given as $t_{\rm cool} = U/\Lambda_{\rm net}$.   The transition to the radiative phase is defined by the condition that the post-shock cooling time becomes comparable to the dynamical time \citep{CioffiMcKeeBertschinger1988,Draine2011}. Following \citealt{Draine2011} the post-shock temperature is $T = \frac{3}{16}\frac{\mu m_p}{k_B} v_s^2$ (where $v_{\rm s}$ is the shock velocity). This gives,
\begin{equation}
t_{\rm cool} = \frac{9}{16}\frac{\mu m_p}{n \Lambda(T)} v_s^2. \label{eq:t_cool}
\end{equation}

In the temperature range $T \sim 10^5\text{--}10^6\ \mathrm{K}$, the cooling function may be approximated as \citep{SutherlandDopita1993}
\begin{equation}
\Lambda_{\rm N}(T) \approx 2 \times 10^{-22} 
\left(\frac{T}{10^6\ \mathrm{K}}\right)^{-0.7}
\ \mathrm{erg\,cm^3\,s^{-1}},
\end{equation}
which when substituted in equation~(\ref{eq:t_cool}) gives
\begin{equation}
t_{\rm cool} \approx 1.1 \times 10^4\ 
n_0^{-1}
\left(\frac{v_s}{200\ \mathrm{km\,s^{-1}}}\right)^{3.4}
\ \mathrm{yr}.
\end{equation}

Equating $t_{\rm cool} \sim t$, one obtains a characteristic transition velocity
\begin{equation}
v_{\rm rad} \approx 200\ 
E_{51}^{1/15} n_0^{2/15}\ \mathrm{km\,s^{-1}},
\end{equation}
in agreement with detailed radiative blastwave calculations \citep{CioffiMcKeeBertschinger1988}.
The preceding evolution follows the Sedov--Taylor solution \citep{Sedov1959,Taylor1950},
\begin{equation}
v_s(t) = 523\ E_{51}^{1/5} n_0^{-1/5} t_{\rm 10kyr}^{-3/5}\ \mathrm{km\,s^{-1}},
\qquad
R(t) = 13.4\ E_{51}^{1/5} n_0^{-1/5} t_{\rm10kyr}^{2/5}\ \mathrm{pc}.
\end{equation}
where $t_{\rm 10 kyr} = t/ 10$ kyr. 
Setting $v_s = v_{\rm rad}$, the transition time is
\begin{equation}
t_{\rm rad} \approx 5 \times 10^4\ 
E_{51}^{4/17} n_0^{-9/17}\ \mathrm{yr},
\end{equation}
and the corresponding radius is
\begin{equation}
R_{\rm rad} \approx 25.4\ 
E_{51}^{5/17} n_0^{-7/17}\ \mathrm{pc}.
\end{equation}

For a canonical explosion with $E_{51}=1$, the transition properties for representative ambient densities are:
\begin{equation}
\begin{aligned}
n = 1\ \mathrm{cm^{-3}}: \quad 
& t_{\rm rad} \approx 5 \times 10^4\ \mathrm{yr}, 
\quad R_{\rm rad} \approx 25.4\ \mathrm{pc}, \\
n = 0.1\ \mathrm{cm^{-3}}: \quad 
& t_{\rm rad} \approx 1.8 \times 10^5\ \mathrm{yr}, 
\quad R_{\rm rad} \approx 67.1\ \mathrm{pc}, \\
n = 0.01\ \mathrm{cm^{-3}}: \quad 
& t_{\rm rad} \approx 6.4 \times 10^5\ \mathrm{yr}, 
\quad R_{\rm rad} \approx 177\ \mathrm{pc}.
\end{aligned}
\end{equation}
These results show that while the transition velocity is nearly universal, the corresponding time and radius depend sensitively on the ambient density, increasing substantially in environments of lower-density \citep{Draine2011}.

\section{Normalization Factors in Blast-Wave Dynamics}
\label{app:blastwave_correction}
\setcounter{equation}{0}

Equation~(\ref{eq:E(u,R)}) captures the correct
dynamical scaling but neglects order-unity normalization factors associated
with the detailed structure of the self-similar blast wave.  Here we summarize the exact
corrections in the two asymptotic regimes.

\subsection{Ultra-relativistic regime}

For $u\gg1$, the shock evolution is described by the Blandford--McKee (\citealt{BM76}; hereafter BM76)
self-similar solution. In this regime, the normalized proper speed approaches
the Lorentz factor of the shock front,
\begin{equation}
u\simeq \Gamma\;.
\end{equation}
The BM76 solution gives
\begin{equation}
E=
\frac{8\pi}{17-4k}\rho_1\Gamma^2R^3c^2 \;,
\end{equation}
where the external density profile is assumed to be
\begin{equation}
\rho_1(R)\propto R^{-k}\;.
\end{equation}
The swept-up mass is
\begin{equation}
M(<R)=\int_0^R4\pi r^2\rho_1(r)dr
=\frac{4\pi}{3-k}\rho_1R^3\;.
\end{equation}
Combining these expressions gives
\begin{equation}
u^2=
\frac{17-4k}{2(3-k)}
\frac{E}{M(<R)c^2}\;.
\end{equation}
Therefore, the correction factor in the normalized proper speed relative to the
simple estimate is
\begin{equation}
C_u^{\rm BM}
=
\left[\frac{17-4k}{2(3-k)}\right]^{1/2} = \begin{cases}
1.68, & k=0\;,\\
1.80, & k=1\;,\\
2.12, & k=2\;.
\end{cases}
\end{equation}
Since the shock radius satisfies $R\simeq ct$ in the ultra-relativistic limit,
the same factors correspond approximately to the normalization correction in
$R(t)$ at a fixed lab-frame time.

\subsection{Newtonian regime}

At late times, when $u\ll1$, the blast wave approaches the Sedov--Taylor
solution. The shock velocity can be written as \citep{Petruk2000}

\begin{equation}
v^2=
\frac{16\pi}
{(5-k)^2(3-k)\alpha_A}
\frac{E}{M(<R)}\;,
\end{equation}
where for $\alpha_A$ we use the approximation of \citet{OM88} for a spherical blastwave \citep[eq.~(44) of][]{Petruk2000},
\begin{equation}\label{eq:alpha_A}
\alpha_A = \frac{16\pi}{3(5-k)^2}\cdot\frac{11\hat{\gamma}-5-k(\hat{\gamma}-1)}{(\hat{\gamma}^2-1)[5\hat{\gamma}+1-k(\hat{\gamma}+1)]}\;,
\end{equation}

Taking $\hat{\gamma}=5/3$ as the adiabatic index of a non-relativistic gas, we obtain 
the correction factor in the normalized proper
speed,
\begin{equation}
C_u^{\rm ST}
=
\left[
\frac{16\pi}
{(5-k)^2(3-k)\alpha_A}
\right]^{1/2} =
\begin{cases}
1.116, & k=0\;,\\
1.29, & k=1\;,\\
1.63, & k=2\;.
\end{cases}
\end{equation}
The correction to the shock radius is smaller because the Sedov--Taylor
solution gives
\begin{equation}
R\propto t^{2/(5-k)}.
\end{equation}
Therefore, the radius correction is
\begin{equation}
C_R=(C_u^{\rm ST})^{2/(5-k)},
\end{equation}
yielding
\begin{equation}
C_R=
\begin{cases}
1.045, & k=0,\\
1.136, & k=1,\\
1.387, & k=2 .
\end{cases}
\end{equation}
Thus, the exact self-similar solutions modify the approximate blast-wave
normalization by factors of order unity. For the radius at a fixed lab-frame
time, the corresponding correction factors are approximately
\begin{equation}
C_R\simeq
\begin{cases}
1.68,\;1.80,\;2.12, & u\gg1,\\
1.76,\;1.83,\;2.00, & u\ll1,
\end{cases}
\end{equation}
for $k=0,1,2$, respectively.
These corrections modify the normalization of the blast-wave solution but do
not change its fundamental scaling.

\section{Useful expressions in the DN regime}\label{app:expr}
\setcounter{equation}{0}

In this appendix we give useful expressions for typical dynamical and spectral quantities. Like the main text we use a value of $\gamma_\mathrm{dn} = \sqrt{2}$ and $p=2.5$ throughout the text.

The deceleration radius $R_\mathrm{dec}$ is given by (for $u_{0} \gg 1$) as  
\begin{equation}
\begin{split}
   R_\mathrm{dec} &\ = \left[ \frac{(3-k) E}{4 \pi u_{\rm 0}^2 A c^2} \right]^{\frac{1}{(3-k)}}    \approx
   \begin{cases}
      \;1.2 \times 10^{17}   \; u_{2}^{-\frac{2}{3}}\; E_{\rm iso,53}^{\frac{1}{3}} \; n_{0}^{-\frac{1}{3}}\; \; \text{cm} \hspace{0.2cm} (k=0)\;, \; \vspace{0.2cm}\\      
     \;1.8 \times 10^{15}    \; u_{2}^{-2}\; E_{\rm iso,53} \; A_{\star}^{-1} \;  \; \text{cm} \hspace{0.2cm} (k=2)\;, \; \\  
   \end{cases}
\end{split}
\end{equation}
where we use the typical Lorentz factor and isotropic kinetic energy of outflow for GRB afterglows. Here, $u_{2} = u/10^2$, $E_{\rm iso,53} = E_{\rm iso} /10^{53}\; \text{erg}$ , $n_{0} = n/1\; \text{cm$^{-3}$} $ and $A_{\star} = A/ 5 \times 10^{11} \; \text{gm cm$^{-1}$}$ \citep{Chevalier99}. The observed deceleration time for $u_{\rm 0} \gg 1$ with $E_{\rm iso}$ can be written as
\begin{equation}
\begin{split}
t_{\rm dec} &\ \approx \frac{R_{\rm dec}}{2 u_{0}^2 c}     \approx 
\begin{cases}
    \; 194.4 \; u_{2}^{-\frac{8}{3}}\; E_{\rm iso,53}^{\frac{1}{3}} \; n_{0}^{-\frac{1}{3}}\; \text{sec} \hspace{0.2cm} (k=0)\;, \; \vspace{0.2cm}\\
    \;  2.9 \; u_{2}^{-4}\; E_{\rm iso,53} \; A_{\star}^{-1} \; \text{sec} \hspace{0.2cm} (k=2)\;.    
\end{cases}
\end{split} 
\end{equation}
The Newtonian transition radius  can be expressed as (for $u_\mathrm{0} > u_{\rm n} = 1$)
\begin{equation}
\begin{split}
R_{\rm N} &\ =  \left[ \frac{(3-k)E}{4\pi A u_{\rm N}^2 c^2} \right]^{\frac{1}{(3-k)}}                   \approx 
    \begin{cases}
        \;  0.2   \; E_{51}^{\frac{1}{3}} \; n_{0}^{-\frac{1}{3}}\; \; \text{pc}    \hspace{0.2cm}        (k=0)\;,           \vspace{0.2cm}\\
        \;  0.06  \; E_{51} \; A_{\star}^{-1} \;  \; \text{pc}         \; \hspace{0.1cm}          (k=2)\;.         
    \end{cases}
\end{split}
\end{equation}
The DN transition radius  is (for $u_{\rm 0} > u_{\rm DN}$)
\begin{equation}
\begin{split}
    &\ R_{\rm DN}  =  \left[ \frac{(3-k)E}{4\pi A u_{\rm DN}^2 c^2} \right]^{\frac{1}{(3-k)}}      \approx 
    \begin{cases}
        \; 0.53 \; n_{0}^{-\frac{1}{3}}\; E_{51}^{\frac{1}{3}}  \; \tilde{f}(p)^{\frac{1}{3}} \; \xi_{\rm e0}^{-\frac{1}{3}} \; \epsilon_{\rm e,-1}^{\frac{1}{3}} \;       \; \text{pc}         \; \hspace{0.1cm}          (k=0)\;,          \vspace{0.2cm}\\
        \;  1.62 \;   A_{\star}^{-1}\; E_{51} \; \tilde{f}(p) \; \xi_{\rm e0}^{-1} \; \epsilon_{\rm e,-1} \;     \; \text{pc}         \; \hspace{0.1cm}          (k=2)\;.           
    \end{cases}
\end{split}
\end{equation}

The cyclotron frequency in the DN regime is given as (for $t>t_{\rm DN}$)
\begin{equation}
    \begin{split}
      &\ \nu_{\rm B} \equiv \frac{q_{\rm e} B}{2 \pi m_{\rm e} c} = \frac{q_{\rm e}}{2 \pi m_{\rm e} c} \sqrt{16 \pi \epsilon_{\rm B} A R_{\rm DN}^{-k} \; u_{\rm DN}^2 c^2} \left( \frac{t}{t_{\rm DN}} \right)^{-\frac{3}{5-k}}  \\ &\ \approx
       \begin{cases}
           \;   14.5 \;  n_{0}^{\frac{1}{2}} \; \tilde{f}(p)^{-\frac{1}{2}}  \;  \xi_{e0}^{\frac{1}{2}}  \; \epsilon_{\rm e,-1}^{-\frac{1}{2}}  \; \epsilon_{\rm B,-2}^{\frac{1}{2}} \left( \frac{t}{t_{\rm DN}} \right)^{-\frac{3}{5}}    \; \; \text{kHz} \; \hspace{0.2cm} (k=0,p=2.5)\;,  \vspace{0.25cm}\\ 
           \; 1.6 \;  E_{51}^{-1} \; A_{\star}^{\frac{1}{2}}\;  \tilde{f}(p)^{-\frac{3}{2}}  \;  \xi_{e0}^{\frac{3}{2}} \; \epsilon_{\rm e,-1}^{-\frac{3}{2}}  \; \epsilon_{\rm B,-2}^{\frac{1}{2}} \left( \frac{t}{t_{\rm DN}} \right)^{-1}    \text{kHz} \;     \hspace{0.2cm}             (k=2,p=2.5)\;,
       \end{cases}
    \end{split}
\end{equation}

The DN cooling LF $\gamma_{\rm c}$ is given by  is (for $t>t_{\rm DN}$)
\begin{equation}\label{eq:DN_cool}
\begin{split}
    &\ \gamma_{\rm c}  \equiv \frac{6 \pi m_{e} c}{\sigma_{\rm T} B^2 t} = \frac{3 m_{\rm e}\left( u_{\rm DN} E \right)^{\frac{k-1}{3-k}} }{8 \sigma_{\rm T} \epsilon_{\rm B} A }  \left( \frac{t}{t_{\rm DN}} \right)^{\frac{k+1}{5-k}}  \approx  \begin{cases}
        2.5 \times 10^{5}  \;     E_{51}^{\frac{1}{3}} \; n_{0}^{-1}\;  \left[ \frac{G(p)}{G(2.5)} \right]^{\frac{1}{6}} \; \xi_{e0}^{-\frac{1}{6}}  \; \epsilon_{\rm e,-1}^{\frac{1}{6}} \; \epsilon_{\rm B,-2}^{-1} \; \left( \frac{t}{t_{\rm DN}} \right)^{\frac{1}{5}}  \hspace{0.2cm}   (k=0)\,,   \vspace{0.2cm}\\ 
        4.1\!\times\!10^{6} \;   E_{51} \; A_{\star}^{-1} \left[ \frac{G(p)}{G(2.5)} \right]^{-1} \, \xi_{e0} \, \epsilon_{\rm e,-1}^{-1} \epsilon_{\rm B,-2}^{-1} \,   \left( \frac{t}{t_{\rm DN}} \right) \hspace{0.2cm} (k=2)\,. 
    \end{cases}
\end{split}
\end{equation}

The cooling break $\nu_{\rm c}$ during the DN phase is (for $t>t_{\rm DN}$) 
\begin{equation}
    \begin{split}
        &\ h \nu_{\rm c} \equiv h \gamma_{\rm c}^2 \nu_{\rm B}  \approx  \begin{cases}
            \; 3.6  \; E_{51}^{\frac{2}{3}} \; n_{0}^{-1} \; \tilde{f}(p)^{- \frac{1}{6}} \xi_{\rm e0}^{\frac{1}{6}} \; \epsilon_{\rm e,-1}^{-\frac{1}{6}} \; \epsilon_{\rm B,-2}^{-\frac{3}{2}} \; \left( \frac{t}{t_{\rm DN}} \right)^{-\frac{1
            }{5}} \text{eV}      \;   \hspace{0.2cm}   (k=0, p =2.5)\; , \vspace{0.25cm}\\ 
            \;  108.9\;E_{51}^{\frac{3}{2}} \; A_{\star}^{-\frac{3}{2}} \; \tilde{f}(p)^{-\frac{7}{2}} \; \xi_{\rm e0}^{\frac{7}{2}} \; \epsilon_{\rm e,-1}^{-\frac{7}{2}} \; \epsilon_{\rm B,-2}^{\frac{3}{2}}\; \left( \frac{t}{t_{\rm DN}}\right)    \; \text{eV}  \; \hspace{0.2cm}  (k=2, p =2.5)\;,  \\ 
        \end{cases}  
    \end{split}
\end{equation}
which shows that in the DN regime for a constant density medium ($k=0$) the cooling break is just below the optical band while for a wind-like medium ($k=2$) it lies in the ultraviolet band.  
The minimal frequency $\nu_{\rm m}$ in the DN regime (for $t> t_{\rm DN}$)
\begin{equation}
    \begin{split}
       &\  \nu_{\rm m}  \equiv \gamma_{\rm dn}^2 \nu_{\rm B} \approx 
        \begin{cases}
            \;  28.9 \; 
            n_{0}^{\frac{1}{2}} \; \tilde{f}(p)^{-\frac{1}{2}}  \;  \xi_{e0}^{\frac{1}{2}}  \; \epsilon_{\rm e,-1}^{-\frac{1}{2}}  \; \epsilon_{\rm B,-2}^{\frac{1}{2}} \left( \frac{t}{t_{\rm DN}} \right)^{-\frac{3}{5}}   \; \;  \text{kHz} \;   \hspace{0.2cm}     (k=0,p=2.5)\; ,  \vspace{0.25cm}\\ 
            \; 3.2 \; E_{51}^{-1} \; A_{\star}^{\frac{1}{2}}\;  \tilde{f}(p)^{-\frac{3}{2}}  \;  \xi_{e0}^{\frac{3}{2}} \; \epsilon_{\rm e,-1}^{-\frac{3}{2}}  \; \epsilon_{\rm B,-2}^{\frac{1}{2}} \left( \frac{t}{t_{\rm DN}} \right)^{-1}    \; \; \text{kHz} \;   \hspace{0.2cm}       (k=2,p=2.5)\;.
        \end{cases}
    \end{split}
\end{equation}
The maximum spectral luminosity in the DN phase (for $t>t_{\rm DN}$)
\begin{equation}
\begin{split}
        &\ L_{\rm \nu,max} \equiv N_{\rm rel} \frac{m_{\rm e} c^2}{3 q_{\rm e}} \sigma_{\rm T} B  \approx 
        \begin{cases}
          \; 3.7 \times 10^{31} \;  E_{51} \; n_{0}^{\frac{1}{2}}\; \xi_{\rm e0}^{-\frac{2}{3}}\;  \epsilon_{\rm e,-1}^{\frac{1}{3}} \; \epsilon_{\rm B,-2}^{\frac{1}{2}} \; \left( \frac{t}{t_{\rm DN}} \right)^{-\frac{3}{5}}\text{erg s$^{-1}$ Hz$^{-1}$}  \; \hspace{0.2cm} (k=0)\;,                \vspace{0.2cm}\\ 
          \;  4.0 \times 10^{30} \; A_{\star}^{\frac{3}{2}} \; \xi_{\rm e0}^{-1} \;\epsilon_{\rm B,-2}^{\frac{1}{2}} \; \left( \frac{t}{t_{\rm DN}} \right)^{-1}\; \text{erg s$^{-1}$ Hz$^{-1}$}  \; \hspace{0.2cm} (k=2)\;.                  \\ 
        \end{cases}
\end{split}
\end{equation}
The maximum spectral luminosity $L_{\rm \nu,max}$ is not the peak spectral luminosity as the ordering $\nu_{\rm sa} < \nu_{\rm m}$ cannot be obtained for a broad parameter space (see Appendix \ref{app:Spec1}).  

\section{Implausibility of realizing the ordering  $\nu_{\rm \lowercase{sa}} < \nu_{\rm \lowercase{m} }$ in the DN phase}\label{app:Spec1}
\setcounter{equation}{0}

Assuming that $\nu_{\rm m} < \nu_{\rm sa}$ can be realized in the DN regime, the self-absorption frequency $\nu_{\rm sa}$ can be written as 
\begin{equation}
    \frac{2 k_\mathrm{B} T \nu_{\rm sa}^2}{c^2} = \frac{L_{\rm \nu,max}}{4 \pi^2 R^2 } \left( \frac{\nu_{\rm sa}}{\nu_{\rm m}} \right)^{\frac{1}{3}} \Rightarrow  \frac{\nu_{\rm sa}}{\nu_{\rm m}}   = \left( \frac{1}{m_{\rm e}} \; \frac{L_{\rm \nu,max}}{8 \pi^2 R^2 } \frac{1}{\gamma_{\rm dn}^{5}} \frac{1}{\nu_{\rm B}^2}\right)^{\frac{3}{5}}  
\end{equation}
where $k_\mathrm{B} T \approx \gamma_{\rm dn} m_{\rm e} c^2$ and $\nu_{\rm m} = \gamma_{\rm dn}^2 \nu_{\rm B} $ and $L_{\rm \nu,max} = N_{\rm rel} P_{\rm \nu,max} = \xi_{\rm e0} N_{\rm e} \frac{u^2}{u_{\rm DN}^2}\frac{m_{\rm e} c^2}{3 q_{\rm e}} \sigma_{\rm T} B$ where $N_{\rm e}$ is the total number of electrons in the shocked fluid and $B$ is the post-shock magnetic field (such that $\nu_{\rm B} = \frac{q_\mathrm{e} B}{2 \pi m_{\rm e} c}$ is the cyclotron frequency). 
In the coasting phase, 
\begin{equation}\label{eq:spec1_coast}
\begin{split}
  \frac{\nu_{\rm sa}}{\nu_{\rm m}} &\ =  \left[ \frac{f(p)\,\epsilon_{\rm e}\,\sigma_{\rm T}\,m_{\rm e}\,c^{2}}{48\,\pi^{3/2}\,q_{\rm e}^{3}\,\sqrt{\epsilon_{\rm B}\,A}} \right]^{3/5}
\; 2^{\frac{3k-18}{20}}
\; E^{\frac{3(k-2)}{20}}
\; M^{\frac{3(6-k)}{20}}
\; t^{\frac{3(k-4)}{10}}     \; \hspace{2cm} \; \text{for $t < t_{\rm dec}$}  \\
&\ \approx  \begin{cases}
     &\ 1.6 \times 10^{7} \; \tilde{f}(p)^{\frac{3}{5}}  \; \epsilon_\mathrm{e,-1}^{\frac{3}{5}} \; \epsilon_\mathrm{B,-2}^{-\frac{3}{10}} \; n_{0}^{-\frac{9}{10}}\;E_{51}^{-\frac{3}{10}} \; M_{3}^{\frac{9}{10}} \; t_{100}^{-\frac{6}{5}}       \hspace{1cm} \text{(k=0)}        \\
     &\  9.6 \times 10^{5} \;  \tilde{f}(p)^{\frac{3}{5}}  \; \epsilon_\mathrm{e,-1}^{\frac{3}{5}} \; \epsilon_\mathrm{B,-2}^{-\frac{3}{10}} \; A_{\star}^{-\frac{3}{10}} \; M_{3}^{\frac{6}{5}} \; t_{100}^{-\frac{3}{5}}         \hspace{2cm} \text{(k=2)}   
 \end{cases}
\end{split}
\end{equation}
where $t_{100} = t/100$ days.
In the adiabatic phase, 
\begin{equation}\label{eq:spec1_adia}
\begin{split}
      \frac{\nu_{\rm sa}}{\nu_{\rm m}} &\ = \left[ f(p)\;  \frac{ m_{\rm e} c^2 \sigma_{\rm T}}{24 q_{\rm e}^3}   \frac{\epsilon_{\rm e}}{\epsilon_{\rm B}^{\frac{1}{2}}} \right]^{\frac{3}{5}}  \left( \frac{E}{R} \right)^{\frac{3}{10}} \\ 
&\ \approx 2^{\frac{3}{10(k-3)}} 3^{\frac{3(k-4)}{10(3-k)}} \left( \frac{m_{\rm e}}{m_{\rm p}} \right)^{\frac{3(7-2k)}{10(3-k)}} \left( \frac{m_{\rm e} \sigma_{\rm T} c^2}{12 q_{\rm e}^3} \right)^{\frac{3}{5}}  f(p)^{\frac{3}{10(k-3)}} \epsilon_{\rm e}^{\frac{3}{10(k-3)}} \epsilon_{\rm B}^{-\frac{3}{10}} \xi_{\rm e0}^{\frac{3(7-2k)}{10(3-k)}} (4 \pi A c^2)^{\frac{3}{10(3-k)}} E^{\frac{3(2-k)}{10(3-k)}}  \left( \frac{u}{u_{\rm dn}} \right)^{\frac{3}{5(3-k)}} \\       
 &\ \begin{cases}
     &\   6.3 \times 10^{5} \;     \tilde{f}(p)^{-\frac{1}{10}}  \; \epsilon_\mathrm{e,-1}^{-\frac{1}{10}} \; \epsilon_\mathrm{B,-2}^{-\frac{3}{10} } \; \xi_{e0}^{\frac{7}{10}} \;  n_{0}^{\frac{1}{10}} \; E^{\frac{1}{5}}  \; \left( \frac{u}{u_{\rm dn}} \right)^{\frac{1}{5}}    \hspace{1cm} \text{(k=0)}        \\
     &\ 2.6 \times 10^{5} \; \tilde{f}(p)^{\frac{3}{5}}  \; \epsilon_\mathrm{e,-1}^{\frac{3}{5}} \; \epsilon_\mathrm{B,-2}^{-\frac{3}{10}} \; \xi_{e0}^{\frac{9}{10}} \; A_{\star}^{\frac{3}{10}} \; \left( \frac{u}{u_{\rm dn}} \right)^{\frac{3}{5}} \hspace{2cm}            \text{(k=2)}  
 \end{cases}
\end{split}
\end{equation}
Equations~\ref{eq:spec1_coast}--\ref{eq:spec1_adia} show that   for any 
reasonable choice of microphysical parameters and bulk energetics $\nu_{1}/\nu_{\rm m} < 1$ is not realized.
 This implies that the ordering $\nu_{1}/\nu_{\rm m} < 1$, cannot be achieved over a 
broad region of parameter space in the DN phase. Physically, this is expected since in this regime $\nu_{\rm m}$ lies 
close to the cyclotron frequency. Achieving $\nu_{1} < \nu_{\rm m}$ 
would therefore require the self-absorption frequency to be pushed 
below the cyclotron frequency, which is difficult to realize in practice.

\section{Useful Tables}
\setcounter{equation}{0}
In this appendix, we provide few useful tables. Table \ref{tab:gen_dynRad} summarizes the temporal scalings of the various physical quantities in the adiabatic phase for $u_\mathrm{0} > u_\mathrm{DN}$, while Table \ref{tab:dn_dynRad} lists their values at the deep Newtonian transition time, $t_{\mathrm{DN}}$.

\begin{table*}[ht]
\centering
\renewcommand{\arraystretch}{1.5}
\caption{Time-dependent scaling relations for key dynamical and radiative quantities in the Newtonian and the DN regime in a spherical blast wave expanding into a stratified external medium with density profile $\rho_{\rm ext} \propto R^{-k}$. The evolution is presented for two regimes: the Newtonian phase ($t < t_{\rm DN}$) and the deep Newtonian phase ($t > t_{\rm DN}$). Here, $u$ is the proper velocity of the shocked fluid, $\xi_{\rm e}$ is the fraction of accelerated electrons (with pre-deep Newtonian value $\xi_{e0}$), $N_p$ and $N_e$ are the number of swept-up protons and non-thermal electrons, $e_{\rm int}$ is the internal energy density, and $\nu_B$ is the cyclotron frequency. $\gamma_m$ and $\gamma_c$ are the minimum and cooling Lorentz factors of the electron distribution, associated with synchrotron frequencies $\nu_m$ and $\nu_c$, respectively. $L_{\nu,\rm max}$ is the maximum synchrotron spectral luminosity. All quantities  are expressed in terms of their values at the deep Newtonian transition time $t_{\rm DN}$.}
\begin{tabular}{lll}
\toprule
Quantity & $t_{\rm N}<\; t < t_{\rm DN}$ & $t > t_{\rm DN}$ \\
\hline 
$u$ & $u_{\rm DN} \left( \frac{t}{t_{\rm DN}} \right)^{\frac{3-k}{k-5}}$ & $u_{\rm DN} \left( \frac{t}{t_{\rm DN}} \right)^{\frac{3-k}{k-5}}$ \\
$\xi_{\rm e}$ & $\xi_{e0}$ & $\xi_{e0} \left( \frac{t}{t_{\rm DN}} \right)^{\frac{2(3-k)}{k-5}}$ \\
$N_p$ & $N_{{\rm p,DN}} \left( \frac{t}{t_{\rm DN}} \right)^{\frac{2(3-k)}{5-k}}$ & $N_{{\rm p,DN}} \left( \frac{t}{t_{\rm DN}} \right)^{\frac{2(3-k)}{5-k}}$ \\
$N_e$ & $N_{{\rm e,DN}} \left( \frac{t}{t_{\rm DN}} \right)^{\frac{2(3-k)}{5-k}}$ & $N_{{\rm e,DN}}$ \\
$e_{\rm int}$ & $e_{{\rm int,DN}} \left( \frac{t}{t_{\rm DN}} \right)^{-\frac{6}{5-k}}$ & $e_{{\rm int,DN}} \left( \frac{t}{t_{\rm DN}} \right)^{-\frac{6}{5-k}}$ \\
$\nu_B$ & $\nu_{{\rm B,DN}} \left( \frac{t}{t_{\rm DN}} \right)^{-\frac{3}{5-k}}$ & $\nu_{{\rm B,DN}} \left( \frac{t}{t_{\rm DN}} \right)^{-\frac{3}{5-k}}$ \\
$\gamma_m$ & $\gamma_{\rm dn} \left( \frac{t}{t_{\rm DN}} \right)^{-\frac{2(3-k)}{5-k}}$ & $\gamma_{\rm dn}$ \\
$\gamma_c$ & $\gamma_{{\rm c,DN}} \left( \frac{t}{t_{\rm DN}} \right)^{\frac{1+k}{5-k}}$ & $\gamma_{{\rm c,DN}} \left( \frac{t}{t_{\rm DN}} \right)^{\frac{1+k}{5-k}}$ \\
$\nu_m$ & $\gamma_{\rm dn}^2 \nu_{\rm B,DN} \left( \frac{t}{t_{\rm DN}} \right)^{-\frac{15 - 4k}{5-k}}$ & $\gamma_{\rm dn}^2 \nu_{\rm B,DN} \left( \frac{t}{t_{\rm DN}} \right)^{-\frac{3}{5-k}}$ \\
$\nu_c$ & $\gamma_{\rm c,DN}^2 \nu_{\rm B,DN} \left( \frac{t}{t_{\rm DN}} \right)^{\frac{2k-1}{5-k}}$ & $\gamma_{\rm c,DN}^2 \nu_{\rm B,DN} \left( \frac{t}{t_{\rm DN}} \right)^{\frac{2k-1}{5-k}}$ \\
$L_{\nu, \max}$ & $L_{\nu, {\rm max,DN}} \left( \frac{t}{t_{\rm DN}} \right)^{\frac{3-2k}{5-k}}$ & $L_{\nu, {\rm max,DN}} \left( \frac{t}{t_{\rm DN}} \right)^{-\frac{3}{5-k}}$ \\
\hline 
\end{tabular}\label{tab:gen_dynRad}
\end{table*}

\begin{table*}[ht]
\centering
\renewcommand{\arraystretch}{1.6}
\caption{Key dynamical and radiation quantities relevant to the deep Newtonian (DN) transition of a non-relativistic blast wave propagating into an external medium with density $\rho_{\rm ext} \propto R^{-k}$. The DN transition occurs at radius $R_{\rm DN}$ and time $t_{\rm DN}$. The parameter $\xi_{e0}$ denotes the fraction of electrons accelerated to relativistic energies at the transition. Quantities are expressed in terms of the total isotropic equivalent energy $E$, microphysical parameters $\epsilon_{\rm e}, \epsilon_{\rm B}$, power-law index $p$ of the electron distribution, and ambient density normalization $A = m_{\rm p} n(R_0) R_0^k$ at a fiducial radius $R_0$.}
\label{tab:dn_table}
\begin{tabular}{cc}
\toprule
\textbf{Dynamics} & \textbf{Radiation} \\
\hline 
\multicolumn{1}{c}{$\displaystyle A = m_{\rm p} \, n(R_0) \, R_0^k$} & \multicolumn{1}{c}{$\displaystyle N_{\rm DN} = \xi_{e0} \cdot \dfrac{4 \pi A R_{\rm DN}^{3-k}}{3-k}$} \\[8pt]
\multicolumn{1}{c}{$\displaystyle R_{\rm nr} = \left[ \dfrac{(3-k)E}{4 \pi A c^2} \right]^{\frac{1}{3-k}}$} & \multicolumn{1}{c}{$\displaystyle e_{\rm int,DN} = 4 A c^2 u_{\rm DN}^2 R_{\rm DN}^{-k}$} \\[8pt]
\multicolumn{1}{c}{$\displaystyle R_{\rm DN} = R_{\rm nr} \left[ \dfrac{1}{\gamma_{\rm dn}} \cdot \dfrac{(p-2)}{2(p-1)} \cdot \dfrac{m_{\rm p}}{m_{\rm e}} \cdot \dfrac{\epsilon_{\rm e}}{\xi_{e0}} \right]^{\frac{1}{3-k}}$} & \multicolumn{1}{c}{$\displaystyle \nu_{\rm B,DN} = \dfrac{q_{\rm e}}{2 \pi m_{\rm e} c} \sqrt{8 \pi \epsilon_{\rm B} e_{\rm int,DN}}$} \\[8pt]
\multicolumn{1}{c}{$\displaystyle u_{\rm DN} = \sqrt{2 \gamma_{\rm dn}} \left[ \dfrac{m_{\rm p}}{m_{\rm e}} \cdot \dfrac{(p-2)}{(p-1)} \cdot \dfrac{\epsilon_{\rm e}}{\xi_{e0}} \right]^{-\frac{1}{2}}$} & \multicolumn{1}{c}{$\displaystyle t_{\rm c1,DN} = \dfrac{3 m_{\rm e} c}{4 \epsilon_{\rm B} e_{\rm int,DN}}$} \\[8pt]
\multicolumn{1}{c}{$\displaystyle t_{\rm DN} = \dfrac{R_{\rm DN}}{u_{\rm DN} c}$} & \multicolumn{1}{c}{$\displaystyle \gamma_{\rm c,DN} = \dfrac{t_{\rm c1,DN}}{t_{\rm DN}}$} \\[8pt]
& \multicolumn{1}{c}{$\displaystyle L_{\nu,{\rm max,DN}} = N_{\rm e,DN} \cdot \dfrac{m_{\rm e} c^2}{3 q_{\rm e}} \cdot \sigma_{\rm T} \cdot \sqrt{8 \pi \epsilon_{\rm B} e_{\rm int,DN}}$} \\
\hline 
\end{tabular}
\label{tab:dn_dynRad}
\end{table*}

\section{Relating the observed time and emission radius}\label{app:t-R}
\setcounter{equation}{0}

\begin{figure}
    \centering
    \includegraphics[scale=0.6]{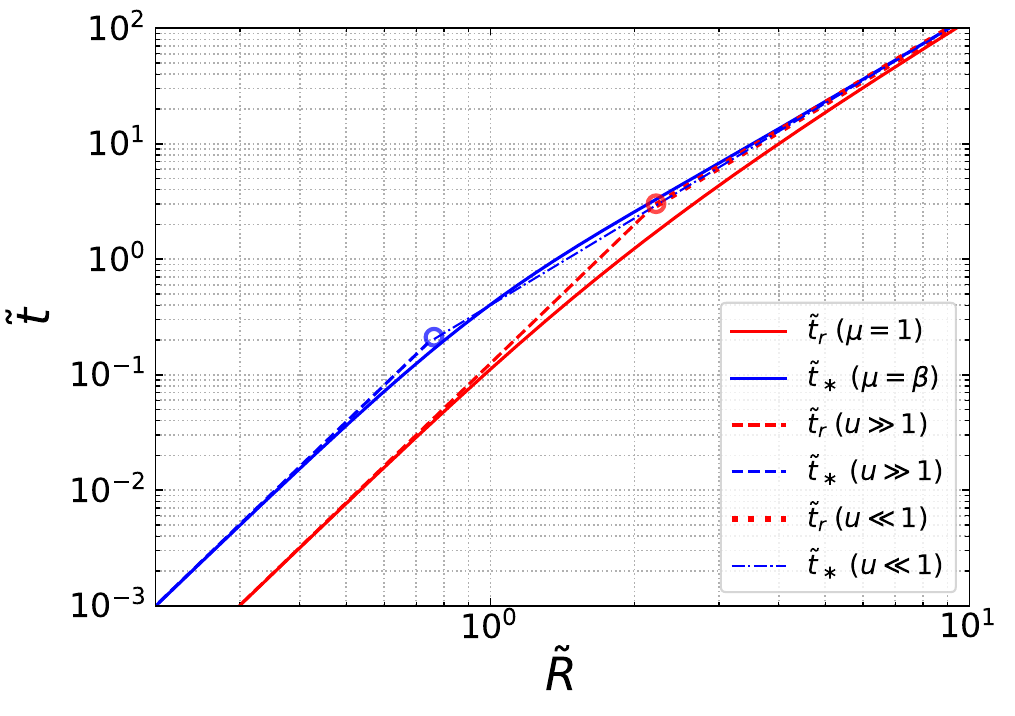}
    \caption{
Illustration of the observed normalized time $\tilde{t}$ as a function of the normalized radius $\tilde{R}$ for a spherical blast wave expanding into a constant-density medium ($k=0$).
The blue and red solid curves correspond to the radial normalized time $\tilde{t}_{\rm r}$ (for $\mu=1$) and the angular normalized time $\tilde{t}_{\ast}$ (for $\mu=\beta$), respectively.
The dashed blue and red curves show the ultra-relativistic ($u\gg1$) asymptotic limits for $\tilde{t}_{\ast}$ and $\tilde{t}_{\rm r}$, respectively.
The dot-dashed blue and dotted red curves represent the Newtonian ($u\ll1$) asymptotic limits for $\tilde{t}_{\ast}$ and $\tilde{t}_{\rm r}$, which coincide, indicating that line-of-sight effects become negligible in the Newtonian regime. The intersection points of the ultra-relativistic and Newtonian asymptotes are indicated by open circles (red for $\tilde{t}_{\rm r}$ and blue for $\tilde{t}_{\ast}$) (see text for details).
}
    \label{fig:EATS_time}
\end{figure}
The differential relation between the lab frame time and radius is given by $dt_\mathrm{lab} = \frac{dR}{\beta c}$, implying
$t_\mathrm{lab}(R)=\int_0^{R}\frac{dr}{\beta(r) c}$. For our simple spherical dynamics $u=\tilde{R}^{(k-3)/2}$ where 
$\tilde{R}=R/R_N$ is the radius normalized by the Newtonian transition radius $R_N$. Similarly defining a normalized time $\tilde{t}=\frac{ct}{R_N}$ we have $d\tilde{t}_\mathrm{lab} = \frac{d\tilde{R}}{\beta c}$ where $\beta(\tilde{R})=(1+\tilde{R}^{3-k})^{-1/2}$ leading to
\begin{equation}
\tilde{t}_{\rm lab}(\tilde{R}) = \int_0^{\tilde{R}}dx\sqrt{1+x^{3-k}} = \tilde{R}\,\times\,_2F_1\left(-\frac{1}{2},\frac{1}{3-k},\frac{4-k}{3-k},-\tilde{R}^{3-k}\right)\;,
\end{equation}
where $_2F_1$ is the hypergeometric function. For $k=2$ it reduces to a simple expression, $\tilde{t}_{\rm lab}(\tilde{R};k\!=\!2) = \frac{2}{3}[(1+\tilde{R})^{3/2}-1]$.

The observed time is given by $t(R,\mu) = t_{\rm lab}(R) - \frac{R\mu}{c}$ where $\mu=\cos\theta$ and $\theta$ is the angle from our line of sight. Along our line of sight ($\mu=1$), it reduces to the so called radial time, $t_r = t_{\rm lab} - \frac{R}{c}$, which in our case is given by
\begin{equation}
\tilde{t}_r(\tilde{R}) = \tilde{R}\,\left[\,_2F_1\left(-\frac{1}{2},\frac{1}{3-k},\frac{4-k}{3-k},-\tilde{R}^{3-k}\right)-1\right]\;.
\end{equation}
However, at a given observed time $t$ the emission does not arrive only from the line of sight, but from all of the equal arrival time surface (EATS), corresponding to a range of radii. Therefore, one would like to choose a representative radius along the EATS. For example, in the relativistic regime $\theta=1/\Gamma$ is chosen \citep[e.g.][]{SPN98}. More generally, this corresponds to the widest point of the EATS, $R_*$ that corresponds to the outer edge of the image, where $\mu_*=\beta_*$ \citep{GranotSari02,GRL05}, where the values of quantities at this point are denotes by an asterisk subscript. Adopting this representative point on the EATS leads to
\begin{equation}
\tilde{t}_{*}(\tilde{R}) = \tilde{R}\,\left[\,_2F_1\left(-\frac{1}{2},\frac{1}{3-k},\frac{4-k}{3-k},-\tilde{R}^{3-k}\right)-(1+\tilde{R}^{3-k})^{-1/2}\right]\;.
\end{equation}

Figure \ref{fig:EATS_time} shows $\tilde{t}_r$ and $\tilde{t}_*$ versus $\tilde{R}$, along with their asymptotic limits. For $\tilde{t}\ll1$ ($u\gg1$) we have $\tilde{t}_r\approx\frac{1}{8-2k}\tilde{R}^{4-k}$ and $\tilde{t}_*\approx\frac{5-k}{8-2k}\tilde{R}^{4-k}$ while $\tilde{t}\gg1$ ($u\ll1$) we have $\tilde{t}_r\approx\tilde{t}_*\approx\frac{2}{5-k}\tilde{R}^{(5-k)/2}$. The more realistic $\tilde{t}_*$ (\textit{solid blue line}) nicely agrees with our approximations (\textit{dashed} and \textit{dashed-dotted} blue lines), where we primarily use the $u\ll1$ approximation ($\tilde{t}_r\approx\tilde{t}_*\approx\frac{2}{5-k}\tilde{R}^{(5-k)/2}$) for $t\geq t_N\Leftrightarrow\tilde{t}\geq\frac{2}{5-k}$ ($u\leq1\Leftrightarrow\tilde{R}\geq1$).

\section{Derivation of DN scalings in the adiabatic phase}\label{app:adia}
\setcounter{equation}{0}

In this appendix, we consider the DN phase in the adiabatic phase. We consider the adiabatic expansion of a spherical blast wave post-deceleration in a cold and stratified medium described by profile $A R^{-k}$. 
As the flow speed drops below the critical proper $u<u_{\rm DN}$, the minimal LF of the relativistic electrons saturates to the value $\gamma_{\rm dn}$.
Let relativistic electrons be a fraction $\xi_{\rm e}$ that carry a fraction $\epsilon_{\rm e}$ of the internal energy density, while the post-shock magnetic field carries $\epsilon_{\rm B}$ of the internal energy density. 
The adiabatic evolution of the blastwave for $R > R_{\rm dec}$ is given as
\begin{equation}
    M(<R) c^2 u^2 = C_\mathrm{u}^2  E   \hspace{1cm} \text{for $R \geq R_{\rm dec}$}
\end{equation}
where $E$ is the total energy in the outflow. Here we assume the normalization $C_\mathrm{u}$ to be unity for convenience (for exact numerical value see Appendix \ref{app:blastwave_correction}). 

The minimal LF  $\gamma_{\rm m}$ is given as 
\begin{equation}
    \frac{\gamma_{\rm m}}{G(p)} - 1\approx  \frac{\epsilon_{\rm e}}{\xi_{\rm e}} \frac{m_{\rm p}}{m_{\rm e}} \frac{u^2}{2}   \hspace{1cm} \text{for $u \ll 1$}
\end{equation}
where $G(p) = \frac{(p-2)}{(p-1)}$ and $p$ is the power-law index of the relativistic electrons.

The critical proper speed $u_{\rm DN}$ can be estimated as 
\begin{equation}
    u_{\rm DN}^2 =   \frac{2}{f(p)} \frac{m_{\rm e}}{m_{\rm p}} \frac{\xi_{\rm e0}}{\epsilon_{\rm e}}.  
\end{equation}
which shows that the proper critical speed $u_{\rm DN}$ is not a hydrodynamical transition but is decided by the microphysics of particle acceleration. Here $f (p) \equiv G(p)/(\gamma_\mathrm{dn} - G(p))$.

The DN radius $R_{\rm DN}$ and the corresponding DN time $t_{\rm DN}$ are 
\begin{equation}
    R_{\rm DN}  \equiv R (u = u_{\rm DN})=  \left( \frac{(3-k) E}{4 \pi A c^2} \right)^{\frac{1}{3-k}} u^{\frac{2}{k-3}}_{\rm DN}  \quad , \quad t_{\rm DN} \equiv f_{\rm k} \frac{R_{\rm DN}}{u_{\rm DN} c} = f_{\rm k} \frac{1}{c} \left( \frac{(3-k) E}{4 \pi A c^2} \right)^{\frac{1}{3-k}} u_{\rm DN}^{\frac{5-k}{k-3}}
\end{equation}
where $f_\mathrm{k} = 2/(5-k)$ as motivated in Appendix \ref{app:t-R}. For the rest of the analysis we will assume this factor to be unity as our primary objective in this appendix is to obtain the scalings.

The fraction of power-law electrons in relativistic electrons is given as 
\begin{equation}
    \xi_{\rm e} = \xi_{e0} \frac{u^2}{u^2_{\rm DN}}. 
\end{equation}

\noindent\textbf{Relativistic electrons $N_{\rm rel}$}:

\noindent
The total number of electrons in the shocked fluid is
\begin{equation}
    N_{\rm e} = \frac{M(<\!R)}{m_{\rm p}} = \frac{E}{m_{\rm p} c ^2 u^2} 
\end{equation}
The number of relativistic electrons in the shocked fluid
\begin{equation}
    N_{\rm rel} = \xi_{\rm e}  N_{\rm e} = \xi_{\rm e0} \; N_{\rm e} \frac{u^2}{u^2_{\rm DN}} = \frac{\xi_{\rm e0} E}{m_{\rm p} u_{\rm DN}^2 c^2} = \frac{f(p) \epsilon_{\rm e} E}{2  m_{\rm e} c^2}
\end{equation}
which shows that in the DN regime the number of relativistic electrons saturates and is equal to the number of relativistic electrons at the appropriate critical speed $u = u_{\rm DN}$. Interestingly, it does not depend on $\xi_{\rm e,0}$.
The proper speed in the deep Newtonian regime can be expressed as
\begin{equation}
    \frac{u^2}{u^2_{\rm DN}} = \left( \frac{R_{\rm DN}}{R} \right)^{3-k} = \left( \frac{u_{\rm DN} c }{u c} \frac{t_{\rm DN}}{t} \right)^{3-k} < 1        \hspace{1cm} \text{for $u< u_{\rm DN}$}
\end{equation}
which gives 
\begin{equation}
   \frac{u}{u_{\rm DN}} = \left( \frac{R}{R_{\rm DN}} \right)^{\frac{k-3}{2}}  = \left( \frac{t}{t_{\rm DN}} \right)^{\frac{k-3}{5-k}}  
\end{equation}

\noindent\textbf{Cyclotron frequency $\nu_{\rm B}$}:\\
The internal energy density is  $e_{\rm int} = 4 \Gamma (\Gamma -1) Ac^2 R^{-k}$ given as 
\begin{equation}
    e_{\rm int}  \approx 2 u^2 c^2 A R^{-k} =  e_{\rm int,DN} \; \left( \frac{R}{R_{\rm DN}} \right)^{-3} = e_{\rm int,DN} \left( \frac{t}{t_{\rm DN}} \right)^{\frac{6}{k-5}}
\end{equation}
where $e_{\rm int,DN} =2 A u_{\rm DN}^2 c^2 R_{\rm DN}^{-k} = \frac{(3-k)^{\frac{k}{k-3}}}{2 \pi} (4 \pi A c^2)^{\frac{3}{3-k}} E^{\frac{k}{k-3}} u_{\rm DN}^{\frac{6}{3-k}}$.  The post-shock magnetic field strength can be written as 
\begin{equation}
    B = \sqrt{8 \pi \epsilon_{\rm B} e_{\rm int}} = \sqrt{8 \pi \epsilon_{\rm B} e_{\rm int,DN}} \left( \frac{t}{t_{\rm DN}} \right)^{\frac{3}{k-5}} = B_{\rm DN} \left( \frac{t}{t_{\rm DN}} \right)^{\frac{3}{k-5}}
\end{equation}
where $    B_{\rm DN} =  \sqrt{8 \pi \epsilon_{\rm B} e_{\rm int,DN}}$ can be expressed in terms of the base parameters as
\begin{equation}
    B_{\rm DN} = 2^{\frac{9-2k}{6-2 k}}   (3-k)^{\frac{k}{2k-6}} f(p)^{\frac{3}{2k-6}}   \left( \frac{m_{\rm e}}{m_{\rm p}} \right)^{\frac{3}{6-2k}}       \epsilon_{\rm B}^{\frac{1}{2}} \epsilon_{\rm e}^{\frac{3}{2k -6}} \xi_{\rm e0}^{\frac{3}{6- 2 k}}    (4 \pi A c^2)^{\frac{3}{6-2k}} E^{\frac{k}{2k-6}}
\end{equation}
The cyclotron frequency can be written as 
\begin{equation}
    \nu_{\rm B} \equiv \frac{q_{\rm e} B}{2 \pi m_{\rm e} c} = \nu_{\rm B,DN} \left( \frac{t}{t_{\rm DN}} \right)^{\frac{3}{k-5}}
\end{equation}
where $\nu_{\rm B,DN} = \frac{q_\mathrm{e} B_{\rm DN}}{2 \pi m_{\rm e} c}$ can be expressed in terms of the basic parameters as 
\begin{equation}
    \nu_{\rm B,DN} =\left( \frac{q_{\rm e}}{2 \pi m_{\rm e} c}\right)  2^{\frac{9-2k}{6-2 k}}   (3-k)^{\frac{k}{2k-6}} f(p)^{\frac{3}{2k-6}}   \left( \frac{m_{\rm e}}{m_{\rm p}} \right)^{\frac{3}{6-2k}}       \epsilon_{\rm B}^{\frac{1}{2}} \epsilon_{\rm e}^{\frac{3}{2k -6}} \xi_{\rm e0}^{\frac{3}{6- 2 k}}    (4 \pi A c^2)^{\frac{3}{6-2k}} E^{\frac{k}{2k-6}}
\end{equation}

\noindent\textbf{Maximum spectral luminosity $L_{\rm \nu,max}$}:\\
The maximum spectral luminosity due to relativistic electrons is given by
\begin{equation}
    L_{\rm \nu,max} =  N_{\rm rel} P_{\rm \nu,max} =   N_{\rm rel}   \left( \frac{m_{\rm e} c^2}{3 q_{\rm e}} \sigma_{T} B \right) = L_{\rm \nu,max,DN} \left( \frac{t}{t_{\rm DN}} \right)^{\frac{3}{k-5}}
\end{equation}
where 
\begin{equation}
    L_{\rm \nu,max,DN} = \frac{f(p) \epsilon_{\rm e} E}{2 m_{\rm e} c^2} \left(  \sqrt{8 \pi \epsilon_{\rm B} e_{\rm int,DN}} \frac{m_{\rm e} c^2}{3 q_{\rm e}} \sigma_{T}  \right) 
\end{equation}
which can be expressed in terms of the basic parameters as  
\begin{equation}
    L_{\rm \nu, {max},DN} =    \frac{\sigma_{\rm T}}{3 q_{\rm e}}2^{\frac{3}{6-2 k}}  (3-k)^{\frac{k}{2k-6}} \left( \frac{m_{e}}{m_{p}} \right)^{\frac{3}{6 - 2 k}} \; f(p)^{\frac{3-2k}{6-2k}}\; \epsilon_{\rm B}^{\frac{1}{2}} \epsilon_{\rm e}^{\frac{3-2k}{6 - 2k}} \xi_{\rm e0}^{\frac{3}{6 - 2 k}} (4 \pi A c^2)^{\frac{3}{6 - 2k}} E^{\frac{3k-6}{2k-6}}
\end{equation}

\noindent\textbf{The minimal frequency $\nu_{\rm m}$}:

\noindent
The minimal frequency is given as 
\begin{equation}
    \nu_{\rm m} \equiv \gamma^2_{\rm dn} \nu_{\rm B} = \nu_{\rm m,DN} \left( \frac{t}{t_{\rm DN}} \right)^{\frac{3}{k-5}} \hspace{1cm} \text{where $    \nu_{\rm m,DN} = \gamma_{\rm dn}^2  \nu_{\rm B,DN}$}
\end{equation}

\noindent
\textbf{The cooling frequency $\nu_{\rm c}$}:

\noindent
The cooling LF is $ \gamma_{\rm c} = \frac{6 \pi m_{\rm e} c}{\sigma_{T} B^2 t }$ while the cyclotron frequency can be written as
\begin{equation}
   \nu_{\rm c} \equiv \gamma_{\rm c}^2 \nu_{\rm B} = \nu_{\rm c,DN} \left( \frac{t}{t_{\rm DN}} \right)^{\frac{1-2k}{k-5}}  \hspace{1cm} \text{where $\nu_{\rm c,DN} = \left( \frac{q_{\rm e}}{2 \pi m_{\rm e} c}\right) \left( \frac{6 \pi m_{\rm e} c}{\sigma_{\rm T}} \right)^2 \frac{1}{B_{\rm DN}^3 t_{\rm DN}^2}$}
\end{equation}

\noindent
\textbf{The self-absorption frequency $\nu_{\rm sa}$}:
\noindent
The self-absorption frequency $\nu_{\rm sa}$ in Spectrum 2 ($\nu_{\rm m} < \nu_{\rm sa}$) can be obtained from the relationship,
\begin{equation}
    \frac{2 k_\mathrm{B} T \nu_{\rm sa}^2}{c^2} \times \frac{\pi R^2}{d^2} = \frac{L_{\rm \nu,max}}{4 \pi d^2} \left( \frac{\nu_{\rm sa}}{\nu_{\rm m}}\right)^{\frac{1-p}{2}}
\end{equation}
where $k_{\rm B} T = \gamma_{\rm sa} m_{\rm e} c^2 = \sqrt{\frac{\nu_{\rm sa}}{\nu_{\rm B}}} m_{\rm e} c^2$ (Here $k_{\rm B}$ is the Boltzmann's constant). 
The self-absorption frequency in Spectrum 2  can be written as 
\begin{equation}
    \nu_{\rm sa} = \left[ \frac{1}{m_{\rm e}} \frac{L_{\rm \nu,max}}{8 \pi^2 R^2} \;  \gamma_{\rm dn}^{p-1} \nu_{\rm B}^{\frac{p}{2}}\right]^{\frac{2}{p+4}} = \nu_{\rm sa,DN} \; \left( \frac{t}{t_{\rm DN}}\right)^{\frac{3p+14}{(k - 5)(p+4)}}
\end{equation}
where
\begin{equation}
\begin{split}
\nu_{\rm sa,DN} &\  = \left[\frac{1}{m_{\rm e}}  \frac{L_{\rm \nu,max,DN}}{8 \pi^2 R_{\rm DN}^2 } \gamma_{\rm dn}^{p-1} \nu_{\rm B,DN}^{\frac{p}{2}} \right]^{\frac{2}{p+4}} \\
&\ = K_1 \;  A^{\frac{14 + 3p}{2(3-k)(p+4)}} 
E^{\frac{4-6k - pk}{2(3-k)(p+4)}}
\gamma_{\rm dn}^{\frac{2(p-1)}{(p+4)}}
f(p)^{\frac{-4k-2 - 3p}{2(3-k)(p+4)}}
\xi_{\rm e0}^{\frac{14 + 3p}{2(3-k)(p+4)}}
\epsilon_{\rm e}^{\frac{-4k-2 - 3p}{2(3-k)(p+4)}}
\epsilon_B^{\frac{p+2}{2(p+4)}}        \\
&\ = 
\begin{cases}
    K_1( k=0 ) \; A^{\frac{14 + 3p}{6(p+4)}} 
E^{\frac{2}{3(p+4)}}
\gamma_{\rm dn}^{\frac{2(p-1)}{(p+4)}}
f(p)^{\frac{-2 - 3p}{6(p+4)}}
\xi_{\rm e0}^{\frac{14 + 3p}{6(p+4)}}
\epsilon_{\rm e}^{\frac{-2 - 3p}{6(p+4)}}
\epsilon_B^{\frac{p+2}{2(p+4)}}  \hspace{1cm} (k=0),          \\
    K_1 (k=2)  \;  A^{\frac{14+3p}{2(p+4)}} 
E^{-1}
\gamma_{\rm dn}^{\frac{2(p-1)}{(p+4)}}
f(p)^{\frac{-10 - 3p}{2(p+4)}}
\xi_{\rm e0}^{\frac{14 + 3p}{2(p+4)}}
\epsilon_{\rm e}^{\frac{-10 - 3p}{2(p+4)}}
\epsilon_B^{\frac{p+2}{2(p+4)}}   \hspace{1cm} (k=2), 
\end{cases}
\end{split}
\end{equation}
where the constant $K_1$ is
\begin{eqnarray}
\begin{array}{l}
{K}_{1}={\left[\frac{{\sigma}_{T}{q}_{\rm e}^{p/2-1}}{48{\pi}^{2}{m}_{\rm e}^{1+p/2}}\right]}^{\frac{2}{p+4}}
\quad \cdot {2}^{\frac{9p+54-4k}{2(3-k)(p+4)}}\cdot  {\pi}^{\frac{2kp-3p+14}{2(3-k)(p+4)}}\
\quad \cdot {c}^{\frac{kp+14}{(3-k)(p+4)}}\cdot {(3-k)}^{-\frac{8+k(p+2)}{2(3-k)(p+4)}}\cdot {\left(\frac{{m}_{\rm e}}{{m}_{\rm p}}\right)}^{\frac{3p+14}{2(3-k)(p+4)}}.
\end{array}
\end{eqnarray}

\noindent
\textbf{Spectral luminosity $L_{\nu_{\rm sa}}$ at the self-absorption frequency}:

\noindent
The spectral luminosity at the self-absorption frequency can be estimated as
\begin{equation}
    L_{\nu_{\rm sa}} = L_{\rm \nu_{\rm sa,DN}} \left( \frac{t}{t_{\rm DN}}\right)^{\frac{2p+13}{(k-5)(p+4)}} 
\end{equation}
where 
\begin{equation}
\begin{split}
   L_{\nu_{\rm sa,DN}} &\ = L_{\rm \nu,max,DN} \; \nu_{\rm sa,DN}^{\frac{1-p}{2}} \; \nu_{\rm m,DN}^{\frac{p-1}{2}}    \\
   &\ = K_{2}  \; A^{\frac{2p+13}{2(3-k)(p+4)}}
E^{\frac{26-13k+4p-2kp}{2(3-k)(p+4)}}
\gamma_{\rm dn}^{\frac{5(p-1)}{(p+4)}}
f(p)^{\frac{17-10k-2p}{2(3-k)(p+4)}}
\xi_{\rm e0}^{\frac{2p+13}{2(3-k)(p+4)}}
\epsilon_{\rm e}^{\frac{17-10k-2p}{2(3-k)(p+4)}}
\epsilon_B^{\frac{2p+3}{2(p+4)}}  \\
&\ = \begin{cases}
    &\  K_2(k=0)  \;  A^{\frac{2p+13}{6(p+4)}}
E^{\frac{13+2p}{3(p+4)}}
\gamma_{\rm dn}^{\frac{5(p-1)}{(p+4)}}
f(p)^{\frac{17-2p}{6(p+4)}}
\xi_{\rm e0}^{\frac{2p+13}{6(p+4)}}
\epsilon_{\rm e}^{\frac{17-2p}{6(p+4)}}
\epsilon_B^{\frac{2p+3}{2(p+4)}}                      \hspace{1cm} (k=0) \\ 
    &\  K_2(k=2) \; A^{\frac{2p + 13}{2(p+4)}}
E^{0} 
\gamma_{\rm dn}^{\frac{5(p-1)}{(p+4)}}
f(p)^{-\frac{(3 + 2p)}{2(p+4)}}
\xi_{\rm e0}^{\frac{2p+13}{2(p+4)}}
\epsilon_{\rm e}^{-\frac{(3 + 2p)}{2(p+4)}}
\epsilon_B^{\frac{2p+3}{2(p+4)}} \hspace{1cm}                       (k=2) \\
\end{cases}
\end{split}
\end{equation}
where the constant $K_2$  is 
\begin{equation}
    K_2 = \sigma_T^{\frac{5}{p+4}} \; q_e^{\frac{2p-7}{p+4}} {(3-k)}^{\frac{4p-4-2kp-3k}{2(3-k)(p+4)}}
\Bigl( 2^{a_2} \; 3^{b_2} \; \pi^{c_2} \; c^{2F_c} \; m_e^{m_2/2} \; m_p^{p_2/2} \Bigr)^{\frac{1}{2(3-k)(p+4)}}
\end{equation}
where
\[
\begin{aligned}
a_2 &= 21 + 6k + 24p - 6kp ,\\
b_2 &= -30+10k,\\
c_2 &= 2p + 13,\\
F_c &= 19-2k -4p +2kp,\\
m_2 &= 38 - 4k -8p +4kp,\\
p_2 &= -4p - 26.
\end{aligned}
\]

\section{DN scalings in the coasting phase}\label{app:coast}
\setcounter{equation}{0}

In this section we consider a newtonian outflow of energy $E$ and mass $M$ launched at a single velocity $u_{0} < u_{\rm DN}$. In this scenario the DN regime is realized at launch itself. 
The coasting speed $u_{0}$ is
\begin{equation}
    u_{0} = \frac{1}{c} \sqrt{\frac{2 E}{M}}.
\end{equation}
The critical proper speed $u_{\rm DN}$ is  
\begin{equation}
    u_{\rm DN}^2 =   \frac{2}{f(p)} \frac{m_{\rm e}}{m_{\rm p}} \frac{\xi_{\rm e0}}{\epsilon_{\rm e}}.  
\end{equation}
The radius is related to the coasting velocity as
\begin{equation}
    R = u_{0} c t.
\end{equation}
The number of relativistic electrons is 
\begin{equation}
    N_{\rm rel} = \xi_{\rm e0} \frac{u_{0}^2}{u_{\rm DN}^2} \frac{4 \pi A R^{3-k}}{(3-k) m_{\rm p}}. 
\end{equation}
The post-shock magnetic field is
\begin{equation}
    B = \sqrt{16 \pi \epsilon_{\rm B} u^2_{\rm 0} A c^2 R^{-k} }
\end{equation}
The cyclotron frequency is given as
\begin{equation}
    \nu_{\rm B} = \frac{q_{\rm e}}{2 \pi m_{\rm e} c} \sqrt{16 \pi \epsilon_{\rm B} u^2_{\rm 0} A c^2 R^{-k} }. 
\end{equation}
The minimal frequency is given as 
\begin{equation}
    \nu_{\rm m} = \gamma_{\rm dn}^2 \nu_{\rm B} = \gamma_{\rm dn}^2 \; \frac{q_{\rm e}}{2 \pi m_{\rm e} c} \sqrt{16 \pi \epsilon_{\rm B} u^2_{\rm 0} A c^2 R^{-k} }.
\end{equation}
The maximum spectral luminosity is given as (($t<t_{\rm dec}$))
\begin{equation}
    \begin{split}
        L_{\nu,\max} &\ =  N_{\rm rel} \frac{m_{\rm e} c^2}{3 q_{\rm e}} \sigma_{T} B  = \frac{16 \pi^{3/2} \sigma_T }{3 q_e (3-k) } \; 
f(p) \epsilon_{\rm e} \epsilon_{\rm B}^{\frac{1}{2}} \; A^{3/2} \; 
2^{\frac{8-3k}{4}} \left( \frac{E}{M} \right)^{\frac{3(4-k)}{4}} \; 
t^{3 - \frac{3k}{2}} \\
      &\ = \begin{cases}
          &\  \frac{64 \pi^{3/2} \sigma_T }{9 q_e } \; 
f(p) \epsilon_{\rm e} \epsilon_{\rm B}^{\frac{1}{2}} \; A^{3/2} \;
\left( \frac{E}{M} \right)^{3} t^{3} \hspace{0.6cm} (k=0), \\
          &\ \frac{16 \sqrt{2} \pi^{3/2} \sigma_T}{3 q_e } \; 
f(p) \epsilon_{\rm e} \epsilon_{\rm B}^{\frac{1}{2}} \; A^{3/2} \;
\left( \frac{E}{M} \right)^{3/2} \hspace{0.6cm} (k=2).           \\ 
      \end{cases}  
    \end{split}
\end{equation}%

The self-absorption frequency $\nu_{\rm sa}$ is given as ($t<t_{\rm dec}$)
\begin{equation}
    \begin{split}
     \nu_{\rm pk} &\ =   \nu_{\rm sa} = \left[ \frac{1}{m_{\rm e}} \frac{L_{\rm \nu,max}}{8 \pi^2 R^2} \;  \gamma_{\rm dn}^{p-1} \nu_{\rm B}^{\frac{p}{2}}\right]^{\frac{2}{p+4}}  \\
     &\ = \left[ \frac{\sigma_{\rm T} }{3(3-k)} \cdot 
\frac{ 2^{\frac{16 - 6k + 6p - kp}{8}} \; \pi^{-\frac{1}{2} - \frac{p}{4}} }
{ m_e^{1 + \frac{p}{2}} \; c^{\frac{p}{2}} \; q_e^{1 - \frac{p}{2}} } \right]^{\frac{2}{p+4}} \, \gamma_{\rm dn}^{\frac{2(p-1)}{p+4}} \, f(p)^{\frac{2}{p+4}} \, 
A^{\frac{6+p}{2(p+4)}} \, \epsilon_{\rm e}^{\frac{2}{p+4}} \,
\epsilon_B^{\frac{2+p}{2(p+4)}} \,
\left( \frac{E}{M} \right)^{\frac{16 - 6k + p(2-k)}{4(p+4)}} \,
t^{\frac{4 - 6k - kp}{2(p+4)}}         \\
        &\ = 
        \begin{cases}
            &\ \left[ \frac{\sigma_T \,}{9} \cdot 
\frac{ 2^{\frac{16 + 6p}{8}} \; \pi^{-\frac{1}{2} - \frac{p}{4}} }
{ m_e^{1 + \frac{p}{2}} \; c^{\frac{p}{2}} \; q_e^{1 - \frac{p}{2}} } \right]^{\frac{2}{p+4}} \; \gamma_{\rm dn}^{\frac{2(p-1)}{p+4}} \, f(p)^{\frac{2}{p+4}} \, A^{\frac{6+p}{2(p+4)}} \, \epsilon_{\rm e}^{\frac{2}{p+4}} \,
\epsilon_B^{\frac{2+p}{2(p+4)}} \,
\left( \frac{E}{M} \right)^{\frac{8 + p}{2(p+4)}} \,
t^{\frac{2}{p+4}}  \hspace{0.5cm} (k=0) ,         \\
            &\  \left[ \frac{\sigma_T }{3} \cdot 
\frac{ 2^{\frac{1+p}{2}} \; \pi^{-\frac{1}{2} - \frac{p}{4}} }
{ m_e^{1 + \frac{p}{2}} \; c^{\frac{p}{2}} \; q_e^{1 - \frac{p}{2}} } \right]^{\frac{2}{p+4}} \;   \gamma_{\rm dn}^{\frac{2(p-1)}{p+4}} \, f(p)^{\frac{2}{p+4}} \,
A^{\frac{6+p}{2(p+4)}} \, \epsilon_{\rm e}^{\frac{2}{p+4}} \,
\epsilon_B^{\frac{2+p}{2(p+4)}} \,
\left( \frac{E}{M} \right)^{\frac{1}{p+4}} \,
t^{-1}  \hspace{0.5cm} (k=2).
        \end{cases}
    \end{split}
\end{equation}
The peak/ self-absorption spectral luminosity is given as ($t<t_{\rm dec}$)
\begin{equation}
    \begin{split}
      L_{\rm \nu,sa} &\ = L_{\rm \nu,max} \left( \frac{\nu_{\rm sa}}{\nu_{\rm m}} \right)^{\frac{(1-p)}{2}} \;
      \\ &\ = K(k,p)\;   f(p)^{\frac{5}{p+4}} \,
\gamma_{\rm dn}^{\frac{5(p - 1)}{p+4}} \,
A^{\frac{2p+13}{2(p+4)}} \, \epsilon_{\rm e}^{\frac{5}{p+4}} \,
\epsilon_B^{\frac{2p+3}{2(p+4)}} \,
\left( \frac{E}{M} \right)^{\frac{(4-k)(2p+13)}{4(p+4)}} \,
t^{\frac{4p + 26 - 2kp -13k}{2(p+4)}} \\
&\ =  \begin{cases}
    &\          K(0,p) \; 
\gamma_{\rm dn}^{\frac{5(p - 1)}{p+4}} \,f(p)^{\frac{5}{p+4}} \,
\epsilon_{\rm e}^{\frac{5}{p+4}} \,
A^{\frac{2p+13}{2(p+4)}} \,
\epsilon_B^{\frac{2p+3}{2(p+4)}} \,
\left( \frac{E}{M} \right)^{\frac{2p+13}{p+4}} \,
t^{\frac{2p+13}{p+4}}     \hspace{0.6cm}  (k=0) ,    \\
    &\    K(2,p) \;  
\gamma_{\rm dn}^{\frac{5(p - 1)}{p+4}} \, f(p)^{\frac{5}{p+4}} \,
\epsilon_{\rm e}^{\frac{5}{p+4}} \,
A^{\frac{2p+13}{2(p+4)}} \,
\epsilon_B^{\frac{2p+3}{2(p+4)}} \,
\left( \frac{E}{M} \right)^{\frac{2p+13}{2(p+4)}} \,
t^{0}         \hspace{0.6cm}     (k=2). 
\end{cases}
    \end{split}
\end{equation}
where
\begin{eqnarray}
    &\ K(k,p) = 16\,
\left[3(3-k)\right]^{-\frac{5}{p+4}}\,
2^{\alpha(k,p)}\,
\pi^{\frac{4p+26}{4(p+4)}}\,
m_e^{\frac{1-p}{p+4}}\,
c^{\frac{2(1-p)}{p+4}}\,
q_e^{\frac{2p-7}{p+4}}\,
\sigma_T^{\frac{5}{p+4}} \\
    &\ \alpha(k,p) = \frac{8-3k}{4}
-
\frac{(1-p)(4+k)}{4(p+4)}
 \\
    &\ K(0,p) = 16 \cdot \,
2^{\frac{3p+7}{p+4}}\,
3^{-\frac{10}{p+4}}\,
\pi^{\frac{2p+13}{2(p+4)}}\,
m_e^{\frac{1-p}{p+4}}\,
c^{\frac{2(1-p)}{p+4}}\,
q_e^{\frac{2p-7}{p+4}}\,
\sigma_T^{\frac{5}{p+4}} \\ 
    &\ K(2,p) =16 \cdot \,
2^{\frac{4p+1}{2(p+4)}}\,
3^{-\frac{5}{p+4}}\,
\pi^{\frac{2p+13}{2(p+4)}}\,
m_e^{\frac{1-p}{p+4}}\,
c^{\frac{2(1-p)}{p+4}}\,
q_e^{\frac{2p-7}{p+4}}\,
\sigma_T^{\frac{5}{p+4}}
\end{eqnarray}

\section{Velocity stratification in SNR}\label{app:strat_SNR}
\setcounter{equation}{0}

We consider the density profile for homologously expanding supernova ejecta, commonly parameterized as \citep{Chevalier82,Chevalier94}
\begin{equation}
\rho(r, t) = A \, r^{-n} t^{\,n-3}
\label{eq:density_profile}
\end{equation}
where \( r \) is radius, \( t \) is time, and \( A \) is a constant. Homologous expansion implies \( r = v t \), where \( v \) is the constant velocity of a given mass shell. Substituting \( r = v t \) into the density profile gives:
\begin{equation}
\rho(v, t) = A \, (v t)^{-n} t^{\,n-3} = A \, v^{-n} t^{-3}.    
\end{equation}
Thus, at a fixed time, the density depends on velocity as \( \rho(v) \propto v^{-n} \). The mass per unit velocity interval is:
\begin{equation}
\frac{dm}{dv} = 4 \pi r^2 \rho(r,t) \frac{dr}{dv} = 4\pi A \, (v^2 t^2) \, (v^{-n} t^{-3}) \, t = 4\pi A \, v^{\,2-n},    
\end{equation}
where we used \( r = v t \), \( \frac{dr}{dv} = t \), and \( \rho(v,t) = A v^{-n} t^{-3} \)
The kinetic energy per unit velocity is:
\begin{equation}
   \frac{dE}{dv} \propto v^2 \cdot v^{\,2-n} = v^{\,4-n}. 
\end{equation}
The cumulative kinetic energy in material with velocity greater than \( v \) is:
\begin{equation}
    E(>v) = \int_v^{v_{\text{max}}} \frac{dE}{dv'} dv' \propto \int_v^{v_{\text{max}}} v'^{\,4-n} dv'.
\end{equation}
For \( n > 5 \), the integral converges and the upper limit \( v_{\text{max}}^{\,5-n} \to 0 \), yielding:
\begin{equation}
 E(>v) \approx \frac{2\pi A}{n-5} \; v^{-s} \hspace{1cm} \text{where $s = n-5$}. 
\end{equation}
For $ n \approx 9 $--$ 10 $, the stratification index $s = (4,5)$.

\section{ $\Sigma-D$  relation for resolved Galactic supernova remnants with $p>2$}\label{app:SigmaD}
\setcounter{equation}{0}
The spectral luminosity in PLS G of Spectrum 2 (for $\nu_{\rm G} > \nu_{\rm sa}$) can be written as
\begin{equation}
    L_{\rm \nu,G}  = L_{\rm \nu,sa} \left( \frac{\nu_{\rm G}}{\nu_{\rm sa}}\right)^{\frac{1-p}{2}} = L_{\rm \nu,max} \left( \frac{\nu_{\rm sa}}{\nu_{\rm m}}\right)^{\frac{1-p}{2}}  \left( \frac{\nu_{\rm G}}{\nu_{\rm sa}}\right)^{\frac{1-p}{2}} =  L_{\rm \nu,max} \left( \frac{\nu_{\rm G}}{\nu_{\rm m}}\right)^{\frac{1-p}{2}} 
\end{equation}
which shows that the spectral luminosity in PLS G can be obtained by extrapolation of the maximum spectral luminosity at $\nu_{\rm G}$. 
Post-deceleration the proper speed $u$ can be written as 
\begin{equation}
    u^2 = \frac{(3-k) E R^{k-3}}{4 \pi A c^2} \hspace{2cm} \text{for $R> R_{\rm dec}$}
\end{equation}
such that the internal energy density for $R > R_{\rm dec}$ can be obtained as
\begin{equation}
\begin{split}
      e_{\rm int} &\ = 4 \Gamma (\Gamma -1) A R^{-k} c^2 \approx 2 u^2 c^2 A R^{-k}  = 2 A c^2 R^{-k} \left( \frac{(3-k) E}{4 \pi A c^2} R^{k -3} \right)
       = \frac{(3-k) E R^{-3}}{2 \pi} = 2\left(1 - \frac{k}{3} \right) \left( \frac{E}{\frac{4\pi}{3} R^3} \right) 
\end{split}
\end{equation}
where we used $\Gamma - 1 \approx \frac{u^2}{2} \ll 1$. It can be seen that in the post-deceleration phase the internal energy density is  weakly dependent on the stratification of the ambient medium $k$ and independent of the normalization constant of the ambient density $A$. In fact, for a constant density medium it is equal to twice the average energy density within a sphere of radius $R$. 

In the DN regime post-deceleration, the maximum spectral luminosity post-deceleration can be written as 
\begin{equation}
\begin{split}
   L_{\rm \nu,max} &\ = \xi_{\rm e} N_{\rm e} P_{\rm \nu,max}= \left( \xi_{\rm e0} \frac{u^2}{u_{\rm DN}^2}\right)  \left[ \frac{4 \pi A R^{3-k}}{(3 - k) m_{\rm p}} \right] P_{\rm \nu,max}  = \frac{\sigma_{\rm T}}{3 q_{\rm e}} \left( \frac{m_{\rm e}}{m_{\rm p}} \right)  \left( \frac{\xi_{e0} E }{ u_{\rm DN}^2} \right) \sqrt{8 \pi \epsilon_{\rm B} e_{\rm int}} 
\end{split}
\end{equation}
such that the spectral luminosity in PLS  G can be obtained as  
\begin{equation}
\begin{split}
        L_{\rm \nu, G} &\ = L_{\rm \nu,max} \;  \nu_{\rm m}^{\frac{p-1}{2}} \; \nu_{\rm G}^{\frac{1-p}{2}} =  2^{\frac{3(p+1)}{4}} \left( \frac{3-k}{2 \pi} \right)^{\frac{p+1}{4}} \left( \frac{q_{\rm e}}{2 \pi m_{\rm e} c } \right)^{\frac{p-1}{2}} \frac{\gamma_{\rm dn}^{p-1}}{u_{\rm DN}^2} \left( \frac{\sigma_{\rm T}}{3 q_{\rm e}} \right) \left( \frac{m_{\rm e}}{m_{\rm p}} \right)  \left(8 \pi \epsilon_{\rm B} \right)^{\frac{p+1}{4}} \xi_{\rm e0} E^{\frac{p+5}{4}} \; \nu_{\rm G}^{\frac{1-p}{2}} \; D^{-\frac{3(p+1)}{4}}, 
\end{split}
\end{equation}
where $D = 2R$ is the diameter of the source.
The surface-brightness in PLS G can be written as 
\begin{equation}
\begin{split}
  \Sigma_{\rm G} &\  = \frac{F_{\rm \nu, G}}{\Omega} = \frac{L_{\rm G}}{4 \pi d^2} \frac{1}{\Omega} = \frac{L_{\rm G}}{4 \pi d^2} \frac{1}{\frac{\pi R^2}{d^2}}  = \frac{L_{\rm G}}{\pi^2 D^2}  = \mathcal{K} \; \epsilon_{\rm e} \epsilon_{\rm B}^{\frac{p+1}{4}} E^{\frac{p+5}{4}} D^{-\frac{(3p+11 )}{4}}   \hspace{2cm} \text{for $D > 2 R_{\rm dec}$}
\end{split}
\end{equation}
where the numerical constant $\mathcal{K}$ is given as  (for $p>2$)
\begin{equation}
    \mathcal{K} =  \left( \frac{\sigma_{\rm T}}{3 q_{\rm e}} \right) \left( \frac{q_{\rm e}}{2 \pi m_{\rm e} c} \right)^{\frac{p-1}{2}} f(p) \left(8 \pi  \right)^{\frac{p+1}{4}}  \frac{2^{\frac{3p-1}{4}}}{\pi^2} \left( \frac{3-k}{2 \pi} \right)^{\frac{p+1}{4}}  \gamma_{\rm dn}^{p-1} \nu_{\rm G}^{\frac{1-p}{2}}  
\end{equation}
Thus, the surface-brightness of SNR scales as  $\Sigma_{\rm G} \propto D^{-\frac{3p+11}{4}}$ and the normalization $\mathcal{K}$ depends weakly on the stratification index $k$.  Table \ref{tab:SigmaD} summarizes the typical diameters, surface brightnesses, and spectral indices ($\alpha$) of selected Galactic supernova remnants in the power-law segment (PLS G). The electron energy distribution power-law index is inferred from the spectral index using $p=2\alpha+1$, and the table includes only sources with $p>2$.

\begin{table}[h!]
\centering
\caption{Supernova remnants with spectral index $\alpha > 0.5$. Surface brightness $\Sigma$ is in units of $10^{-21}\ \mathrm{W\ m^{-2}\ Hz^{-1}\ sr^{-1}}$, diameter $D$ in pc, and $p = 2\alpha + 1$. The data is compiled from \citet{Pavlovic14} and \citet{Green25}. The common names of the remnants are given in brackets. }
\begin{tabular}{lcccc}
\hline\hline
Name & $\Sigma$ & $D$  & $\alpha$ & $p$ \\
 & ($10^{-21}\ \mathrm{W\ m^{-2}\ Hz^{-1}\ sr^{-1}}$) & (pc) & & \\
\hline
G4.5+6.8 (Kepler)         & 318.00   & 5.2   & 0.64 & 2.28 \\
G21.8$-$0.6 (Kes 69)      & 26.00    & 30.3  & 0.56 & 2.12 \\
G27.4+0.4 (Kes 73)        & 56.40    & 10.1  & 0.68 & 2.36 \\
G33.6+0.1 (Kes 79)        & 33.10    & 20.4  & 0.51 & 2.02 \\
G46.8$-$0.3 (HC30)        & 9.50     & 33.7  & 0.54 & 2.08 \\
G65.1+0.6                 & 0.18     & 175.6 & 0.61 & 2.22 \\
G78.2+2.1 ($\gamma$ Cygni)& 13.40    & 20.9  & 0.51 & 2.02 \\
G93.7$-$0.2 (CTB 104A)    & 1.50     & 34.9  & 0.65 & 2.30 \\
G96.0+2.0                 & 0.07     & 30.3  & 0.60 & 2.20 \\
G111.7$-$2.1 (Cassiopeia A)& 16400.00 & 4.8   & 0.77 & 2.54 \\
G116.9+0.2 (CTB 1)        & 1.04     & 15.8  & 0.57 & 2.14 \\
G120.1+1.4 (Tycho)        & 132.00   & 5.8   & 0.58 & 2.16 \\
G132.7+1.3 (HB3)          & 1.06     & 51.2  & 0.60 & 2.20 \\
G152.4$-$2.1               & 0.06     & 31.1  & 0.70 & 2.40 \\
G160.9+2.6 (HB9)          & 0.98     & 30.2  & 0.64 & 2.28 \\
G190.9$-$2.2               & 0.05     & 18.8  & 0.70 & 2.40 \\
G296.8$-$0.3 (1156-62)    & 4.80     & 46.7  & 0.60 & 2.20 \\
G315.4$-$2.3 (RCW 86)     & 4.20     & 28.1  & 0.60 & 2.20 \\
G327.4+0.4 (Kes 27)       & 10.20    & 29.6  & 0.60 & 2.20 \\
G327.6+14.6 (SN1006)      & 3.20     & 14.8  & 0.70 & 2.40 \\
G337.0$-$0.1 (CTB 33)     & 134.00   & 4.2   & 0.60 & 2.20 \\
G352.7$-$0.1               & 12.50    & 15.1  & 0.60 & 2.20 \\
\hline
\end{tabular}\label{tab:SigmaD}
\end{table}

\subsection{Correction factors from exact self-similar solutions}

The exact Blandford--McKee and Sedov--Taylor self-similar solutions introduce order-unity corrections to the fiducial post-deceleration blast-wave evolution. Following \cite{Petruk2000}, we write
\begin{equation}
u^2=C_u^2u_0^2,\qquad R=C_RR_0,\qquad D=C_RD_0,
\end{equation}
where
\begin{equation}
C_u=C_u^{\rm ST}=\left[\frac{16\pi}{(5-k)^2(3-k)\alpha_A}\right]^{1/2},\qquad C_R=C_u^{2/(5-k)}.
\end{equation}

Using Eq.~(\ref{eq:alpha_A}) for $\alpha_A$ and 
for $\gamma=5/3$, $C_u^{\rm ST}\simeq1.116,1.29,1.63$ for $k=0,1,2$, respectively.
The corrected proper speed can be written in terms of the physical radius as
\begin{equation}
u^2=C_u^2\frac{(3-k)ER^{k-3}}{4\pi Ac^2},\qquad R>R_{\rm dec},
\end{equation}
such that
\begin{equation}
e_{\rm int}\simeq2u^2c^2AR^{-k}=C_u^2\frac{(3-k)ER^{-3}}{2\pi}.
\end{equation}

In the DN regime, $\xi_{\rm e}\propto u^2$, $N_{\rm e}\propto R^{3-k}$, and $P_{\nu,\max}\propto\sqrt{e_{\rm int}}$, giving
\begin{equation}
L_{\nu,\max}\propto\left(C_u^2R^{k-3}\right)\left(R^{3-k}\right)\left(C_uR^{-3/2}\right)=C_u^3R^{-3/2}.
\end{equation}
Similarly, since $\nu_{\rm m}\propto\gamma_{\rm m}^2B$, with $\gamma_{\rm m} = \gamma_\mathrm{dn}$ and $B\propto\sqrt{e_{\rm int}}$,
\begin{equation}
\nu_{\rm m} \propto C_u R^{-3/2}.
\end{equation}
For $\nu_{\rm G}>\nu_{\rm sa}$, the spectral luminosity is therefore
\begin{equation}
L_{\nu,\rm G}\propto L_{\nu,\max}\nu_{\rm m}^{(p-1)/2}\nu_{\rm G}^{(1-p)/2}\propto C_u^{(p+5)/2}R^{-[3(p+1)]/4}\nu_{\rm G}^{(1-p)/2}.
\end{equation}
Using $R=C_u^{2/(5-k)}R_0$ and $D_0=2R_0$, we obtain
\begin{equation}
L_{\nu,\rm G}\propto\left(C_u^{\rm ST}\right)^{\frac{(2-k)p+22-5k}{2(5-k)}}D_0^{-\frac{3(p+1)}{4}}.
\end{equation}
The corresponding surface brightness is
\begin{equation}
\Sigma_{\rm G}=\frac{L_{\nu,\rm G}}{\pi^2D^2},
\end{equation}
and hence
\begin{equation}
\Sigma_{\rm G}\propto C_u^{(p+5)/2}R^{[-(3p+11)]/4}.
\end{equation}
Substituting $R=C_u^{2/(5-k)}R_0$ gives
\begin{equation}
\Sigma_{\rm G}\propto\left(C_u^{\rm ST}\right)^{\frac{(2-k)p+14-5k}{2(5-k)}}D_0^{-\frac{(3p+11)}{4}}.
\end{equation}

Thus, the self-similar correction modifies only the normalization of the $L_{\nu,\rm G}$--$D_0$ and $\Sigma_{\rm G}$--$D_0$ relations, while retaining their power-law slopes. For $k=0$ and $p=2.3$, the self-similar correction factor $C_u$ increases the surface brightness by a factor of $\sim 1.2$ which is again an order of unity effect.

\end{document}